\documentclass[12pt]{article}
\usepackage{jheppub}

\usepackage{amsmath}
\usepackage{dsfont}
\usepackage{amsfonts}
\usepackage{amssymb}
\usepackage[utf8]{inputenc}
\usepackage{color}
\usepackage{graphicx}

\DeclareMathOperator{\Tr}{Tr}
\renewcommand{\Re}{\operatorname{Re}}
\renewcommand{\Im}{\operatorname{Im}}

\title{Integrable Kondo problems}

\author[a]{Davide Gaiotto,}
\author[a]{Ji Hoon Lee,}
\author[a]{Jingxiang Wu}

\affiliation[a]{Perimeter Institute for Theoretical Physics, Waterloo, Ontario, Canada N2L 2Y5}

\abstract{We discuss the integrability and wall-crossing properties of Kondo problems, where an 1d impurity is coupled to a 2d chiral CFT and triggers a defect RG flow. 
We review several new and old examples inspired by constructions in four-dimensional Chern-Simons theory and by affine Gaudin models.}
\begin{document}
\maketitle
\section{Introduction and Motivations}
The objective of this paper is to study the integrability properties of Kondo problems in two dimensional CFTs.
We use the moniker ``Kondo problem'' to denote a broad class of problems where some local impurity is coupled to a 2d CFTs
to produce a line defect which breaks conformal symmetry and has a non-trivial RG flow. The study of this subject was one of the first applications of RG flow methods and 
has been a source of inspiration for several theoretical developments \cite{Kondo:1964nea,Wilson:1974mb,Andrei:1980fv,Cardy:1989ir,Saleur:1998hq,Saleur:2000gp,Affleck:1990by,Affleck:1990iv,Affleck:1991yq,Affleck:1995ge,Nakagawa_2018,tsvelick1985exact,PhysRevLett.52.364,Fendley:1995kj,doi:10.1080/00018738300101581,Andrei:1982cr, PhysRevB.46.10812}.
See \cite{Bachas:2004sy} for several more references and a good introduction to the problems discussed in this paper. 

We use the moniker ``Kondo problem'' to denote a situation where the local impurity coupling only involves chiral local operators in the 2d CFT, so that the resulting defect is transparent to the anti-chiral degrees of freedom. A special property of these defects is translation invariance in the direction transverse to the defect \cite{Konik:1997gx}. In particular, a Kondo defect wrapping the space circle will commute with the Hamiltonian 
and define a continuous family of conserved charges, labelled by the RG flow scale.

A surprising observation is that in many important examples these conserved charges will commute with each other, revealing a hidden integrability structure in the underlying CFT \cite{Bazhanov:1994ft,Bazhanov:1996dr,Runkel:2007wd}. In many situations, integrability comes together with rich extra structures such as Yangian symmetry, Hirota recursion relations, Thermodynamic Bethe Ansatz equations and more.
The emergence of these structures is accompanied by other unexpected relations such as the ODE/IM correspondence \cite{Dorey:2007zx}, which identifies the expectation values of transfer matrices with the transport data of certain ordinary differential equations.

Recently, a four-dimensional version of Chern-Simons theory has emerged as a general organizing principle for many integrable problems \cite{Costello:2013zra,Costello:2013sla,Costello:2017dso,Costello:2018gyb,Costello:2019tri,Costello:2018txb,Vicedo:2019dej,Delduc:2019whp}, including integrable field theories. This paper is part of a multi-pronged exploration of that construction: we collect here 2d CFT calculations and novel applications of the 
ODE/IM correspondence which will be used in separate upcoming works making contact with four-dimensional Chern-Simons theory and the affine Gauden models discussed e.g. in \cite{Delduc:2018hty}. 

As a physical bonus, we also describe how to combine the ODE/IM tools with a careful WKB analysis \cite{VOROS1983,GMN:2009hg,Gaiotto:2012rg} to derive the IR RG flow endpoints 
of a large variety of Kondo problems. 

\subsection{The $\mathfrak{su}(2)_1$ Kondo problem}

The prototypical example 
of a Kondo problem involves a single qubit impurity coupled to a doublet of chiral complex fermions by an $SU(2)$ invariant local coupling 
\begin{equation}
g \,\vec S \cdot \vec J(t,0) \label{eq:Kondocoupling}
\end{equation}
Here $\vec S$ are the Pauli matrices acting on the qubit impurity and 
\begin{equation}
\vec J(t,0) \equiv \psi^\dagger(t,0) \vec \sigma \psi(t,0)
\end{equation}
are the $\mathfrak{su}(2)_1$ WZW currents built out of the complex fermions. 

This model provides one of the simplest, best studied examples of (defect) RG flow. The perturbative coupling $g$ is classically marginal and is marginally relevant for $g>0$. 
The deformation thus defines a UV-complete line defect $L_{\frac12}[\theta]$ in the chiral CFT, equipped with a dynamically generated non-perturbative scale $\mu \equiv e^\theta$ which breaks scale invariance. 
The IR endpoint of the RG flow is conjecturally known \cite{Affleck:1995ge,Bachas:2004sy}: it is the non-trivial topological defect ${\cal L}_{\frac12}$ whose Cardy label is 
the spin $\frac12$ primary in the WZW model. 

We can define operators $\hat T[\theta]$ by Wick-rotating the line defects $L[\theta]$ to wrap a space circle of unit radius. 
Almost by construction, the $\hat T[\theta]$ are operators acting on the Hilbert space of the CFT which commute with the Hamiltonian. 
They are renormalized path-ordered exponentials of the chiral currents $\vec J$, which were computed at the first few orders of perturbation theory in 
\cite{Bachas:2004sy}. 

The basic integrability claim is that they commute with each other:
\begin{equation}
\left[\hat T[\theta], \hat T[\theta'] \right] =0
\end{equation}

We actually expect a stronger statement to be true. Consider Kondo defects of spin $j$ in the same CFT. These are defined in the same manner as the 
basic Kondo defect, except that $\vec S$ are taken to be $\mathfrak{su}(2)$ generators acting on a spin $j$ irreducible representation. Global $SU(2)$ invariance insures that 
renormalization can only affect the overall coupling $g$, which is is again marginally relevant (when positive) and gives rise to a family of line defects $L_{j}[\theta]$. 
Conjecturally, the RG flow ends on IR free line defects, defined as spin $j - \frac12$ Kondo defects with negative coupling.

Define ``transfer matrix'' operators $\hat T_{j}[\theta]$ as above. Then we claim that 
\begin{equation}
\left[\hat T_{j}[\theta], \hat T_{j'}[\theta'] \right] =0
\end{equation}
and that in an appropriate renormalization scheme (See Appendix \ref{app:HirotaSU2},) a Hirota fusion-like relation holds true:\cite{KLUMPER1992304,baxter2016exactly} (See also the review article \cite{Kuniba:2010ir} and references therein for a more modern exposition.) 
\begin{equation}
\hat T_{j}\left[\theta+ \frac{i \pi}{2}\right]\hat T_{j}\left[\theta- \frac{i \pi}{2}\right] =1 + \hat T_{j-\frac12}[\theta]\hat T_{j+\frac12}[\theta]
\end{equation}
Combined with general physical considerations, the Hirota relations lead to a TBA framework to compute the $\hat T_j[\theta]$ eigenvalues,
which is the ``conformal limit'' of the one for the chiral Gross-Neveu model (See e.g. \cite{vanTongeren:2016hhc} for a review of the TBA framework). 

The Hirota relations also lead us to simple ODE/IM relation\footnote{The vacuum module at $k=1$ and other related ODEs have been proposed and studied in \cite{Bazhanov_2003,Lukyanov_2004,Lukyanov_2004_1,Lukyanov_2006,Lukyanov_2007,Lukyanov_2007_1,Lukyanov_2013}}: 
the expectation values of $\hat T_{j}[\theta]$ on a spin $l$ primary state match the transport data for the second order differential equation 
\begin{equation}
\partial_x^2 \psi(x) = \left[e^{ 2 \theta} (1 + g x) e^{2x}  +\frac{l(l+1) g^2}{(1 + g x)^2}\right] \psi(x) \label{eq:ODEexample}
\end{equation}
for $\mathfrak{su}(2)_1$ and a simple modification for twisted sectors and higher WZW levels. We discuss this model in detail in Section \ref{sec:wzw}, 
leaving a full description of the excited state ODE/IM for a separate publication \cite{Gaiotto:2020dhf}.

\subsection{Multichannel $\mathfrak{su}(2)$ Kondo problem}
We can generalize the basic Kondo problem by coupling the impurity to multiple copies of the chiral fermion theory, with a coupling 
\begin{equation}
 \,\vec S \cdot \sum_i g_i \vec J^{(i)}(t,0)
\end{equation}
involving $n$ decoupled $\mathfrak{su}(2)_{1}$ WZW currents built out of the complex fermions
\footnote{If any $k$ couplings coincide, say $g_i = g_{i+1} =\cdots = g_{i+k-1}$, the defect only couples to the diagonal $\mathfrak{su}(2)_k$ WZW current
$J^{(i)} + J^{(i+1)}+ \cdots  J^{(i+k-1)}$. That means this setup includes as a special case the coupling of an impurity to any collection of $\mathfrak{su}(2)_{k_i}$ WZW currents.}.

The RG flow is now potentially much richer, as it takes place in an $n$-dimensional space of couplings. Four-dimensional Chern-Simons considerations suggest an important simplifying feature:  
in an appropriate RG scheme, the RG flow should preserve the differences $g_i^{-1}-g_j^{-1}$ between the inverse couplings. If we set, say, 
\begin{equation}
g_i^{-1} = g^{-1} + z_i,
\end{equation}  
say with $\sum_i z_i=0$, then the RG flow should only change the overall coupling $g$ \footnote{This is compatible with the fact that the RG flow must fix the loci $g_i = g_j$, see the previous footnote}
and the $z_i$ should label RG flow trajectories. 
We will test this conjecture at the first few orders in perturbation theory. 

Furthermore, we conjecture the following integrability relation: 
\begin{equation}
\left[\hat T[\theta;z_i], \hat T[\theta';z_i] \right] =0
\end{equation}
which should only hold for line defects in the {\it same} RG flow trajectory. 

We also conjecture that with an appropriate labelling of RG trajectories, higher spin impurities will give other commuting transfer matrices $\hat T_{j}[\theta;z_i]$, still satisfying Hirota relations. 
We will formulate an ODE/IM statement involving the transport data for the second order differential equation 
\begin{equation}
\partial_x^2 \psi(x) = \left[e^{ 2 \theta} e^{2x} \prod_i (1 + g_i x)   +\sum_i \frac{l_i(l_i+1) g_i^2}{(1 + g_i x)^2} + \sum_{i<j}\frac{2 l_i l_j g_i g_j}{(1 + g_i x)(1 + g_j x)}  \right] \psi(x) \label{eq:ODEexample2}
\end{equation}
for $\prod_i \mathfrak{su}(2)_1$.

We discuss this model in detail in Section \ref{sec:multichannelSU2}.

\subsection{Generalizations and future directions}
The ideas of this paper can be extended to a wide variety of integrable Kondo problems associated to 4d Chern-Simons theory. 
In Section \ref{sec:coset} we look briefly at another well studied example \cite{Dorey:2007ti}, involving integrable deformations of topological line defects in 
$\frac{\prod_i \mathfrak{su}(2)_{k_i} }{\mathfrak{su}(2)_{\sum_i k_i} }$ coset models, such as Virasoro minimal models. 

The corresponding conjectural ODE is a polynomial potential: 
\begin{equation}
\partial_x^2 \psi(x) = \left[e^{ 2 \theta} \prod_i (x-z_i)^{k_i}   +\sum_i \frac{l_i(l_i+1)}{(x-z_i)^2} + \sum_{i<j}\frac{2 l_i l_j}{(x-z_i)(x-z_j)}  \right] \psi(x) \label{eq:ODEcoset}
\end{equation}
with RG flow acting as a common rescaling of the $z_i$. 

We also briefly comment on further extensions along the direction of the integrable CFTs discussed in \cite{Costello:2019tri}
and to transfer matrices for integrable deformations of these CFTs. For recent studies of polynomial potentials along the directions of this work, see e.g. \cite{ito2019tba}.

For simplicity, we only discuss models associated to the $SU(2)$ group. Broad generalizations to other groups $G$ are possible and mostly straightforward. 
The biggest subtlety is that the space of endomorphisms of an irreducible representation may contain multiple copies of the adjoint representation, so that 
the RG flow may deform $\vec S$ away from the generators of the Lie algebra $\mathfrak{g}$ for $G$ even if we impose global $G$ invariance. 

4d Chern-Simons theory constructions predict the existence of a specific integrable RG trajectory in the space of couplings for any irrep
which can be extended to a representation of the Yangian for $\mathfrak{g}$. It would be very interesting to see how 
such a restriction arises in the Kondo problem, at least perturbatively.

We plan to come back to these problems in future work. 

\section{Generalities of chiral defects} \label{sec:chiral}

A ``chiral line defect'' in a 2d CFT is a line defect $L$ which is invariant under translations 
along the direction of the defect and transparent to the anti-holomorphic part $\bar T$ of the stress tensor.
All line defects in a chiral CFT are obviously chiral. Deformations of topological line defects by chiral operators also give rise to 
chiral defects. 

A chiral defect does {\it not} have to preserve conformal symmetry or scale invariance. Indeed, a conformal invariant 
chiral line defect would be actually topological. See Appendix \ref{app:line} for a review of relevant materials.

We are interested in chiral line defects which are not topological, and thus must depend on some intrinsic scale $\mu$. 
We will write $\mu = \mu_0 e^\theta$ and label the corresponding RG flow family of line defects as $L[\theta]$. Although 
$\theta$ starts its life as a real, positive parameter, it makes sense to analytically continue $L[\theta]$ 
to general complex $\theta$. This deformation breaks reflection positivity, but will unlock important features. 

A useful perspective\footnote{Another nice perspective is studied in \cite{Nakagawa_2018}. Essentially, complexifying $\theta$ is equivalent to complexifying the Kondo coupling $g$ in \eqref{eq:Kondocoupling}. The resulting non-Hermitian extension of the Kondo problem has been studied in \cite{Nakagawa_2018} to model the inelastic scattering and atom losses, where some neat physical interpretations of the wall-crossing behaviors we discuss in Section \ref{sec:IRWKBanalysis} are given. We thank 
Masaya Nakagawa for the correspondence on this point. } on the analytic continuation is that infinitesimal variations of $\theta$ are implemented by an exactly marginal 
local operator: the defect stress tensor $t^{00}$. This operator enters the local energy conservation law for 
the defect
\begin{equation}
T^{01} |_{x^1=0^+}- T^{01} |_{x^1=0^-} = \partial_{x^0} t^{00} \label{eq:wardidentity}
\end{equation}
and measures the local violation of scale invariance for a chiral line defect placed along $x^0$. 

Analytic continuation in $\theta$ is thus achieved infinitesimally by adding $t^{00}$ to the defect action with a complex coefficient. 
It is important to observe that the line defect $L[\theta]$ is generically {\it not} periodic under shifts $\theta \to \theta + 2 i \pi$. 
It is an entire function of the $\theta$ plane. 

A chiral line defect can be freely translated in a direction perpendicular to the defect. 
Such a translation is implemented infinitesimally by
\begin{equation}
\frac{1}{2 i} \int dx^0 \left[T - \bar T\right]|_{x^1=0^+} - \left[T - \bar T\right]|_{x^1=0^-} 
\end{equation} 
The argument inside the integral equals
\begin{equation}
\left[T + \bar T\right]|_{x^1=0^+} - \left[T + \bar T\right]|_{x^1=0^-} 
\end{equation} 
which is proportional to the total derivative $\partial_{x^0} t^{00}$ and integrates by parts to zero for a rigid rotation. \footnote{We also see that a more general 
deformation of the defect will be possible at the price of introducing a position-dependent $\theta$ along the defect. This is analogous to the ``framing anomaly'' encountered in \cite{Costello:2017dso}, which plays an important role in understanding the shifts of $\theta$ which occur in Hirota-like relations.}

A chiral line defect wrapping a space circle gives rise to a conserved charge, as it commutes with time translations. 
We denote the corresponding operator on the Hilbert space of the theory as $\hat T_L[\theta]$.

Before moving on, we would like to clarify a notational issue. The vev of the line defect depends also on the radius of the space circle. As $\mu$ and $R^{-1}$ are the only energy scales in the problem, 
the operator $\hat T_L[\theta]$ can only depend on the combination $2 \pi R \mu \equiv 2 \pi R \mu_0 e^\theta$. Without loss of generality, 
we can thus do calculations either at fixed $R$ or at fixed $\mu$. In most of the expressions below we will do the former: fix the radius to a convenient value $2 \pi R=\mu_0^{-1}=1$ 
and write answers as a function of $e^{\theta}$. However in explicit calculations, for example in Appendix \ref{app:Isingloopcal} and \ref{app:wzwKondoPert}, it is often useful to keep $R$ generic
and set $\theta=0$. The $\theta$ dependence can be easily restored.

In perturbative situations, where the line defect is labelled by some renormalized coupling(s) $g$, one can also absorb the $\theta$ dependence into an effective coupling $g_{\mathrm{eff}}(\theta)$, so that 
\begin{equation}
\langle L_g \rangle_{e^\theta R} \equiv \langle L_g[\theta] \rangle_{R} \equiv \langle L_{g_{\mathrm{eff}}(\theta)} \rangle_{R} \label{eq:exchangeMuR}
\end{equation}
For example, in the WZW case below we define the renormalized coupling through a dimensionally transmuted scale $\mu_0 = g^{\frac{k}{2}}e^{-\frac{1}{g}}$
and the operator $\hat T_g[\theta]$ is a function of 
\begin{equation}
2 \pi R g^{\frac{k}{2}}e^{-\frac{1}{g}} e^\theta \equiv 2 \pi R g_{\mathrm{eff}}(\theta)^{\frac{k}{2}}e^{-\frac{1}{g_{\mathrm{eff}}(\theta)}}.
\end{equation}

The ground state of the theory is automatically an eigenstate of $\hat T_L[\theta]$ (not to be confused with the stress tensor!),
with an eigenvalue we can denote as $T_L[\theta]$. This can be identified with the (exponential of the) ``g-function'' of the defect \cite{AffleckLudwig}. 
For real $\theta$, when the line defect is unitary/reflection positive, $T_L[\theta]$ varies monotonically along the RG flow \cite{Friedan:2003yc}. 

More generally, $\hat T_L[\theta]$ only mixes states in the CFT within the same chiral algebra module and 
with the same $L_0$ eigenvalues. The corresponding $\hat T_L[\theta]$ eigenvalues will be also studied below.

\subsection{RG flow of chiral defects and wall-crossing}
In the far IR, a chiral line defects should flow to a conformal invariant chiral line defects and thus become topological. 
In a given renormalization scheme, the IR topological defect will be dressed by a constant local counterterm, the 
ground state energy $E_L$ of the line defect. In an Euclidean setting, that appears as a prefactor $e^{- 2 \pi R e^\theta E_L}$ in 
front of $\hat T_L[\theta]$. In particular, we learn the asymptotic behaviour of $\hat T_L[\theta]$ for large real positive $\theta$:
\begin{equation}
T_L[\theta] \sim e^{- 2 \pi R E_L e^{\theta}} g_{IR}
\end{equation}
Here we denote as $g_{IR}$ the (exponential of the) $g$-function of the topological line defect in the IR. \footnote{More precisely, $g$-function is originally\cite{AffleckLudwig} defined to a boundary state $|B\rangle$ in $\mathrm{CFT}\otimes\overline{\mathrm{CFT}}$ via the folding trick. And $\log g$ is referred to as the boundary entropy of $|B\rangle$. However the notion is naturally extended to defect lines. See, for example \cite{Chang:2018iay,Kormos:2009sk} for related discussions.}  

The IR behaviour of the line defect $L[\theta]$ is obviously invariant under real shifts of $\theta$. As we explore the imaginary $\theta$ direction, 
though, or as we vary other continuous parameters, the IR behaviour may jump at walls of first order phase transitions. 
At the level of the vevs $T_L[\theta]$, two exponential contributions will exchange dominance at these walls. 
This can happen when $(E_L - E'_L) e^{\theta}$ is purely imaginary, which typically means that the imaginary part of $\theta$ is 
$(n+\frac12)  \pi$ with integer $n$. 

Such wall-crossing behaviour is not only possible. It is necessary in order to have some interesting physics. 
Indeed, an entire function $T_L[\theta]$ with uniform asymptotics of the form above for large positive real part of $\theta$ and arbitrary 
imaginary part, and reasonable behaviour at negative real $\theta$, would have to essentially 
coincide with the far IR answer $e^{-2 \pi R E_L e^{\theta}} g_{IR}$. 

Interesting line defects will instead have a distinct asymptotic behavior
\begin{equation}
T_L[\theta] \sim e^{-2 \pi R E^{(n)}_L e^{\theta}} g^{(n)}_{IR}
\end{equation}
in each strip 
\begin{equation}
(n-\frac12)  \pi<\mathrm{Im} \, \theta <(n+\frac12)  \pi
\end{equation}

We will see some concrete examples momentarily.

\section{Chiral line defects in the Ising model} \label{sec:Ising}
Chiral line defects in Virasoro minimal models are a canonical example of integrable line defects \cite{Bazhanov:1994ft}. The Ising model is a particularly nice case,
because the Kondo problem is exactly solvable in the free fermion description of the model \cite{Casini:2016fgb}. We will discuss it in this section. 

The integrable minimal model Kondo problems involve relevant deformations of topological line defects which support chiral local operators. 
The solvable Ising model examples involve the deformation by the chiral local operator $\psi(z)$ which is the free fermion in disguise.

Recall that the Ising model has three irreducible topological line defects \cite{Frohlich:2004ef,Frohlich:2006ch,Chang:2018iay}: 
\begin{itemize}
\item The trivial line defect $I$, with Cardy label $1$ and $g_I =\langle 0|I|0\rangle = 1$ 
\item The $\mathbb{Z}_2$ symmetry defect $P$, with Cardy label $\epsilon$ and $g_\epsilon =\langle 0|P|0\rangle = 1$
\item The Kramers-Wannier duality defect $S$, with Cardy label $\sigma$ and $g_\sigma =\langle 0|S|0\rangle  =\sqrt{2}$
\end{itemize}
where we also list their $g$-values $g(L_k) = S_{k0}/S_{00}$ and the vacuum expectation values. They form an Ising fusion category, with $P \times P = I$, $S \times S = I + P$, $S \times P = S$. They are Verlinde lines with the action on the primary state given as follows,
\begin{equation}
\hat{L}_{k}\left|\phi_{i}\right\rangle=\frac{S_{k i}}{S_{0 i}}\left|\phi_{i}\right\rangle
\end{equation}
which reads explicitly
\begin{equation}
\begin{array}{ccc}
\hat{P}|1\rangle = |1\rangle, & \hat{P}|\epsilon\rangle = |\epsilon\rangle, & \hat{P}|\sigma\rangle = -|\sigma\rangle \\
\hat{S}|1\rangle = \sqrt{2}|1\rangle, & \hat{S}|\epsilon\rangle = -\sqrt{2}|\epsilon\rangle, & \hat{S}|\sigma\rangle =0
\end{array}\label{eq:VerlindeAction}
\end{equation}
By evaluating the partition function twisted by topological line defects, one can find the Hilbert space of defect fields living on a topological line defect with Kac label $k$ \cite{Kormos:2009sk,Petkova:2000ip}.
\begin{equation}
\mathcal{H}_{k}^{\mathrm{defect}}=\bigoplus_{i, j }\left(R_{i} \otimes \bar{R}_{j}\right)^{\oplus \sum_{x} N_{i j}^{x} N_{k k}^{x}}
\end{equation}
where $R_i$($\bar{R}_j$) are irreps of Virasoro $\mathrm{Vir}$($\mathrm{\overline{Vir}}$) and $N_{ij}^k$ are the fusion rule coefficients. In particular, the only irreducible line defect which supports $\psi(z)$ as a local operator is $S$.\footnote{One can also consider the superposition 
$I + P$, where $\psi$ appears as a boundary-changing operator. The corresponding RG flow can be obtained from the RG flow for $S$ by fusion with a second, topological $S$ line.}
We thus define a Kondo problem by deforming $S$ by the relevant deformation $\psi$ \cite{Kormos:2009sk}:
\begin{equation}
g \int \psi(x^0,0) dx^0
\end{equation} 
The deformation is clearly transparent to the anti-chiral stress tensor. The result is a chiral line defect $L_S$. 

As $\psi$ has dimension $\frac12$, in natural renormalization schemes the RG flow will simply rescale $g$ by $e^{\frac{\theta}{2}}$. We can 
simply set $g=1$ and parameterize the RG flow by $\theta$. If needed, we can restore $g$ by a shift of $\theta$. 

The line defect $L_S[\theta]$ should coincide with $L_S[\theta+ 4 \pi i n]$ up to the only available counterterm, which is a constant. \footnote{Using RCFT technology one can also see that $L_S[\theta+ 2 \pi i]$ should coincide with $L_S[\theta] \times P$ up to a constant counterterm.}

Due to the $g$ theorem \cite{AffleckLudwig}, the RG flow can only end on topological line defects with a lower $g$ function than $S$. The only possibilities are $I$ and $P$. The sign of the coupling $g$ is expected to determine if the flow ends on $I$ or $P$ \cite{Kormos:2009sk,fendley2009boundary}. Up to some convention ambiguities, we can say that a positive deformation will flow to $I$. 
%here

We define the operator $\hat T_S[\theta]$ by wrapping the deformed line defect $L_S[\theta]$ along a space circle.
As $\hat T_S[\theta]$ commutes with the Hamiltonian, the vacuum is an eigenvector of $\hat T_S[\theta]$. The expectation value of $\hat T_S[\theta]$ on the vacuum is of particular interest. We will denote it as
\begin{equation}
T_S(\theta) \equiv \langle 0|\hat{T}_S[\theta] |0\rangle_{2\pi R=\mu_0^{-1}}
\end{equation}

It is instructive to start with a perturbative UV calculation. We can set $\theta =0$ and restore it later on, but keep the radius $R$ generic.
The leading order answer is the quantum dimension $\sqrt{2}$ of $S$. 
The first subleading correction appears at second order, as the vev of $\psi$ vanishes. The  $\psi(s) \psi(s')$ two-point function on 
the cylinder with vacuum states at the two ends is 
\begin{equation}
\frac{1}{2 R \sin \frac{s-s'}{2 R}}
\end{equation}
As a consequence, the leading perturbative correction to the vev has a log divergence
\begin{equation}
2 \pi R g^2 \log \cot \frac{\epsilon}{4 R}
\end{equation}
which requires a constant counterterm $2 \pi R g^2 \log \epsilon$ in a minimal subtraction scheme. 

In a more general renormalization scheme, we have
\begin{equation}
\langle L_S \rangle_R = \sqrt{2} \left(1 + 2 \pi R g^2 \log (2 \pi R) + 2 \pi R g^2 c + \cdots \right)
\end{equation}
We can adjust $c$ to that the answer is a function of $2 \pi R g^2$ only. Recall that the only renormalization ambiguity in the definition of $L_S$ is a constant counterterm $\delta \int dx_0$, which rescales the above correlator by $e^{2 \pi R \delta}$.

Restoring $\theta$ and setting $2 \pi R g^2=1$, we write 
\begin{equation}
T_S[\theta] = \sqrt{2} \left(1 + \theta e^\theta + c' e^\theta + \cdots \right)
\end{equation}
for some arbitrary $c'$. 

The full answer for $T_S[\theta]$ can be obtained by mapping the problem to the free fermion realization of the Ising model. 
Recall that the Ising model is obtained as the GSO projection of a free fermion (spin-)CFT, inverting the Jordan-Wigner transformation. 
See \cite{Karch:2019lnn} for a recent review. The simplest way to realize the S defect is to stack the free fermion theory with an $\mathrm{Arf}$ 
topological field theory (aka Majorana chain) defined on half of space-time and then GSO project the combined system. 

The $\mathrm{Arf}$ theory is trivial on the bulk, but supports a single Majorana mode $\gamma$ at the boundary. The bilinear combination 
$\gamma \psi$ survives the GSO projection and becomes the ``$\psi$'' operator on the S defect. We employ this description for a straightforward one loop calculation of $T_S[\theta]$, reviewed in Appendix \ref{app:Isingloopcal}. 

The unregularized one loop determinant would give $\prod_{n\geq 0} (n+\frac12 +2 \pi R g^2)$. Restoring $\theta$ and setting $2 \pi R g^2=1$, we write the regularized expression as 
\begin{equation}
T_S(\theta) = \frac{\sqrt{2 \pi} e^{\theta e^{\theta}-e^{\theta}}}{\Gamma(\frac12 +  e^{\theta})} \label{eq:IsingTSFull}
\end{equation}
This interpolates nicely between the perturbative answer in the UV for $e^\theta \ll 1$ 
and an infrared expansion 
\begin{equation}
T_S(\theta) \sim 1 + \frac{1}{24} e^{- \theta} + \cdots
\end{equation}
valid for $e^\theta \gg 1$ as long as the phase of $\theta$ lies strictly between $-\pi$ and $\pi$. \footnote{We choose our $c'$ counter-term in such a way that the 
IR ground state energy is $0$. }

This agrees with the expectation that $L_S$ flows to $I$ or $P$, which both have vev $1$ acting on the vacuum.
The leading correction in the IR is a deformation of $I$ or $P$ by the least irrelevant operator, i.e. the stress tensor. 
The coefficient $\frac{1}{24}$ is $-2$ times the vacuum energy, and we will now test the statement further for excited states. 
According to integrability lore \cite{Bazhanov:1994ft}, higher order terms in the IR expansion of $\hat T_S(\theta)$ should correspond to the higher ``quantum KdV'' charges 
hidden in the Ising CFT. 

If we compute the vev of $L_S$ in a different state $|n_i \rangle$, obtained from the vacuum by acting with chiral fermion momentum modes of momentum 
$n_i + \frac12$ with $n_i \geq 0$ (and any anti-chiral fermions) we obtain a similar one-loop determinant but with some signs switched, leading to 
\begin{equation}
\hat T_S(\theta) |n_i \rangle =\prod_i \frac{e^\theta-n_i -\frac12}{e^\theta+n_i +\frac12} \frac{\sqrt{2 \pi} e^{\theta e^{\theta}-e^{\theta}}}{\Gamma(\frac12 +  e^{\theta})} |n_i \rangle  
\end{equation}
In the UV, the correction factor goes as  
\begin{equation}
\prod_i \left(-1 + \frac{2}{n_i +\frac12}  e^\theta + \cdots \right)
\end{equation}
The leading term gives the sign of the action of $S$ on the vacuum module/$\epsilon$ modules, which is $\sqrt{2}$ /$-\sqrt{2}$. This agrees with  \eqref{eq:VerlindeAction}.

In the IR, we have 
\begin{equation}
\langle n_i| \hat T_S(\theta) |n_i \rangle \sim 1 + \frac{1}{24} e^{- \theta} - \sum_i (2 n_i + 1)  e^{- \theta}  \cdots 
\end{equation}
which shows clearly that the leading correction to the identity line defect is the integral of the stress tensor
along the defect, giving a $-2 L_0 e^{- \theta}$. 

Similarly, in the Ramond ground state/$\sigma$ module for the Ising model we get the regularized determinant 
\begin{equation}
\langle\sigma|\hat T_S(\theta) |\sigma\rangle \equiv T_{S;\sigma}[\theta] = \frac{\sqrt{2 \pi e^{\theta}} e^{\theta e^{\theta}-e^{\theta}}}{\Gamma(1+e^{\theta})} 
\end{equation}

In the UV this goes as 
\begin{equation}
T_{S;\sigma}[\theta] \sim \sqrt{2 \pi e^{\theta}} + \cdots
\end{equation}
which arises at the leading order from a one-point function of $\psi$. Note that $T_{S;\sigma}= 0 = \langle\sigma|S|\sigma\rangle$ at the UV fixed point, as expected, since duality line $S$ annihilate $|\sigma\rangle$. \eqref{eq:VerlindeAction}. In the IR, we have 
\begin{equation}
T_{S;\sigma}[\theta] \sim 1 - \frac{1}{12} e^{- \theta} + \cdots
\end{equation}
which agrees again at leading order with $1-2 L_0 e^{- \theta}$. For excited states, we modify that to 
\begin{equation}
\langle\sigma;n_i|\hat T_S(\theta) |\sigma;n_i\rangle = \prod_i \frac{e^\theta-n_i}{e^\theta+n_i} \frac{\sqrt{2 \pi e^{\theta}} e^{\theta e^{\theta}-e^{\theta}}}{\Gamma(1+e^{\theta})} 
\end{equation}

%Although we set the in coupling of $L_S$ to $1$, for comparison with more general examples it is useful to restore 
%a coupling dependence:
%\begin{equation}
%T_S(g;\theta) = \frac{\sqrt{2 \pi} e^{\theta g^2 e^{\theta}- g^2 e^{\theta}}}{\Gamma(\frac12 + g^2 e^{\theta})} 
%\end{equation}
%The power series expansion in $g$ gives the UV perturbative expansion, while in the IR we find the renormalized ground-state energy 
%\begin{equation}
%T_S(g;\theta) \sim e^{- e^\theta g^2 \log g^2}
%\end{equation}

\subsection{Fusion relations, TBA, Hirota and full IR behaviour}
Before the deformation, the $S$ line defects have a nice fusion relation:
\begin{equation}
S \times S = I+P
\end{equation}
with $P$ being the $Z_2$ symmetry line of the Ising model. 

After the deformation, we claim that the fusion is deformed to something like 
\begin{equation}
L_S\left[\theta- i\frac{\pi}{2}\right]L_S\left[\theta + i\frac{\pi}{2}\right] = 1 + e^{-2 \pi \int dx^0}P
\end{equation}
meaning that there is a ground state energy difference of $2 \pi$ between the superselection sectors associated to the 
identity and $P$ lines. \footnote{This statement can in principle be checked with the RCFT tools from \cite{Runkel:2007wd}, as long as renormalization is treated carefully. }

The claim is supported by the fusion relation 
\begin{equation}
T_{S;0}\left[\theta- i\frac{\pi}{2}\right]T_{S;0}\left[\theta + i\frac{\pi}{2}\right] = 1 + e^{-2 \pi e^{\theta}}
\end{equation}
which leads to the integral formula
\begin{equation}
\log T_{S;0}(\theta) =\frac{1}{2 \pi} \int_{-\infty}^\infty d\theta'\ \frac{1}{\cosh[\theta-\theta']} \log \left[1 + e^{-2 \pi e^{\theta'}}\right]
\end{equation}
valid on a strip of width $\pi$ around the real axis. 

The fusion relation holds equally well for excited states, which have extra sources in the integral equation due to the zeroes in the strip:  
\begin{equation}
\log T_{S;\{n_i\}}(\theta) = \sum_i \log \frac{e^\theta-n_i -\frac12}{e^\theta+n_i +\frac12}  + \frac{1}{2 \pi} \int_{-\infty}^\infty d\theta'\ \frac{1}{\cosh[\theta-\theta']} \log \left[1 + e^{-2 \pi e^{\theta'}}\right]
\end{equation}

Furthermore, we have 
\begin{equation}
T_{S;\sigma}\left[\theta- i\frac{\pi}{2}\right]T_{S;\sigma}\left[\theta + i\frac{\pi}{2}\right] = 1 -e^{-2 \pi e^{\theta}}
\end{equation}
and
\begin{equation}
\log T_{S;\sigma}(\theta) =\frac{1}{2 \pi} \int_{-\infty}^\infty d\theta'\ \frac{1}{\cosh[\theta-\theta']} \log \left[1 - e^{-2 \pi e^{\theta'}}\right]
\end{equation}
valid on a strip of width $\pi$ around the real axis.

The fusion relation suggests that if $e^\theta$ has a positive real part, the identity summand dominates and $L_S\left[\theta- i\frac{\pi}{2}\right]$ and $L_S\left[\theta + i\frac{\pi}{2}\right]$ will both flow to the same line, either $I$ or $P$, in accordance with $I \times I = P \times P = I$. If $e^\theta$ has a negative real part, the $P$ summand dominates and $L_S\left[\theta- i\frac{\pi}{2}\right]$ and $L_S\left[\theta + i\frac{\pi}{2}\right]$
will flow to an opposite choice of line, in accordance with $I \times P = P \times I = P$.

In conclusion, the prediction is that $L_S\left[\theta\right]$ will flow to the identity line in the range $|\Im \theta|<\pi$, 
but will flow to a $P$ line (with renormalized ground state energy) in the ranges $\pi<\Im \theta<3\pi$ and $-3\pi<\Im \theta<-\pi$, etcetera,
with periodicity $4 \pi$ and sharp transitions at $\Im \theta = \pm \pi$ where the line flows to a direct sum of $1$ and $P$ with the
same real part of the ground state energy. 
\begin{figure}
	\centering
	\includegraphics[width=0.7\linewidth]{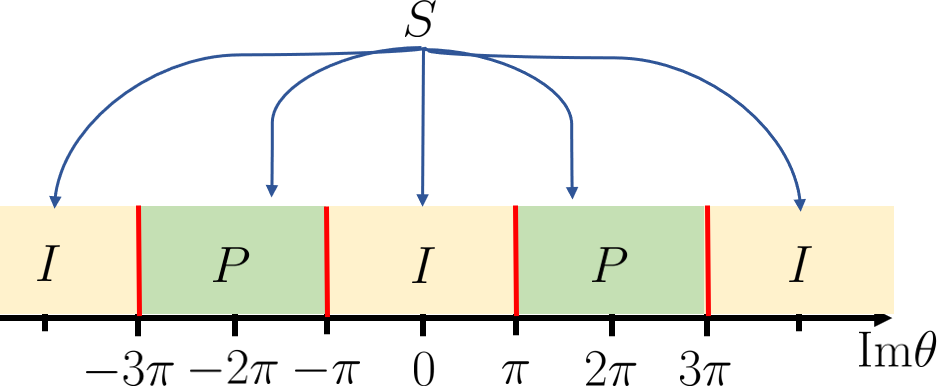}
	\caption{IR fate of the deformed line defect $L_S[\theta]$ for different $\Im\theta$.}
	\label{fig:isingrgendpoint}
\end{figure}

Another sanity check of this prediction Fig.\ref{fig:isingrgendpoint} is that it is compatible with $P\times S = S$. The fusion of $S$ with $P$ must map $\psi$ to a multiple of itself \cite{Graham:2003nc}, so $P$ must act on $L_S[\theta]$ as a shift of $\theta$, up to a constant counterterm shifting the defect Hamiltonian\footnote{Here we are using the standard observation that fusion with topological defects does not affect the local RG flow dynamics. See \cite{Graham:2003nc} for applications of this principle to conformal boundary conditions.}. This obviously agrees with Fig.\ref{fig:isingrgendpoint}, where upon fusing $S$ with $P$, $\Im \theta$ is shifted by $2\pi$, $P\rightarrow I$ and $I\rightarrow P$.

The fusion relation is the simplest example of $\mathfrak{su}(2)$ Hirota dynamics:
\begin{equation}\label{eq:su2hirota}
T_s\left[\theta- i\frac{\pi}{2}\right]T_s\left[\theta + i\frac{\pi}{2}\right] = 1 + T_{s-1} T_{s+1}
\end{equation}
with $T_2=T_S$, $T_1=1$, $T_3 = e^{-2 \pi e^{\theta}}$, $T_0=T_4=0$. Compare with \eqref{eq:HirotaSU2}.

\subsubsection{ODE/IM correspondence}\label{sec:IsingODE}
The function $T_S(\theta)$ coincides with a basic transport coefficient for the harmonic oscillator Schr\"oedinger equation \cite{Dorey:1998pt} 
\begin{equation}
e^{- 2 \theta} \partial_x^2 \psi(x) = (x^2 - 2) \psi(x) \label{eq:IsingODE}
\end{equation}

This equation has four \emph{small} solutions $\psi_n$, uniquely characterized by their exponentially fast decrease along rays of direction $e^{-\frac{\theta}{2}- \frac{i n \pi}{2}}$. 
We can normalize $\psi_0$ so that 
\begin{equation} \label{eq:IsingPsi0asy}
\psi_0 \sim \frac{1}{\sqrt{2 x}} (\sqrt{2 e} x)^{e^\theta} e^{-\frac{\theta}{2}- \frac{e^\theta x^2}{2}}
\end{equation}
and define 
\begin{equation}
\psi_n(x;\theta) = \psi_0(x;\theta + i \pi n)
\end{equation}
The definition can be extended to all integer $n$, with $\psi_{n+4}  = - e^{- 2 \pi i (-1)^n e^\theta} \psi_n$.

The function $\psi_0$ can be given explicitly in terms of parabolic cylinder functions:
\begin{equation}
\psi_0 = \frac{e^{\frac{1}{4} \left(-2 e^\theta (\theta-1)-\theta\right)} D_{\frac{1}{2} \left(-1+2
   e^\theta\right)}\left(\sqrt{2} e^{\theta/2} x\right)}{\sqrt[4]{2}}
\end{equation}
Details can be found in Appendix \ref{app:IsingODEexactsol}.

The Wronskian $(\psi_n,\psi_{n+1}) \equiv \psi_n \partial_x \psi_{n+1}- \psi_{n+1}\partial_x \psi_n $ of consecutive solutions is $-i$. Because of the periodicity, we also have $i(\psi_{-1}, \psi_2) = e^{- 2 \pi i e^\theta}$. 
We have
\begin{equation}
T_S(\theta) = i (\psi_{-1}, \psi_1)
\end{equation} 
The simplest proof of this fact is that the two functions satisfy the same Riemann-Hilbert problem in the $\theta$ plane.

The Hirota recursion \eqref{eq:su2hirota} follows from the Pl\"ucker relation\footnote{This simply follows from the fact that any three vectors $a$, $b$, $c$ in a two dimensional vector space must satisfy a linear relation in the form of $(a,b)c+ (b,c)a + (c,a)b = 0$, where brackets denote exterior product.} between the Wronskians:
\begin{equation}
(\psi_{-1},\psi_1)(\psi_0, \psi_2) =(\psi_{-1},\psi_0)(\psi_1, \psi_2)+ (\psi_0,\psi_1)(\psi_{-1}, \psi_2)
\end{equation}
\begin{figure}
	\centering
	\includegraphics[width=0.7\linewidth]{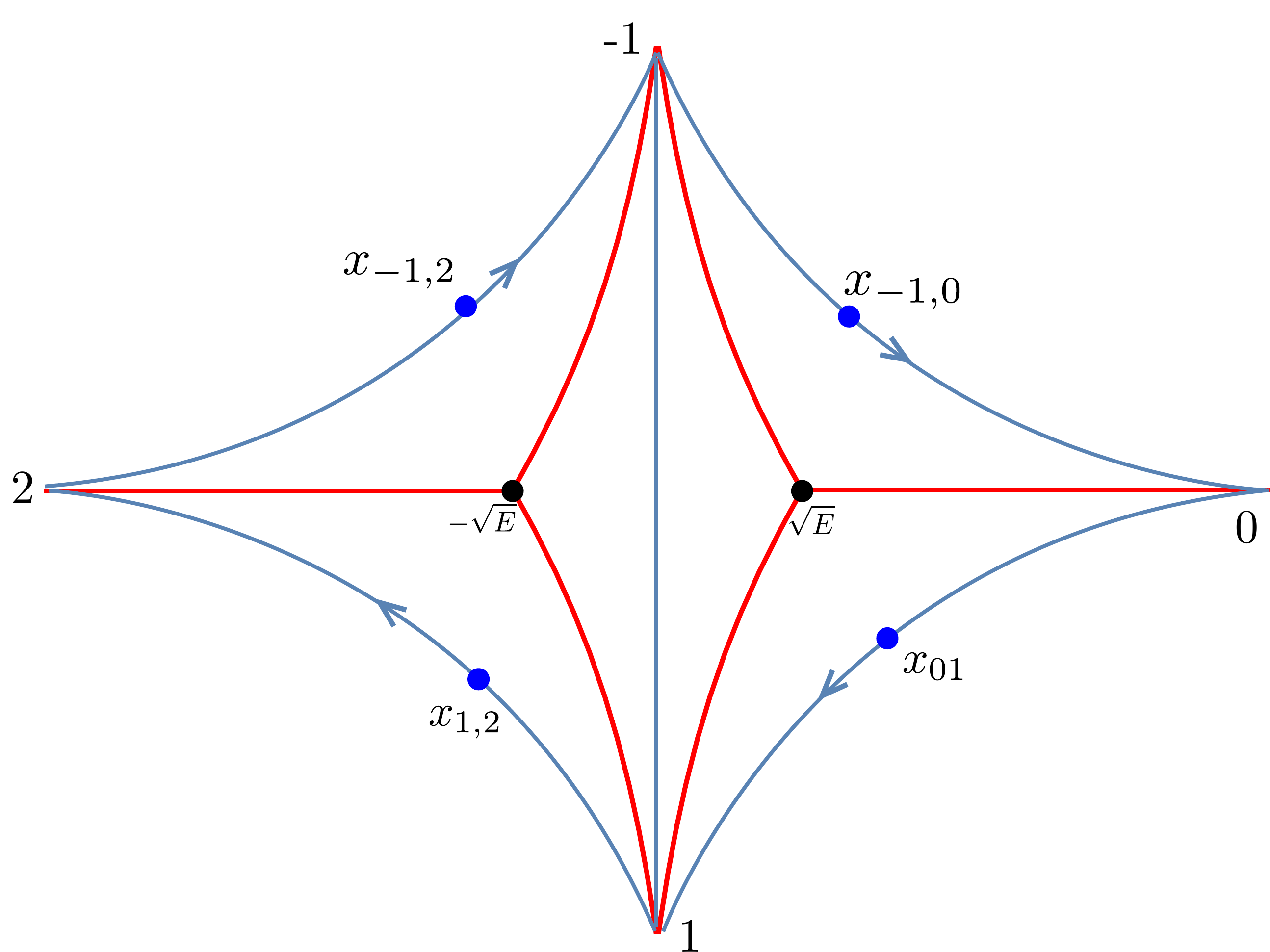}
	\caption{WKB diagram for the differential equation \eqref{eq:IsingODE} defined in \eqref{eq:Isingquadraticdiff}. Generic flow lines and WKB lines are colored blue and red respectively.}
	\label{fig:wkbharmonic}
\end{figure}

A standard WKB analysis as reviewed in Appendix \ref{app:WKB} controls the IR asymptotics. The WKB analysis employs the WKB network, 
namely the union of flow lines\footnote{Various names are used in the literature. The WKB network is often called spectral network or Stokes diagram in the literature, where the WKB line goes under the name of Stokes line or anti-Stokes line.}, along which the WKB differential 
\begin{equation}
\sqrt{x^2-2} e^{\theta} dx  \label{eq:Isingquadraticdiff}
\end{equation} 
is real, shown in Fig. \ref{fig:wkbharmonic}. Contrast to the generic flow lines which end on singularities, there are special lines emanating from a zero of the differential, which we will refer to as WKB lines.

The cross-ratio 
\begin{equation}
\frac{(\psi_0,\psi_1)(\psi_{-1}, \psi_2)}{(\psi_{-1},\psi_0)(\psi_1, \psi_2)} = e^{-2 \pi i e^{\theta}}
\end{equation}
is controlled by the period of $\sqrt{x^2 -2}\, dx$ around a contour wrapping around the cut, while the 
IR asymptotics of the Wronskian are controlled by a (vanishing) regularized period of $\sqrt{x^2 -2} \, dx$ 
from $- i \infty$ to $i \infty$, where the regularization subtracts the reference asymptotics in \eqref{eq:IsingPsi0asy}
\begin{equation}
T_{S}(\theta) \equiv i(\psi_{-1},\psi_{1}) \sim 1+\dots
\end{equation}

The UV asymptotics can be obtained by dropping the constant term on the right hand side of the Schr\"oedinger equation.
Indeed, we can rescale the $x$ variable to get 
\begin{equation}
e^{- 2 \theta} \partial_x^2 \psi(x) = (x^2 - 2 g^2) \psi(x)
\end{equation}
which is amenable to a perturbative expansion in the UV. 

The ODE/IM for excited states will be discussed in a companion paper. \cite{Gaiotto:2020dhf}

\section{The $\mathfrak{su}(2)$ Kondo line defects}\label{sec:wzw}
Consider any CFT equipped with some level $k$ chiral $\widehat{\mathfrak{su}}(2)$ WZW currents $J^a$.
This implies that the CFT is a modular-invariant combination of an $\widehat{\mathfrak{su}}(2)_k$ chiral WZW model 
and some other degrees of freedom \footnote{The obvious choice is an anti-chiral WZW model, but many alternatives are possible. A nice possibility is 
a $\widehat{\mathfrak{u}}(k)_2$ chiral WZW model, which would combine with $\widehat{\mathfrak{su}}(2)_k$ to give a theory of $2 k$ complex chiral fermions,
by level-rank duality. Of course, an universally valid choice is a 3d $SU(2)_k$ Chern-Simons TFT defined on a half-space.}. The line defects we will discuss momentarily only interact with the chiral WZW degrees of freedom and are transparent to everything else. 

We define the Kondo line defects by coupling the theory to a spin $j$ (half integer) quantum-mechanical system by the natural $\mathfrak{su}(2)$-invariant marginally relevant coupling \cite{Bachas:2004sy}
\begin{equation}
g \int \sigma_a J^a dx^0
\end{equation}
with $\sigma_a$ being the matrices representing $\mathfrak{su}(2)$ in the spin $j$ quantum-mechanical system. Dimensional transmutation converts the coupling $g$
into a scale, which we can absorb in the $\theta$ dependence. The result is a family of chiral line defects $L_{j}[\theta]$. 

Gleaning information from the vast literature on integrability, including \cite{Bachas:2004sy,Runkel:2007wd,Dorey:1998pt,Bazhanov:1998wj,dorey1999relation,GMN:2009hg,Gaiotto:2014bza} and more, and adding some judicious guesses
one is presented with the following conjectures: 
\begin{itemize}
\item The Kondo line defects give commuting transfer matrices $\hat T_{2j+1}[\theta]$. These operators commute with the Hamiltonian and 
act within primary towers for the WZW currents. 
\item The Kondo line defects fuse in a manner analogous to representations of the $\mathfrak{su}(2)$ Yangian:
\begin{equation}
\hat T_{2j+1}\left[\theta- i\frac{\pi}{2}\right]\hat T_{2j+1}\left[\theta + i\frac{\pi}{2}\right] = 1 + \hat T_{2j}[\theta] \hat T_{2j+2}[\theta]\label{eq:HirotaSU2}
\end{equation}
\item Expectation values in a generic WZW primary state $|l\rangle$
\begin{equation}
\langle l|\hat T_{2j+1}[\theta] |l\rangle_{2 \pi R=1} \equiv T_{2j+1;l}(\theta)
\end{equation}
or eigenvalues of $\hat T_{2j+1}[\theta]$ acting on descendants give solutions of the Hirota dynamics. 
The vacuum expectation value $T_{2j+1;0}(\theta)$ will just be referred to as $T_{2j+1}(\theta)$
\item The expectation values can be computed as transport coefficients of an auxiliary Schr\"oedinger equation
\begin{equation}
\partial_x^2 \psi(x) = \big[e^{ 2 \theta} e^{2x} (1 + g x)^k +\frac{l(l+1)}{(x+1/g)^2}\big] \psi(x)  
\end{equation}
in the spirit of the ODE/IM correspondence. 
\item The $\hat{T}_{2j+1}[\theta]$ expectation values on the vacuum or other eigenstates are also expected to satisfy certain
TBA equations, which are the conformal limit of the TBA equations for chiral Gross-Neveu models,
i.e. the deformation of a non-chiral WZW model by a $J^a \bar J^a$ marginally relevant interaction.  
\end{itemize}
%The two latter claims should really hold for all common sets of eigenvalues of the $\hat T_{2j+1}[\theta]$.
These claims are hard to prove or even justify in a concise manner directly in 2d. 

\subsection{A perturbative analysis of the Kondo defect vevs}\label{sec:PertOverviewKondo}
Using the definition of the line defects, one can compute in perturbation theory
\begin{equation}
\hat T_n = n + g^2 \hat t_{n,2} + g^3 \hat t_{n,3} + g^4 \hat t_{n,4} + \cdots,
\end{equation}
where $n=2j+1$. The linear term is missing because $\mathrm{Tr} \, \sigma_a=0$.

The calculation requires some careful renormalization, which dimensionally transmutes the coupling into a scale $\mu_0(g)$. 
The RG flow rescales that to $\mu = \mu_0 e^\theta$ and the coupling runs as 
\begin{equation}
\mu_0(g) e^\theta = \mu_0\left(g_{\mathrm{eff}}(\theta)\right)
\end{equation}
The only counter-terms are a constant counterterm and the renormalization of the coupling,
which first appear at order $g^3$.  

Up to a rescaling of coupling, the perturbative RG flow equation takes the form \footnote{The right hand side is the negative of the beta function.}
\begin{equation}
\partial_\theta g_{\mathrm{eff}}(\theta) = g_{\mathrm{eff}}(\theta)^2 + c g_{\mathrm{eff}}(\theta)^3 + \cdots
\end{equation}
where we normalize the coupling such that the leading coefficient is $1$. In this sign convention, a small positive UV coupling will grow in the IR and our line defect will be asymptotically free,
with a typical IR mass scale which is exponentially suppressed at small positive $g_{\mathrm{eff}}$. This is the microscopic definition 
of the $L_n$ line defects we are interested in.  A negative coupling, instead, flows to 0. Such IR free line defects will appear later on as IR outcomes of some of the RG flows we consider, with a typical UV mass scale which is exponentially large at small negative $g_{\mathrm{eff}}$. 

The coefficient $c$ cannot be re-defined away. An explicit calculation in Appendix \ref{app:wzwKondoPert} shows that it is independent of $n$ and equals $-\frac{k}{2}$. The ellipses indicates terms which can be arbitrarily adjusted by a perturbative redefinition of the coupling. This can be checked rather easily. 

We choose to fix the renormalization ambiguities by 
imposing 
\begin{equation}
\partial_\theta g_{\mathrm{eff}}(\theta) = \frac{g_{\mathrm{eff}}(\theta)^2}{1+ \frac{k}{2} g_{\mathrm{eff}}(\theta)} \label{eq:betafunctiongeff}
\end{equation}
i.e.
\begin{equation}
e^{-\frac{1}{g_{\mathrm{eff}}(\theta)}} g_{\mathrm{eff}}(\theta)^{\frac{k}{2}} \equiv  e^{-\frac{1}{g}} g^{\frac{k}{2}} e^{\theta}\label{eq:geffDef}
\end{equation}
or $\mu_0(g) =  e^{-\frac{1}{g}} g^{\frac{k}{2}}$. This choice of RG scheme has the advantage that $0<g_{\mathrm{eff}}<\infty$ parameterizes the full range of scales. It will also agree with the RG scheme implicit in the Hirota relations, ODE/IM correspondence, etc. See Appendix \ref{app:HirotaSU2} and \ref{app:schropert} for more details. Other choices of RG scheme are of course possible and sometimes useful. 

The defect vevs will depend only on the combination $2 \pi R e^{-\frac{1}{g}} g^{\frac{k}{2}} e^{\theta}$.
Perturbatively, that means the $\theta$ dependence of $\hat T_n[\theta]$ is captured by 
\begin{equation}
\hat T_n[\theta] = n + g_{\mathrm{eff}}(\theta)^2 \hat t_{n,2} + g_{\mathrm{eff}}(\theta)^3 \hat t_{n,3} + g_{\mathrm{eff}}(\theta)^4 \hat t_{n,4} + \cdots
\end{equation}
with 
\begin{equation}
g_{\mathrm{eff}}(\theta) = g + \theta g^2 + \theta(\theta-\frac{k}{2}) g^3+ \theta (\theta^2 - \frac54 k \theta + \frac{k^2}{4}) g^4 + \cdots
\end{equation}

The $\hat t_{n,m}$ are complicated expressions of the Fourier modes of WZW currents.  In Appendix \ref{app:wzwKondoPert} we compute the explicit form of $\hat T_n$ up to order $g^4$. Strikingly, the $\hat t_{n,m}$ we computed all commute with each other, confirming that the $\hat T_n[\theta]$ behave as commuting transfer matrices.  

Even more strikingly, we find that our choice of renormalization scheme is such that the $\hat T_n[\theta]$ satisfy Hirota fusion relations 
as long as we fix the reference coupling $g$ to be the same for all defects, at least at the order we could compute. Perturbatively, that requires the relations
\begin{align}
2 n  \hat t_{n,2} &= (n+1) \hat t_{n-1,2}+(n-1) \hat t_{n+1,2} \cr
2 n  \hat t_{n,3} &= (n+1) \hat t_{n-1,3}+(n-1) \hat t_{n+1,3} \cr
2 n  \hat t_{n,4} +  \hat t_{n,2}^2 &= (n+1) \hat t_{n-1,4}+(n-1) \hat t_{n+1,4} +\hat t_{n+1,2}\hat t_{n-1,2}+\frac{3}{2} n \pi^2  \hat t_{n,2}
\end{align}

\subsection{Perturbative and non-perturbative RG flows}
%It is amusing to note that the function $e^{\frac{1}{g}} g^{\frac{k}{2}}$ for positive real $g$ has a minimum at $g=k/2$. 
%As a consequence as we flow to the IR the effective coupling $g_{\mathrm{eff}}(\theta)$ will hit 
%$k/2$ at some finite value of $\theta$. Beyond that, the description in terms of an effective coupling breaks down.
%In any case, the outcome of the RG flow will be dictated by non-perturbative dynamics. 

For physical values of the parameters, perturbation theory is only useful in the UV and non-perturbative dynamics kicks in at low energy. 
If we analytically continue $\theta$ sufficiently away from the real axis, though, we get a surprise: 
under RG flow the effective coupling $g_{\mathrm{eff}}(\theta)$ grows a bit, but then swings back to be small and negative. The imaginary part of 
$\frac{1}{g_{\mathrm{eff}}(\theta)}$ decreases by a finite amount in absolute value, changing by $-\frac{k}{2} \pi$ as the real part flows to large negative values. 

 \begin{figure}
 	\centering
 	\includegraphics[width=0.47\linewidth]{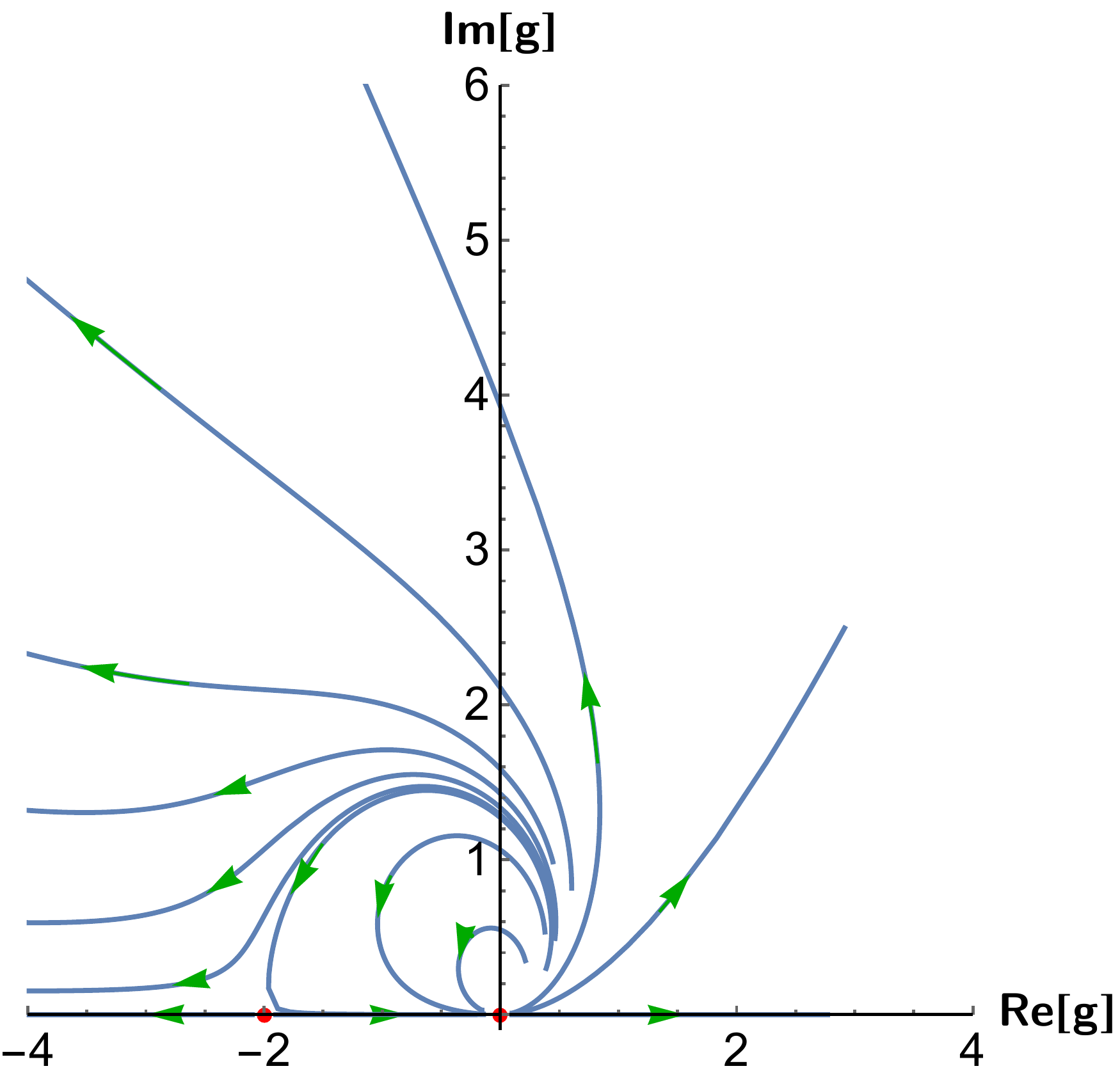}
 	\includegraphics[width=0.47\linewidth]{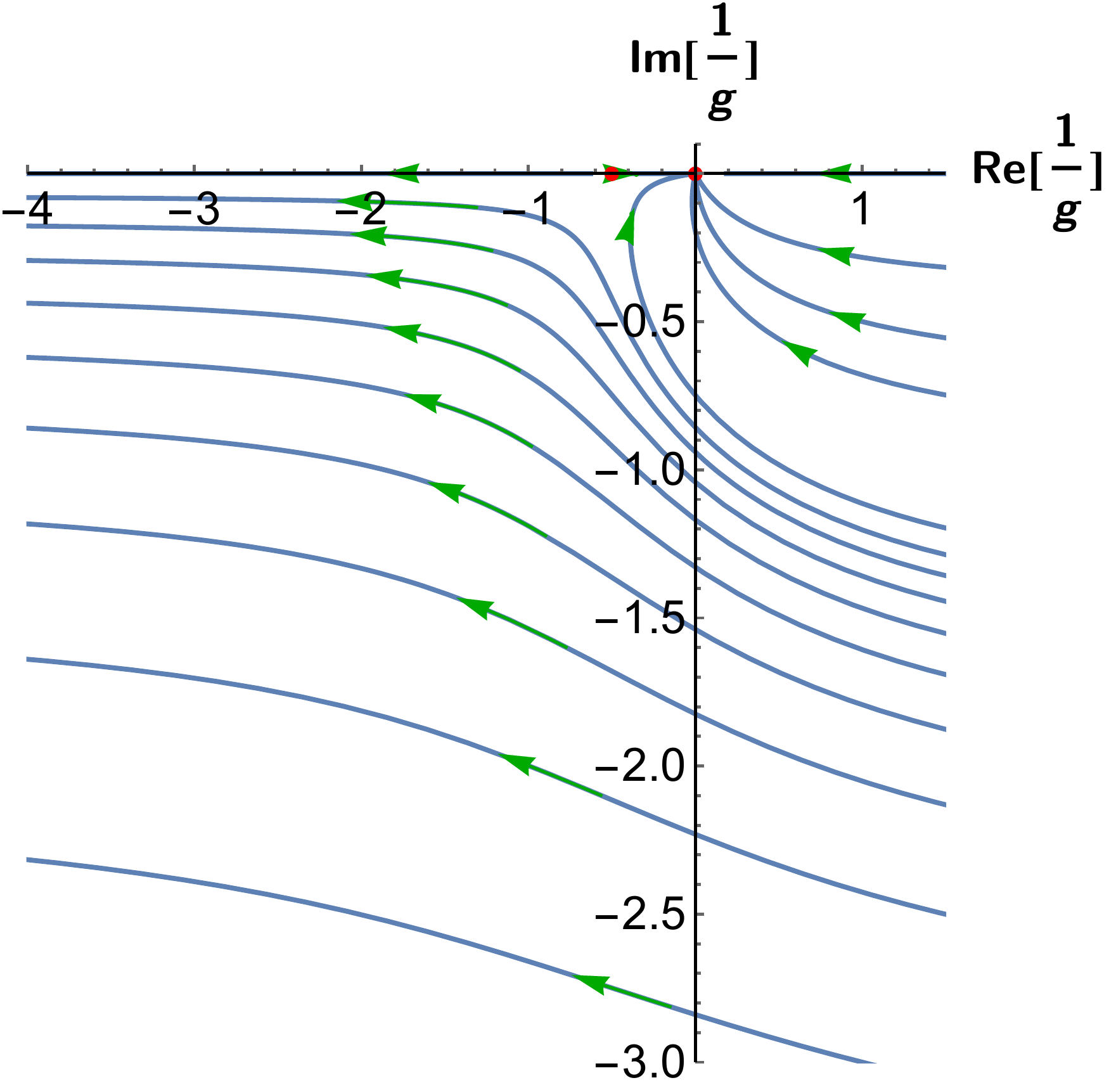}
 	\includegraphics[width=0.47\linewidth]{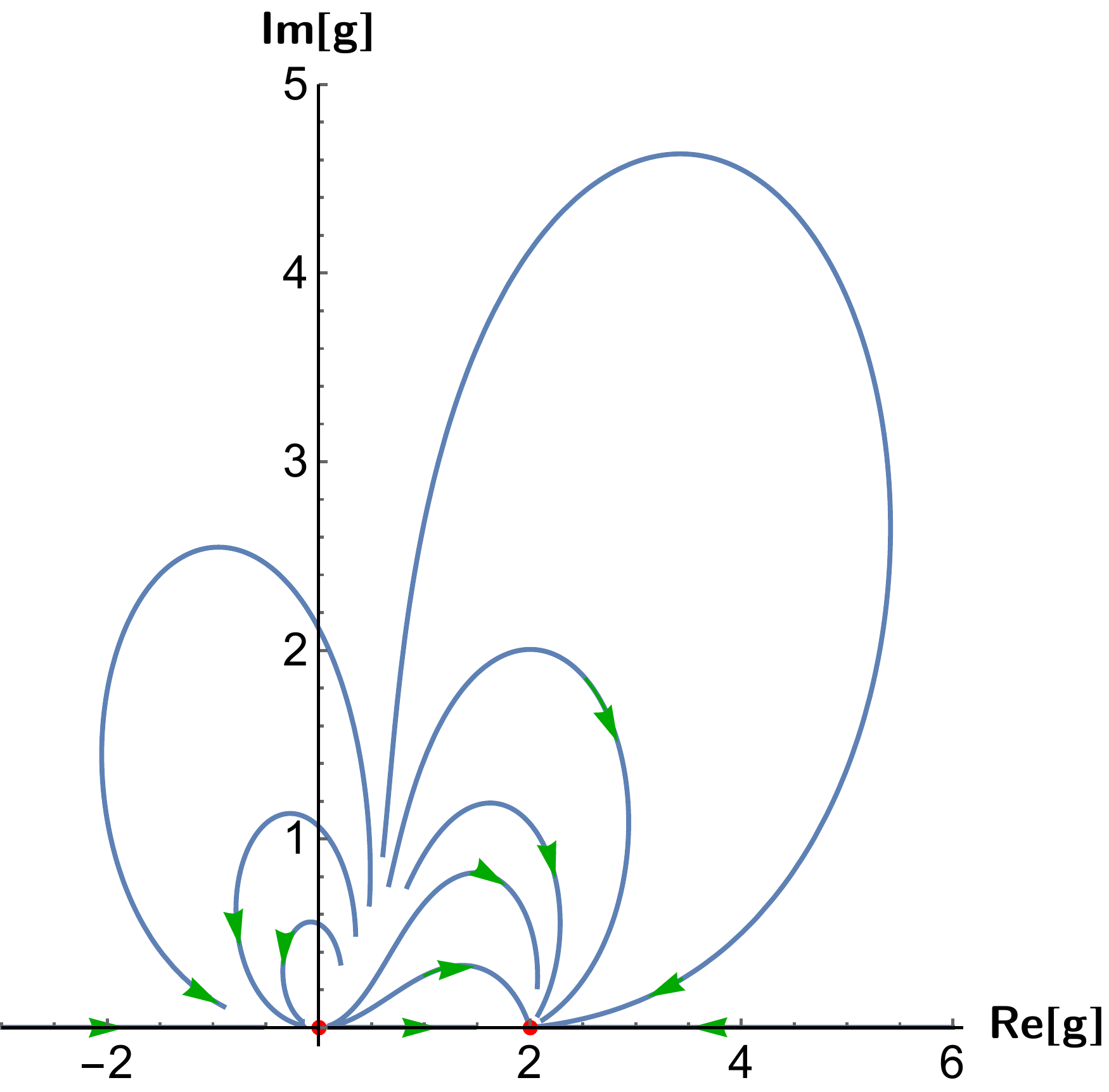}
	\includegraphics[width=0.47\linewidth]{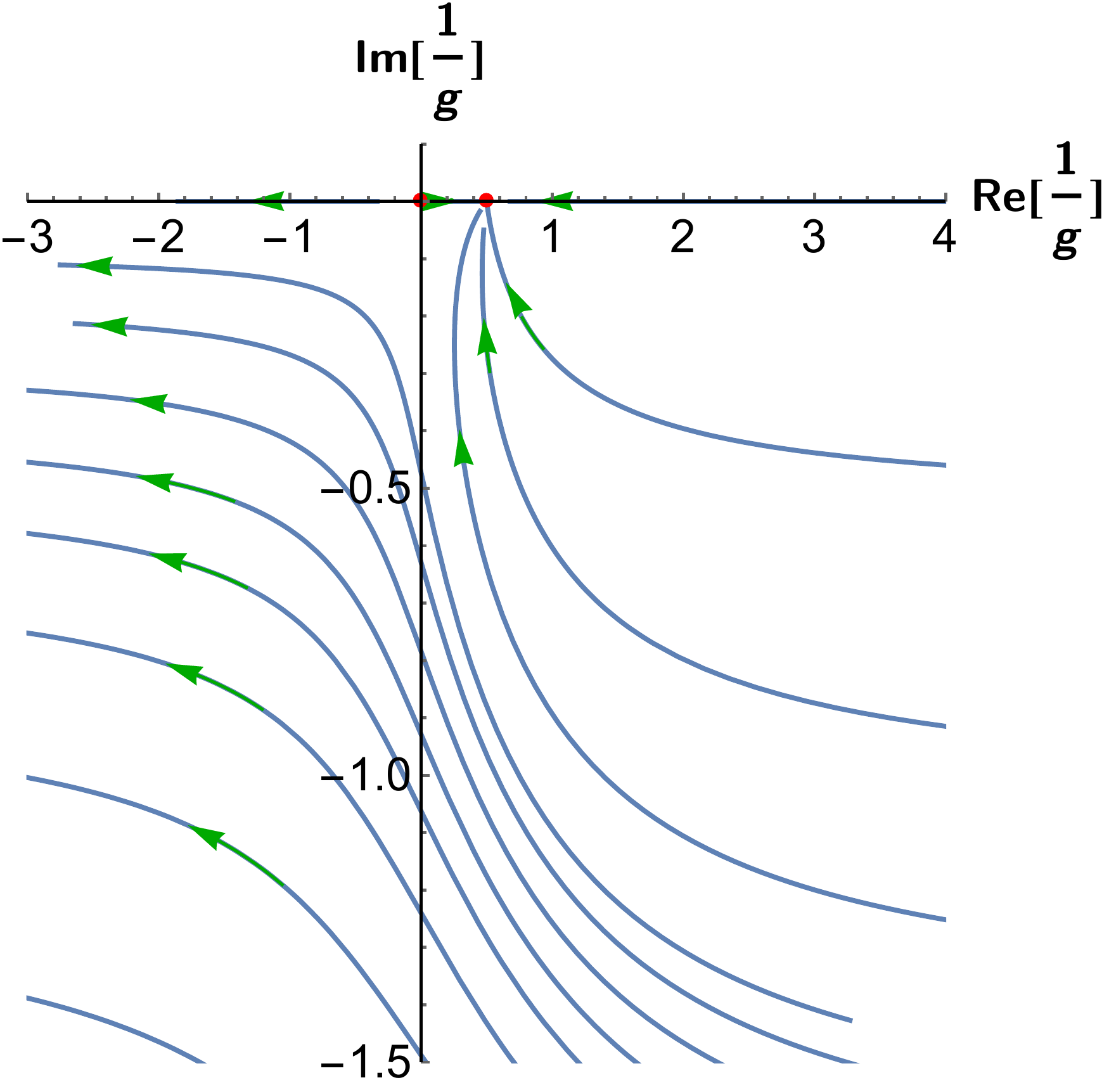}
 	\caption{The RG flow pattern over the complex $g$ plane and the complex $1/g$ plane. The top two and he bottom two are plotted using  the beta function \eqref{eq:betafunctiongeff} and \eqref{eq:betafunctiongeffwithzero} respectively. Note that the lower left figure is the same as Fig. 1 in \cite{Nakagawa_2018}}
 	\label{fig:complexrg}
 \end{figure}
 
That means that the line defects remain perturbative all along the RG flow as long as the initial imaginary part of $\frac{1}{g}$
is sufficiently large!  
The analytically continued line defects are not unitary, so the non-monotonic RG flow is not a contradiction, but it is still 
a bit surprising. 

These perturbative IR limits for large positive and large negative imaginary part of $\frac{1}{g}$ differ, as the two branches of $\frac{1}{g_{\mathrm{eff}}(\theta)}$
differ by $k \pi$. The two perturbative regimes are separated by some intermediate phases, where the RG flow is non-perturbative.

Another important situation where perturbation theory is applicable is large $k$, at least at finite $j$. If we use an alternative RG scheme where  
\begin{equation}
\partial_\theta g'_{\mathrm{eff}}(\theta) = g'_{\mathrm{eff}}(\theta)^2 - \frac{k}{2} g'_{\mathrm{eff}}(\theta)^3 \label{eq:betafunctiongeffwithzero}
\end{equation}
we get a perturbative zero for the $\beta$ function at $g'_r = \frac{2}{k}$. 

That means the RG flow for large $k$ and fixed $j$ must lower the $g$ function by an amount of order $k^{-1}$. 
The leading correction actually comes at order $3$ in perturbation theory and is proportional to $j(j+1)(2 j+1)$.

For $j=\frac12$ our hands are tied: the only topological line defect with quantum dimension slightly lower than $2$ is the 
 topological line ${\cal L}_{\frac12}$ whose Cardy label is the spin $\frac12$ primary field and whose quantum dimension is 
$2 \cos \frac{\pi}{k+2}$. The leading correction is consistent with this. This is a standard result \cite{Affleck:1995ge,Bachas:2004sy}.

The RG flow of $g'_{\mathrm{eff}}(\theta)$ as a function of the imaginary part of $\theta$ is quite interesting. As we increase the imaginary part to 
large values of order $k$, we hit a thin region where the flow reaches strong coupling, and beyond that the perturbative flow back to the 
spin $\frac12$ IR Kondo line discussed before. We interpret this as a phase transition from the flow to ${\cal L}_{\frac12}$ to the flow to
$L^{IR}_{\frac12}$. This will indeed happen in the exact solution proposed below. 

In a similar manner, for sufficiently small $j$, the physical flow of $L_j[\theta]$ should end on the 
 topological line ${\cal L}_{j}$ of quantum dimension 
 \begin{equation}
 d_{2j+1}^{(k)}\equiv \frac{\sin(2j+1)\frac{\pi}{k+2}}{\sin\frac{\pi}{k+2}} \label{eq:quantumDimension}
 \end{equation}
while for sufficiently large imaginary part of $\theta$ it should go back to 
 $L^{IR}_j[\theta]$. Recall that  ${\cal L}_{j}$ is the topological line with Cardy label given by the primary field of spin $j$, where $j=0,\frac{1}{2},1,\dots,\frac{k}{2}$. 
 
We can anticipate here the conjectural behaviour of the physical RG flows for all $k$ and $j$ is that (up to constant counterterms) supported by the ODE/IM solution:
\begin{itemize}
\item For $j \leq \frac{k}{2}$, $L_j[\theta]$ flows to ${\cal L}_{j}$.
\item For $j>\frac{k}{2}$, $L_j[\theta]$ flows to ${\cal L}_{\frac{k}{2}}\times L^{IR}_{j-\frac{k}{2}}[\theta]$
\end{itemize}
These statements are conjecturally valid on a strip of width $2 \pi$ around the real $\theta$ axis. 

Hirota recursion relations determine the IR behaviour of all lines beyond that strip. One finds all sort of combinations of the form 
${\cal L}_{j'}\times L^{IR}_{j-j'}[\theta]$, with $j'$ jumping by $\pm \frac12$ across phase transitions. 

\subsection{The ODE/IM solution for $T_n$ }\label{sec:SU2ODEIM}
We propose to identify the functions $T_n[\theta]$ with the transport data of the Schr\"oedinger equation 
\begin{equation}
\partial_x^2 \psi(x) = e^{ 2 \theta} e^{2x} (1 + g x)^k  \psi(x) \label{eq:ODESU2}
\end{equation}
The first immediate observation is that the shift $x \to x-\frac{1}{g}$ maps the equation to 
\begin{equation}
\partial_x^2 \psi(x) = e^{2 \theta} e^{-\frac{2}{g}}g^{k} e^{2x}x^k \psi(x)
\end{equation}
so that the transport data is only a function of the combination $e^{\theta} e^{-\frac{1}{g}}g^{\frac{k}{2}}$, as in \eqref{eq:geffDef}.

\subsubsection{ODE definition of $T_n$} 
We can define the solution $\psi_0(x;\theta)$ of
\begin{equation}
e^{- 2 \theta} \partial_x^2 \psi(x) = (1 + g x)^k e^{2x} \psi(x)
\end{equation}
for real positive $g$ as the solution which decreases asymptotically fast 
along the line of large real positive $x + \theta$. If we analytically continue in $g$,
the imaginary part of $x + \theta$ has to be accordingly adjusted to keep $e^\theta (1+g x)^{\frac{k}{2}} e^x dx$ real and positive. 

We can normalize $\psi_0$ so that it agrees with WKB asymptotics in that region, as before: 
\begin{equation} \label{eq:asy1}
\psi_0(x;\theta) \sim \frac{1}{\sqrt{2 (1+ g x)^{\frac{k}{2}} e^{x + \theta}}}e^{- e^\theta f_k(x;g)}
\end{equation}
for large positive real $x+\theta$. Here $f_k(x;g)$ is a function defined by 
\begin{equation}
f_k(x;g) = \int^x_{- \frac{1}{g}} e^y (1+ g y)^{\frac{k}{2}} dy = e^{-\frac{1}{g}} g^{\frac{k}{2}} \int^{x+ \frac{1}{g}}_0 e^y y^{\frac{k}{2}} dy
\end{equation}

We then define again an infinite sequence of other solutions 
\begin{equation}
\psi_n(x;\theta) \equiv \psi_0(x;\theta+ i \pi n) 
\end{equation}
which have the above asymptotics for large positive real $x+\theta+i \pi n$.

The transport coefficients of this Schr\"oedinger equation consist of the Wronskians $i (\psi_0, \psi_n)$.
The large positive $x$ asymptotics guarantee $i (\psi_0, \psi_1)=1$, but the other Wronskians are non-trivial functions of $\theta$. 

Adjusting the shifts to match the quantum determinants and Hirota relations in a standard form, we can propose 
\begin{equation}
T_{n;l}(\theta) = i \left(\psi_0(x;\theta - \frac{i \pi n}{2}), \psi_0(x;\theta +\frac{i \pi n}{2}) \right)
\end{equation}

At large negative $x$, the right hand side of the Schr\"oedinger equation decreases exponentially 
and thus we must have 
\begin{equation}
\psi_0(x;\theta) \sim - Q(\theta) (x + \frac{1}{g}) - \tilde Q(\theta)
\end{equation}
up to exponential corrections. We included the $\frac{1}{g}$ shift so that both $Q(\theta)$ and $\tilde Q(\theta)$ are functions of  
$e^{\theta} e^{-\frac{1}{g}}g^{\frac{k}{2}}$ only.  \footnote{Notice that there is an interesting spectral problem where one requires $\psi$ to be finite 
at large negative $x$ and asymptotically decreasing at large positive $x$. The zeroes of the $Q(\theta)$ functions are the solutions of that spectral problem}

The T-functions $T_n$ take the form of quantum determinants built from $Q$ and $\tilde Q$, 
\begin{equation}
T_{n;l}(\theta) = i Q(\theta+ \frac{i\pi n}{2})\tilde{Q}(\theta- \frac{i\pi n}{2})-i \tilde{Q}(\theta+ \frac{i\pi n}{2})Q(\theta- \frac{i\pi n}{2})
\end{equation}
which can be naturally interpreted as the 
two $Q$-functions for the system. \footnote{Finding a direct 2d CFT physical interpretation for the $Q$ functions, or $\psi_0(x;\theta)$ itself, is a long 
standing problem, which we do not address in this paper. }

\subsubsection{The UV fixed point $g=0$}
If we turn off the coupling, we have the simpler equation 
\begin{equation}
e^{- 2 \theta} \partial_x^2 \psi(x) = e^{2x} \psi(x)
\end{equation}

This has a unique solution 
\begin{equation}
\psi_0(x;\theta) = \frac{1}{\sqrt{\pi}}K_0(e^{x + \theta})
\end{equation}
which behaves as 
\begin{equation}
\psi_0(x;\theta) \sim \frac{1}{\sqrt{2 e^{x + \theta}}}e^{- e^{x + \theta}}
\end{equation}
for large positive real $x+\theta$. 

On the other hand, at large negative real part of $x + \theta$ we have 
\begin{equation}
\psi_0(x;\theta) \sim - \frac{1}{\sqrt{\pi}} (x + \theta +\gamma - \log 2 )
\end{equation}
up to exponentially small corrections. 

We can obtain an infinite sequence of other solutions 
\begin{equation}\label{eq:continuing_K0}
\psi_n(x;\theta) \equiv \psi_0(x;\theta+ i \pi n) =  \frac{1}{\sqrt{\pi}}K_0(e^{x + \theta})- \pi i n \frac{1}{\sqrt{\pi}}I_0(e^{x + \theta})
\end{equation}
which have the above asymptotics for large positive real $x+\theta+i \pi n$.
Clearly, at large negative real part of $x + \theta$ we have 
\begin{equation}
\psi_n(x;\theta) \sim - \frac{1}{\sqrt{\pi}} (x + \theta + i \pi n +\gamma - \log 2 )
\end{equation}
so that the Wronskian of two such solutions is exactly 
\begin{equation}
i (\psi_0, \psi_n) = n
\end{equation}
which is the expected UV value of $T_n$. 

\subsubsection{Weak-coupling expansion}\label{sec:weakcouplingExpansion}
When $g$ is sufficiently small and positive, it is reasonable to attempt a perturbative expansion of the solution $\psi_0$ 
around the $g=0$ solution 
\begin{equation}
\psi^{(0)}_0(x;\theta) = \frac{1}{\sqrt{\pi}}K_0(e^{x + \theta})
\end{equation}
The perturbative expansion should be valid for $x \ll \frac{1}{g}$ and match smoothly with the expansion of the asymptotic expression \ref{eq:asy1}
in positive powers of $g$. 

At each order of the perturbative expansion we solve a Schr\"oedinger equation with a source which decreases exponentially at large positive $x$
and select the solution $\psi^{(n)}_0(x;\theta)$ which also decreases exponentially and matches the expansion of the asymptotic expression \ref{eq:asy1} in positive powers of $g$. 

At large negative $x$, the perturbative corrections will systematically correct the Q functions to some 
\begin{align}
Q(\theta) &= \frac{1}{\sqrt{\pi}} \left( 1 + q_1 g_{\mathrm{eff}}(\theta) + \cdots \right)\cr
\tilde Q(\theta) &= \frac{1}{\sqrt{\pi}} \left( - \frac{1}{g_{\mathrm{eff}}} + \tilde{q}_0 + \tilde q_1 g_{\mathrm{eff}}(\theta) + \cdots \right)
\end{align}
where the $g$ and $\theta$ dependence combine into a power series in $g_{\mathrm{eff}}(\theta)$. 

The Wronskian relation $i (\psi_0, \psi_1)=1$ should hold automatically. It actually determines the expansion coefficients of $Q$ in terms of these of $\tilde Q$. 

When we plug the expansion of the Q functions into the quantum determinant expression for $T_n(\theta)$, the result only depends on the
$\tilde{q}_n$ starting from the order $g^4$. The lower orders are fixed uniquely. We have 
\begin{equation}
T_n \sim n -\frac{\pi^{2}}{2}k I_R g^3 
+ \frac{\pi ^2 }{4}  I_R g^4 
\bigg[3 k^2 +2 k (-\tilde{q}_0 -3 \theta)+8 \tilde{q}_1 \bigg] \cdots \label{eq:Tnweakcoupling}
\end{equation}
where $I_R = \frac{1}{6} n\left(-1+n^{2}\right)$ and $\tilde{q}_0 = -\frac{k}{4}+(\gamma-\log 2)$. In particular, in order to match with the explicit line defect calculations we only need the first sub-leading coefficient $\tilde{q}_1$ in the expansion of $\tilde Q$, which can be found in Appendix \ref{app:schropert}. 

\subsubsection{WKB IR expansion}\label{sec:IRWKBanalysis}
The WKB analysis of the Schr\"oedinger equation, valid in the IR limit $e^\theta \to \infty$, 
requires a slightly more refined analysis than the Voros/GMN-style one applicable to meromorphic potentials with simple zeroes \cite{VOROS1983,GMN:2009hg}. In Appendix \ref{app:WKB}, we review the standard analysis and extend it to the case of zeroes of higher degree or exponential 
singularities.

A crucial role is played by the WKB/spectral network, 
which depicts the structure of the WKB lines, along which the leading WKB differential 
\begin{equation}
(1+ g x)^{\frac{k}{2}} e^{x + \theta} dx
\end{equation} 
is real. The main property of WKB lines is that the WKB solutions which are asymptotically growing 
along the WKB lines can be trusted as an approximation for the parallel transport of true solutions.

A GMN-style analysis focusses on generic WKB lines, which join asymptotic directions where some small solutions have been defined. 
The Wronskian of the small solutions at the endpoints of a WKB line can be estimated reliably as the Wronskian of the corresponding 
WKB approximants. The asymptotic approximation is valid in a whole half-plane in the $e^\theta$ plane centered around the 
ray used to draw the WKB network.\footnote{If WKB network is defined by $(1+ g x)^{\frac{k}{2}} e^{x + \theta_0} dx \in \mathbb{R}^+$, then the formal WKB series is an asymptotic series as $e^{-\theta}\rightarrow 0$ within a closed half plane $\mathbb{H}_{\theta_0} = \{\mathrm{Re}(e^{\theta-\theta_0})\geq 0 \}$.}

For Schr\"oedinger equation with a meromorphic potential and simple zeroes and generic $\theta$, the generic WKB lines give estimates for exactly 
enough ``WKB'' Wronskians to fully determine the full transport data. All other Wronskians and monodromies can be reconstructed as Laurent 
polynomials in the WKB Wronskians. 

In more general situations we need to work a bit harder, and use WKB lines which join an asymptotic direction and matching regions near zeroes of higher order or 
where the potential is exponentially small. The WKB lines can still be used to reliably transport the small solutions to the matching regions, where they can be compared with an appropriate basis of local solutions. 
\begin{figure}
	\centering
	\includegraphics[width=0.48\linewidth]{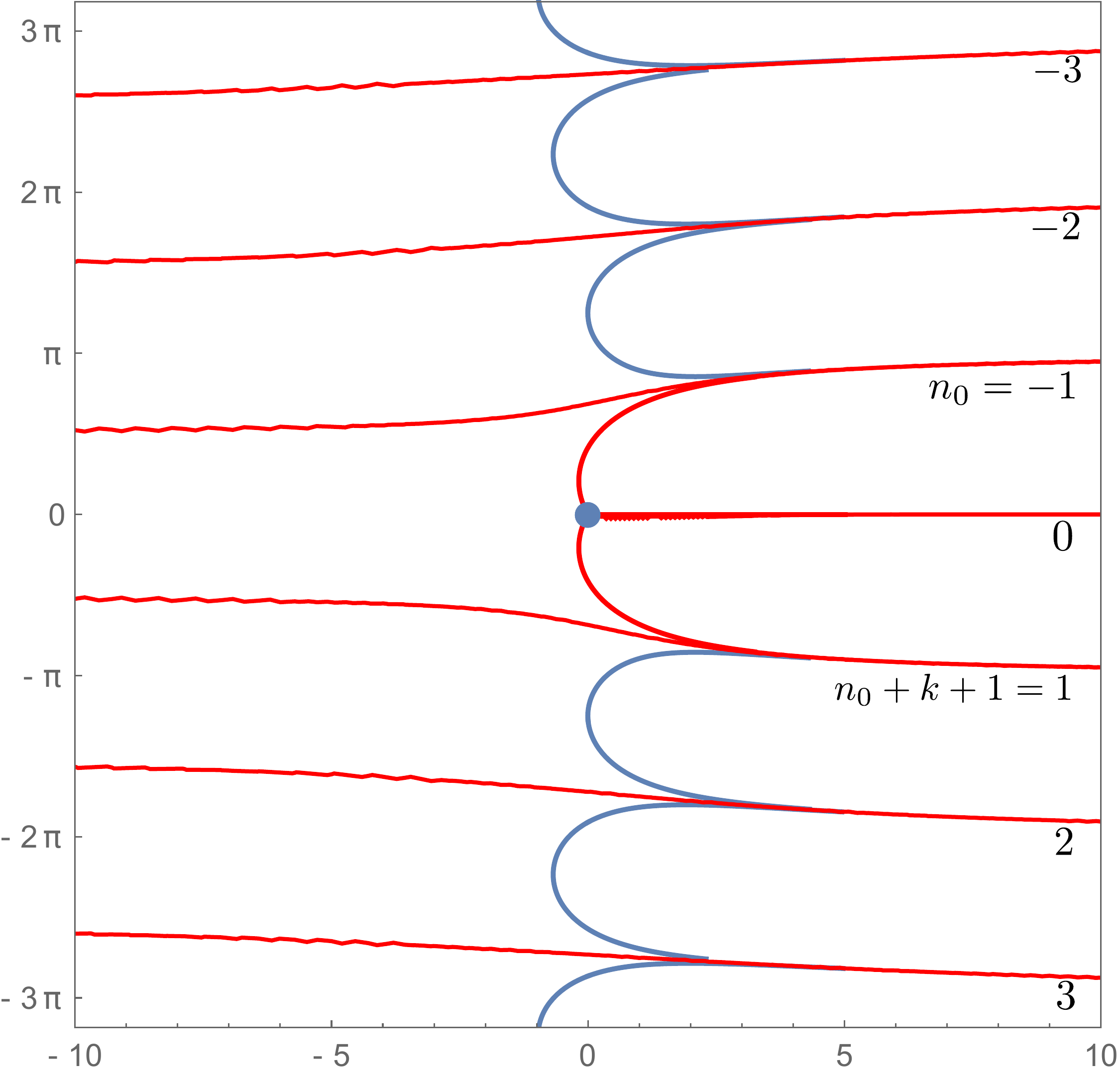}
	\includegraphics[width=0.48\linewidth]{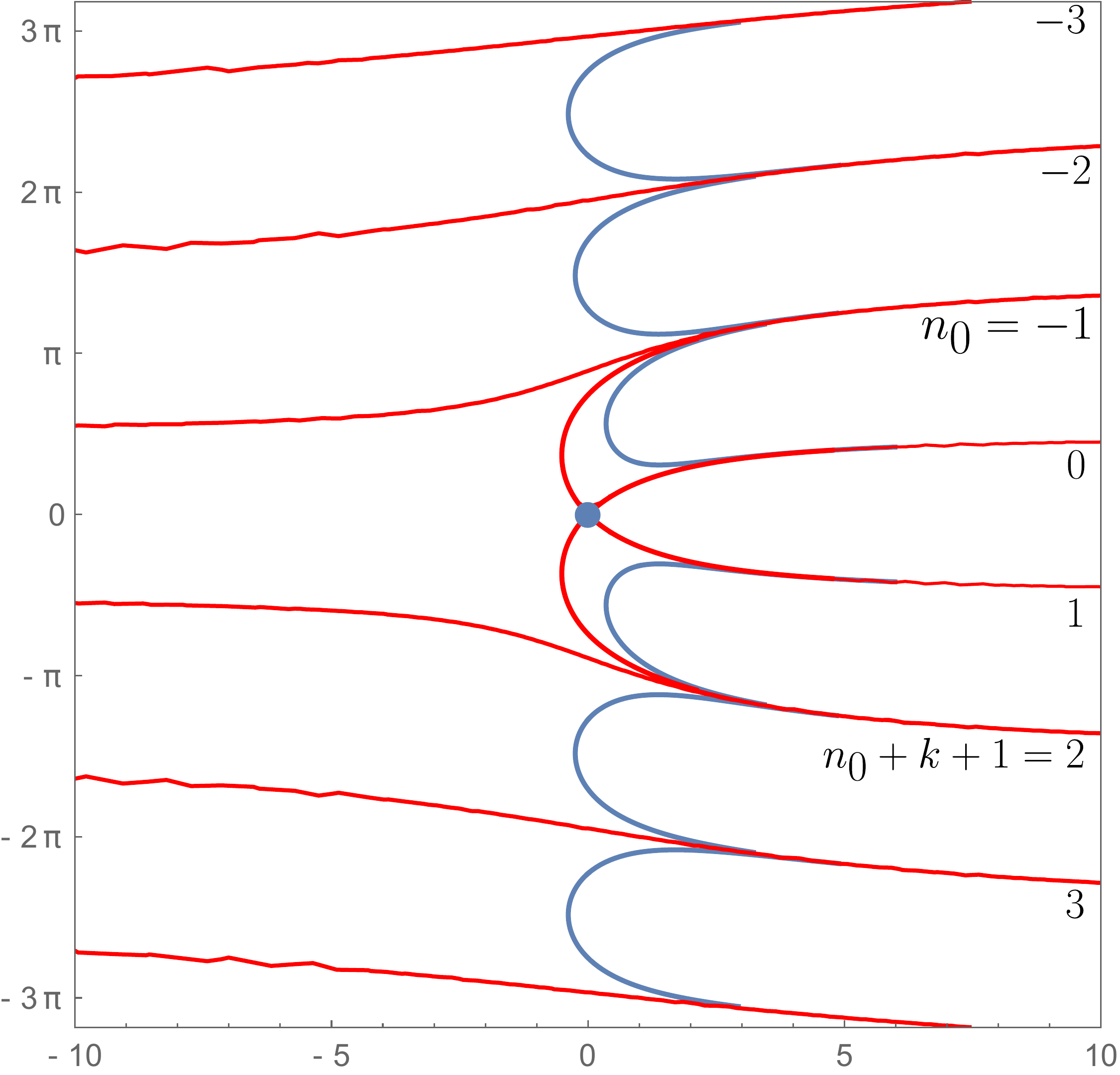}
	\caption{WKB diagram for $k=1$ (left) and $k=2$ (right). Generic flow lines and WKB lines are colored blue and red respectively. $\theta$ is chosen to be $0$ and $-i\frac{\pi}{2}$ respectively. We number the WKB lines on large positve $x$ side increasingly from top. There are $k+2$ WKB lines that are connected to the zero, numbered from $n_0$ to $n_0+k+1$.}
	\label{fig:wkbk1}
\end{figure}
The case at hand is a beautiful example of the generalized analysis. We will present the results here and a more detailed discussion in Appendix \ref{app:HigerOrderZero}. For a generic phase of $e^\theta$ one has that
\begin{itemize}
\item $k+2$ consecutive asymptotic lines at large positive real $x+\theta+i \pi n$ are connected by special WKB lines to the order $k$ zero at $x = - \frac{1}{g}$. Say that $n_0 \leq n \leq n_0 + k + 1$ for some $n_0$ which can be easily determined. This allows a WKB estimate of the Wronskians of pairs of $\psi_n$'s in this interval. With our conventions, it is just $d_{n'-n}^{(k)}$. These Wronskians compute certain $T_{n'-n}$ functions in a specific range of $\mathrm{Im} \theta$. We learn that the corresponding $L_{\frac{n'-n}{2}}[\theta]$ likely flow to ${\cal L}_{\frac{n'-n}{2}}$ topological defects. This expectation will be further solidified by the analysis of the $T_{n'-n;l}$ asymptotics. 
\item The remaining asymptotic lines get connected by special WKB lines to the asymptotic region at large negative real part of $x+\theta$, 
but the imaginary part of $x+\theta$ gets shifted by $\pm \frac{\pi}{2} k$. Say that the imaginary part increases by $\frac{\pi}{2} k$
for $n \geq n_0 + k + 1$ and decreases by $\frac{\pi}{2} k$ for $n \leq n_0$. With our conventions, up to factors of the form $e^{m_{n,n'} e^\theta}$
for some $m_{n,n'}$, one gets WKB estimates $n'-n$ for pairs of lines which are on the same side of $n_0$, $n'-n-k$ otherwise. 
These WKB estimates help us predict RG flows ending on $L^{IR}_{n'-n} \otimes {\cal L}_{\frac{k}{2}}$ or $L^{IR}_{n'-n-k}\otimes {\cal L}_{\frac{k}{2}}$ defects. 
\end{itemize}
This is enough to reconstruct the RG flows near the real $\theta$ axis. The Wronskians that do not fall into these two types can be related to these two types via Pl\"ucker formulae. This is the same as using Hirota relations to explore general $\theta$. 

We refer to Appendix \ref{app:WKBBessel} for details. 

\subsection{ODE/IM for primary fields}

We propose to identify the expectation values $T_{n;l}[\theta]$ on the $\mathfrak{su}(2)$  WZW primary fields $|l\rangle$  with the transport data of the Schr\"oedinger equation 
\begin{equation}
\partial_x^2 \psi(x) = \big[e^{ 2 \theta} e^{2x} (1 + g x)^k +\frac{l(l+1)}{(x+1/g)^2}\big] \psi(x) \label{eq:ODESU2l}
\end{equation}
The second term is the standard angular momentum term \cite{Bazhanov:1998wj}, which accounts for different highest weight modules. Again, the shift $x \to x-\frac{1}{g}$ maps the equation to 
\begin{equation}
\partial_x^2 \psi(x) =\big[ e^{2 \theta} e^{-\frac{2}{g}}g^{k} e^{2x}x^k +\frac{l(l+1)}{x^2} \big]\psi(x)
\end{equation}
so that the transport data is only a function of the combination $e^{\theta} e^{-\frac{1}{g}}g^{\frac{k}{2}}$.

An important observation is in order. The above differential equation makes sense for all values of $l$, and one can define small sections $\psi_n$ at positive infinity and 
their Wronskians $T_{n;l}[\theta]$ as for the $l=0$ case. The regular singularity at $x = -\frac{1}{g}$, though, generically changes the overall monodromy structure of the differential equation: 
solutions are not entire functions of $x$, but have a monodromy around $x = -\frac{1}{g}$. The $T_{n;l}[\theta]$ functions do not exhaust the monodromy data of the differential equation. 

For applications to an $\mathfrak{su}(2)_k$ WZW model, we are interested in integrable modules only, for which $0 \leq l \leq \frac{k}{2}$ and $2l$ is an integer. It turns out that this is a very special 
choice for the differential equation as well. Naively, a regular singularity such that $2l$ is an integer will have a logarithmic monodromy. The order $k$ zero of the regular part of the 
potential, though, forces the monodromy to be simply $(-1)^l$. In other words, the differential equation has a regular singularity of trivial monodromy at $x = -\frac{1}{g}$. 
This guarantees that the differential equation has the same type of monodromy data as the $l=0$, captured fully by the $T_{n;l}[\theta]$ functions. 

The WKB analysis also proceeds in much the same way as for the $l=0$ case, except that the local problem around $x = -\frac{1}{g}$ is modified. 
This has two consequences: 
\begin{itemize}
\item The local Wronskians $d_{2j+1}^{(k)}$ are replaced by 
 \begin{equation}
 d_{2j+1;l}^{(k)}\equiv \frac{\sin(2j+1)(2l+1)\frac{\pi}{k+2}}{\sin(2l+1)\frac{\pi}{k+2}} \label{eq:quantumDimension}
 \end{equation}
 which coincide with the expectation values of ${\mathcal L}_{j}$ in the primary tower of spin $l$.
 \item An extra $(-1)^l$ sign appears in the IR behaviour of certain Wronskians, which we interpret as the expectation values of ${\mathcal L}_{\frac{k}{2}}$ in the primary tower of spin $l$.
\end{itemize}

The perturbative analysis requires some extra care, because the angular momentum term dominates over the exponential for sufficiently negative $x$, leading to a behaviour 
\begin{equation}
\psi_0(x) \sim -\frac{Q_l(g_{\mathrm{eff}})}{2l+1} (x+\frac{1}{g})^{l+1} - \frac{\tilde{Q}_\ell(g_{\mathrm{eff}})}{2l+1} (x+\frac{1}{g})^{-l}. \label{eq:solxgg0}
\end{equation}
In the matching region $\frac{1}{g} \gg -x \gg 0$, the asymptotic behavior becomes
\begin{equation}
\psi_0(x) \sim -\frac{Q_l(g_{\mathrm{eff}})}{2l+1} \frac{1}{g^{l+1}} [1 + (l+1) g x] - \frac{\tilde{Q}_\ell(g_{\mathrm{eff}})}{2l+1} g^l [1 - l g x]. \label{eq:solxgg0}
\end{equation}
Here, we find a perturbative expansion of the two Q-functions:
\begin{align}
Q(\theta) &= \frac{g_{\mathrm{eff}}^l}{\sqrt{\pi}} \left( 1 + q_1 g_{\mathrm{eff}}(\theta) + \cdots \right)\cr
\tilde Q(\theta) &= \frac{g_{\mathrm{eff}}^{-l}}{\sqrt{\pi}} \left( - \frac{1}{g_{\mathrm{eff}}} + \tilde{q}_0 + \tilde q_1 g_{\mathrm{eff}}(\theta) + \cdots \right)
\end{align}
and derive a perturbative expansion for $T_{n;l}$:
\begin{align}
T_{n;l} \sim n -I_R\pi^2 l(l+1)g^2  &+\frac{\pi^{2}}{2}   \big[k\left(-1+l+2 l^{2}\right)-4 l(\theta+l \theta+ \tilde{q}_0)\big]  I_R g^3 \nonumber\\
+ \frac{\pi ^2 }{60}  I_R g^4 
&\bigg[-45 k^2 \left(l^2-1\right)+30 k (\tilde{q}_0 (4 l-1)+\theta  (l+1) (8 l-3)) \big. \nonumber\\
& \big. +\pi ^2 l (l+1) \left(3 n^2 \left(l^2+l+3\right)-(l+2) (7 l+13)\right) \nonumber\\
&-60 l \left(6 \theta  \tilde{q}_0+2 \tilde{q}_0^2+4 \tilde{q}_1+3 \theta ^2 (l+1)\right)+120 \tilde{q}_1 \bigg]
  \cdots \label{eq:Tnweakcoupling}
\end{align}
where $I_R = \frac{1}{6} n\left(-1+n^{2}\right)$ and $\tilde{q}_0 = -\frac{k}{4}+(1+l)(\gamma-\log 2)$. We match it with the explicit line defect calculations in Appendix \ref{app:schropert}. 

\subsection{The 4d Chern-Simons construction}

Classically, the 4d Chern-Simons gauge theory on $\mathbb{C} \times \mathbb{R}^2$ can be minimally coupled to a 2d chiral WZW model, sitting at a point 
$z=0$ in the holomorphic plane and wrapping the $\mathbb{R}^2$ topological directions. The coupling to the 4d CS theory does not induce any 
deformation of the two-dimensional WZW theory, simply because there is no spin $0$ operator in the WZW theory which could describe such a 
deformation. 

The only effect of the coupling is that it allows the WZW model to interact with Wilson lines $W_j[z]$ of the 4d CS theory, lying parallel to the surface defect 
at some separate point in the holomorphic plane. An important property of the 4d CS theory is that 
the interactions are local on the topological plane, so that the Wilson line will appear as a 2d local line defect to the 
2d degrees of freedom. The leading classical interaction takes the form 
\begin{equation}
\int \sigma^a r_{ab}(z) J^b(t) dt
\end{equation}
where $r_{ab}(z)$ is the classical rational R-matrix which takes the role of a propagator in the 4d theory. This is simply the Kondo interaction, with a coupling $g = \frac{\hbar}{i \pi z}$. 

An important quantum correction to this statement is due to the 2d gauge anomaly of the WZW model. This can be cured by a perturbative modification 
of the 1-form $\omega(z)dz$ in the 4d CS action 
\begin{equation}
\int \omega(z)dz \wedge \Omega_{CS}[A]
\end{equation}
which adds a pole at $z=0$: 
\begin{equation}
\frac{dz}{\hbar} \to \frac{dz}{\hbar} + \frac{k}{2\pi i z}dz
\end{equation}
The 4d CS perturbation theory is essentially an expansion in inverse powers of $z$, so this is a sub-leading 
correction to the classical action. 

The Wilson lines of the 4d CS theory automatically satisfy the Yangian fusion relations and, when wrapped along a compact direction in the topological plane, should give vevs which satisfy the $\mathfrak{su}(2)$ Hirota dynamics. In particular, the one form $\omega(z) dz$ controls the precise form of the line defects fusion: when $\omega(z) = 1$ it involves 
shifts of $z$ by multiples of $i\frac{\pi}{2}$, but for general $\omega(z)$ one has to compute the primitive
\begin{equation}
\theta = -i \pi \frac{z}{\hbar} - \frac{k}{2} \log z = -\frac{1}{g} +  \frac{k}{2}\log g \label{eq:primitive}
\end{equation}
such that $d \theta = \omega(z) dz$. Then the fusion relations involve shifts of $\theta$ by multiples of $i\frac{\pi}{2}$.

We therefore identify $\theta$ as the ``spectral parameter'' of the Wilson lines, which is exactly what we found in the purely 2d analysis!

Our analysis is compatible with yet unpublished work \cite{CostelloIV} demonstrating the existence of a renormalization scheme for 4d CS theory coupled to 2d chiral matter,
with the property that the $g = \frac{\hbar}{i \pi z}$ is not renormalized, and RG flow only affects the position of Wilson line defects by a uniform shift of the $\theta$ local coordinate, 
i.e. the beta function for $z$ is proportional to $\omega(z)^{-1}$. 

\section{Expected generalizations}
\subsection{Multichannel Kondo problems}\label{sec:multichannelSU2}

The simplest generalization of the Kondo defects is to consider a theory with multiple $\mathrm{su}(2)_{k_i}$ WZW currents and couple them all to the same line defect by a coupling 
\begin{equation}
\int \sigma_a \sum_i g_i J_i^a dx^0
\end{equation}
which, in 4d CS setup, corresponds to taking multiple chiral WZW surface defects of levels $k_i$ at positions $z_i$. 
This results in a classical coupling of the schematic form 
\begin{equation}
\int \sigma_a \sum_i \frac{1}{z-z_i}J_i^a dx^0
\end{equation}
Assuming the classical couplings are not corrected, we get an immediate prediction: two such line defects should give commuting transfer matrices 
if the couplings can be written as $g_i = \frac{1}{z-z_i}$ and $g'_i = \frac{1}{z'-z_i}$.

A second prediction is that this one-parameter family of commuting defects would be connected by RG flow, 
with the RG flow translating the $z$ parameter according to the 1-form 
\begin{equation}
dz+ \sum_i \frac{k_i}{2 (z-z_i)}dz
\end{equation}
This gives RG flow equations
\begin{equation}
\mu \partial_\mu g_i =  -\frac{g_i^2}{1 + \sum_j \frac{k_j}{2} g_j}\label{eq:betafunctionmultiSU2}
\end{equation}
which can be checked against explicit 2d perturbative calculations. Details can be found in Appendix \ref{app:multisu2}.
We should stress that the perturbative match of the RG flow equations is rather non-trivial in the multi-channel case, as redefinitions of the couplings 
leave invariant infinitely many combinations of beta function coefficients, rather than the single ``$c$'' we found in the single coupling case. 

There is a simple proposal for an ODE/IM solution for the expectation values of the line defect: they should coincide with the transport data for 
the equation
\begin{equation}
\partial_x^2 \psi(x) = \big[e^{2\theta+2x}  \prod_{i} (1+g_ix)^{k_i} + u(x)^2-u'(x) \big]\psi(x)
\end{equation}
where
\begin{equation}
u(x) = \sum_{i} \frac{l_i}{x+1/g_i}
\end{equation}
We can define an overall scaling parameter $g\equiv 1/z$ and set $g_i = g/(1+z_i g)$ where, say, $\sum_i z_i =0$. Performing a translation of $x$, we can rewrite the equation as
\begin{equation}
\partial_x^2 \psi(x) = \Bigg[ e^{2\theta} e^{-2/g} g^{\sum_i k_i} \prod_{i}(1+g z_i)^{-k_i} e^{2 x} \prod_{i}(x+z_i)^{k_i} + \Big( \sum_i \frac{l_i}{x + z_i} \Big)^2 + \sum_{i} \frac{l_i}{(x+z_i)^2} \Bigg]\psi(x)
\end{equation}
and the transport data will only depend on $\{z_i \}$ and the combination
\begin{equation}
 e^{-1/g_{\text{eff}}(\theta)} g_{\text{eff}}(\theta)^{\sum_i k_i/2} \equiv e^{\theta} e^{-1/g} g^{\sum_i k_i/2} \prod_{i}(1+g z_i)^{-k_i/2}
\end{equation}
where we define the effective coupling $ g_{\mathrm{eff}}$, analogous to \eqref{eq:geffDef}.

The transport data can be defined as before in Section \ref{sec:SU2ODEIM}. A primary state $|l_1,l_2,\dots\rangle$ is labeled by a list of half integers, one for each $\mathfrak{su}(2)$ factor. We therefore make the following identification
\begin{equation}
\langle l_1,l_2,\dots|\hat{T}_n|l_1,l_2,\dots\rangle \equiv T_{n;l_i}(\theta) = i \left(\psi_0(x;\theta - \frac{i \pi n}{2}), \psi_0(x;\theta +\frac{i \pi n}{2}) \right)
\end{equation}
In the UV, it will have a perturbative expansion in $g_i$ around $T_{n;l_i} \sim n$. We perform the calculation in detail in Appendix \ref{app:scropertMultiSU2}. The result matches nicely with the direct 2d perturbative calculations.

In the IR, the WKB analysis can be done in a straightforward way, with relatively simple answers for real values of the $g_i$. 

An entertaining check is that if two $g_i$'s coincide, the equation is the same as for a model with one fewer WZW factors. This is reasonable: a coupling involving a sum $J^a_i + J^a_j$ with equal coefficients naturally factors through the WZW model of level $k_i + k_j$ defined by the total currents $J^a_i + J^a_j$, with the remaining coset model decoupling from the line defect. 

Notice that the Schr\"oedinger equation seems to take a universal form 
\begin{equation}
\partial_x^2 \psi(x) =\left[ e^{ 2 \theta}e^{2p(x)} + u(x)^2-u'(x)\right]\psi(x)
\end{equation}
where $\partial_x p(x) = \omega(x)$. We will see momentarily that this statement holds for other examples as well. We expect it to hold universally for any purely chiral 4d $SU(2)$ CS setup.  We will explore this point further, as well as relations to affine Gaudin models and affine Geometric Langlands, in a future publication.\cite{Gaiotto:2020dhf}

\subsection{Coset Kondo lines} \label{sec:coset}
A well studied class of examples of ODE/IM correspondence involves polynomial potentials 
\begin{equation}
\partial^2_x \psi(x) = e^{ 2 \theta} P_n(x) \psi(x)
\end{equation}
where $P_n$ is a polynomial of degree $n$, say with $m$ zeroes of order $k_i$. 

Based on the various examples in the literature e.g. \cite{Dorey:2007ti}, it is easy to guess that this differential equation should control 
the vacuum expectation values of Kondo lines in coset models of the form 
\begin{equation}
\frac{\prod_{i=1}^m \mathrm{su}(2)_{k_i}}{\mathrm{su}(2)_{\sum_i k_i}}
\end{equation}
 
The integrable Kondo defects are deformations of certain topological line defects by the chiral coset primary fields with coset labels 
$[1;3]$. These are the primary fields $\Phi^{(i)}$ which appear in the coset decomposition of the $\mathrm{su}(2)_{k_i}$ currents $J_a^{(i)}$: 
\begin{equation}
J_a^{(i)} = \Phi^{(i)} \otimes \phi^a + \cdots
\end{equation}
with $\phi^a$ being the spin 1 primary of the diagonal $\mathrm{su}(2)_{\sum_i k_i}$. 
There are $m-1$ such coset fields, so the Kondo defects have $m-1$ couplings of the same scaling dimension. They are mapped to the relative positions of the 
zeroes of $P_n(x)$.

The basic topological line defects which support $\phi^a$ local operators are these labelled by primary fields of $\mathrm{su}(2)_{\sum_i k_i}$. 
There are $n-1$ of them, as the identity line or the spin $\frac{n}{2}$ do not support non-trivial primaries. They match nicely the possible Wronskians 
built from the $n+2$ small solutions for the ODE. 

The RG flows admit a perturbative UV description if at least one of the levels is large, so that the scaling dimension of the $\phi^a$ is close to $1$. 
For example, if one of the levels $\kappa$ is large while the others are kept finite, so that we study the coset 
\begin{equation}
\frac{\mathfrak{su}(2)_\kappa \times \prod_i \mathrm{su}(2)_{k_i}}{\mathfrak{su}(2)_{\kappa+ \sum_i k_i}}
\end{equation}
and the ODE will be 
\begin{equation}
\partial^2_x \psi(x) = \left[ \prod_i (1 + g_i x)^{k_i} \right] x^\kappa \psi(x)
\end{equation}

We expect this to be a ``trigonometric'' 4d CS setup, where the holomorphic direction is a $\mathbb{C}^*$ with 
local coordinate $z = e^w$ and the classical differential is 
\begin{equation}
\omega = \frac{\kappa}{2 z}dz
\end{equation}
corrected by the coupling to 2d WZW models to 
\begin{equation}
\omega = \frac{\kappa}{2 z}dz+ \sum_i \frac{k_i}{2 (z-z_i)}dz
\end{equation}

\subsection{WZW vs Kac-Moody}
At the expense of ruining unitarity, we can replace the WZW currents at integral level $k$ with Kac-Moody currents at some generic level $\kappa$. 
At the level of perturbation theory there is no difference. Non-perturbatively, there must be deep differences, as most of the RG statements we made 
for integral $k$ do not have a natural extension to non-integral $\kappa$. 

For large $\kappa$ and finite $j$, the perturbative considerations still indicate the RG flow $L^{UV}_j \to {\cal L}_j$. The topological lines 
${\cal L}_j$ in Kac-Moody exist for all $j$, but non-perturbative effects should kick in as $j \simeq \kappa$. 

As the spin $j$ is integral, one cannot make sense directly of ``${\cal L}_{\frac{\kappa}{2}}$'' which appears in the RG flows at integral $k$.
The spin $\frac{\kappa}{2}$ primary in WZW models, though, has another interpretation: it is the image of the vacuum module under a spectral flow operation. 
This suggests that the large $j$ RG flow in the Kac-Moody theory may land on topological lines associated to spectral flowed modules. 

Another interesting new wrinkle is that once we compute the $\hat T_j$, we can subject the Kac-Moody current modes in them to a spectral flow operation.
As the $\hat T_n$ commute both with $L_0$ and $J_0^3$, the image under $w$ units of spectral flow will also give a conserved operator $\hat T_{n;w}$.
This suggests we should be able to define spectral flow images $L_{j;w}[\theta]$ of the usual $L_{j}[\theta]$ defects. The UV definition may be a bit subtle, 
but the notion should be well-defined. 

We will now propose an ODE/IM interpretation of the $T_{n;w}[\theta]$, which suggests how one may compute the IR image of $L_{j;w}[\theta]$ 
or postulate new sets of Hirota equations controlling their fusion. 

The ODE for general $\kappa$
\begin{equation}
e^{- 2 \theta} \partial_x^2 \psi(x) = (1 + g x)^\kappa e^{2x} \psi(x)
\end{equation}
has a branch cut from $x\to-\infty$ to $x = -\frac{1}{g}$. Consequently, one can take some 
small solution $\psi_n(x;\theta)$, defined in the usual way, analytically continue it $w$ times around $x = -\frac{1}{g}$
to obtain a solution of 
\begin{equation}
e^{- 2 \theta- 2 \pi i \kappa w} \partial_x^2 \psi(x) = (1 + g x)^\kappa e^{2x} \psi(x)
\end{equation}
and take a Wronskian with some $\psi_{n'}(x;\theta+ \pi \kappa w)$. 

Up to picking some convention for the shifts of $\theta$,
this gives a possible definition of $T_{n'-n;w}[\theta]$. Pl\"ucker relations give a slew of new Hirota-like 
formulae controlling the fusion of $T_{n,w}$ functions with all sort of spectral flow amounts and $\theta$ shifts by 
multiples of $i \pi$ and $i \pi \kappa$. 

The WKB analysis of the ODE is straightforward, although the details depend somewhat sensitively on choices such as the sign of the real part of $\kappa$, etc. 
The main novelty is that some of the WKB lines will go across the cut, so that the collection of Wronskians with ``good'' WKB asymptotics 
may include some $T_{n;w}[\theta]$ with $w \neq 0$.

\acknowledgments
We thank K.Costello for participation in early stages of the project and countless in depth discussions. We thank N.Dorey and B.Vicedo for several discussions. 
The research of D.G., J.H.L., and J.W. is supported in part by a grant from the
Krembil foundation by the Perimeter Institute for Theoretical Physics. Research at Perimeter Institute is supported in part by the Government of Canada through the Department of Innovation, Science and Economic Development Canada and by the Province of Ontario through the Ministry of Economic Development, Job Creation and Trade.

\appendix

\section{Exact solutions of the harmonic oscillator  ODE}\label{app:IsingODEexactsol}
In this section we study Wronskians of exact solutions of the ODE for Ising model discussed in  \ref{sec:IsingODE}
\begin{equation}
e^{- 2 \theta} \partial_x^2 \psi(x) = (x^2 - 2) \psi(x)
\end{equation}
which can be solved easily using, say, parabolic cylinder functions $U(a,z)$, the standard solutions of 
\begin{equation}
\frac{\mathrm{d}^{2} w}{\mathrm{d} z^{2}}-\left(\frac{1}{4} z^{2}+a\right) w=0\label{eq:app:parabolicODE}
\end{equation}
It is related to the Whittaker and Watson's parabolic cylinder functions, commonly used in, say, Mathematica, by
$D_{\nu}(z)=U\left(-\frac{1}{2}-\nu, z\right)$. It has the property that if $U(a,z)$ is a solution to \eqref{eq:app:parabolicODE}, $U(a,\pm z)$ and $U(-a,\pm i z)$ are also solutions and all of them are entire functions of $a$ and $z$. Given the large $z$ asymptotics
\begin{equation}
U(a, z) \sim \mathrm{e}^{-\frac{1}{4} z^{2}} z^{-a-\frac{1}{2}}+\dots, \qquad |\mathrm{phase} (z)|< \frac{3}{4}\pi \label{eq:app:UazAsymp}
\end{equation} 
we identify $\psi_0$, the small solution along the ray of $e^{-\theta/2}$ defined in \eqref{eq:IsingPsi0asy}, to be 
\begin{equation}
\psi_0=2^{-1 / 4} e^{\frac{e^\theta}{2}} e^{-\theta(2e^\theta+1)/4} U\left[-e^\theta, \sqrt{2}e^{\theta/2} z\right]
\end{equation}
and infinite other solutions
\begin{equation}
\psi_n(x;\theta) \equiv \psi_0(x;\theta+ i \pi n) 
\end{equation}
With the help of the Wronskians 
\begin{align}
(U(a, z), U(a,-z))&=\frac{\sqrt{2 \pi}}{\Gamma\left(\frac{1}{2}+a\right)}\\
 (U(a, z), U(-a, \pm iz))&=\mp i \exp [{\pm i\pi \left(\frac{1}{2} a+\frac{1}{4}\right)}]
\end{align}
We obtain the results in \ref{sec:IsingODE}
\begin{align}
i(\psi_n,\psi_{n+1}) =1, &\qquad i(\psi_{-1},\psi_{2}) =e^{-2\pi ie^{\theta}}\\
  i(\psi_{-1},\psi_{1}) =& \frac{\sqrt{2 \pi} e^{\theta e^{\theta}-e^{\theta}}}{\Gamma(\frac12 +  e^{\theta})} 
\end{align}
%%%%%%%%%%%%%%%%%%%%%%%%%%%%%%%%%%%%%%%%%%%%%%%%%%%%

\section{WKB analysis}\label{app:WKB}

\subsection{Opers}
Consider a Riemann surface $C$ (equipped with a spin structure) and a classical stress tensor $T(x)$, which transforms between coordinate patches 
as 
\begin{equation}
T(x)= \left(\frac{d\tilde x}{dx}\right)^2 \tilde T(\tilde x) - \frac{2 \frac{d\tilde x}{dx}\frac{d^3 \tilde x}{dx^3}-3 \frac{d^2 \tilde x}{dx^2}\frac{d^2 \tilde x}{dx^2}}{4\left(\frac{d \tilde x}{dx}\right)^2}
\end{equation}
The difference of two classical stress tensors is a quadratic differential. 

A classical stress tensor can be used to define globally a Schr\"oedinger equation
 \begin{equation}
\partial_x^2 \psi(x) - T(x) \psi(x) =0
\end{equation}
which behaves well under coordinate transformations if $\psi$ transforms appropriately:
\begin{equation}
\psi(x) = \frac{1}{\sqrt{\frac{d \tilde x}{dx}}} \tilde \psi(\tilde x)
\end{equation}

The Wronskian of two solutions
\begin{equation}
W(\psi,\psi') = \psi \partial_x \psi' - \psi' \partial_x \psi
\end{equation}
is constant and invariant under coordinate transformations. 

If we know a solution $\psi(x)$, we can get a second independent solution by quadrature:
\begin{equation}
\psi'(x) = \psi(x) \int^{x} \frac{dx'}{\psi(x')^2}
\end{equation}

\subsubsection{Special coordinates}
The ratio of two solutions with Wronskian $1$  
\begin{equation}
s(x) = \frac{\psi'(x)}{\psi(x)}
\end{equation}
gives a map $C \to \mathbb{C}P^1$ defined up to monodromies in $\mathrm{SL}(2,\mathbb{C})$. 
It is also a local coordinate such that the Schr\"oedinger operator reduces to $\partial_s^2$. 

Notice that 
\begin{equation}
\partial_x s(x) = \frac{\psi'(x)}{\psi(x)} = \frac{1}{\psi(x)^2}
\end{equation} 
and thus the map is non-singular.

\subsubsection{Transport coefficients}
The transport coefficients of a Schr\"oedinger equation are defined as the point in an appropriate space of flat $\mathrm{SL}(2,\mathbb{C})$ connections 
defined by parallel transport of the solutions. The data of the flat connections depend on the types of singularities 
of $T(x)$:
\begin{itemize}
\item If $T(x)$ is holomorphic, the transport coefficients encode the monodromy of the solutions around cycles of $C$. 
\item If $T(x)$ has regular singularities, at which 
\begin{equation}
T(x) \sim \frac{m^ 2- \frac14}{(x-x_0)^2}
\end{equation}
then one also has monodromies around the regular punctures. For generic $m$, one can define monodromy eigenvectors 
$\psi^{x_0}_{\pm}(x)$ and express the transport coefficients in terms of Wronskians between eigenvectors transported along various paths on $C$. 
\item If $T(x)$ has irregular singularities, at which 
\begin{equation}
T(x) \sim \frac{c}{(x-x_0)^{r+2}}
\end{equation}
then one also has $r$ Stokes matrices at the irregular puncture. One can define $r$ special solutions $\psi^{x_0}_{i}(x)$
which decay exponentially fast along appropriate rays towards $x_0$. The transport coefficients can be expressed in terms of Wronskians between such solutions transported along various paths on $C$. 
\item We will also be interested in exponential singularities, at which 
\begin{equation}
T(x) \sim e^{\frac{c}{x-x_0}}
\end{equation}
We will see that at such a singularity one can define an infinite sequence of special solutions $\psi^{x_0}_{i}(x)$.
\end{itemize}

\subsection{WKB asymptotics}
Consider now the linear family 
\begin{equation}
T(x) = \frac{U(x)}{\hbar^2} + \frak{t}(x)
\end{equation}
where $\frak{t}(x)$ is a reference stress tensor and $U(x)$ a quadratic differential. When we have singularities, $\frak{t}(x)$ should not be more singular than 
$U(x)$. 

The solutions and transport coefficients for the corresponding Schr\"oedinger equation have a very rich 
asymptotic behaviour as $\hbar \to 0$. This has been studied by a vast literature on the WKB approximation, 
culminating in the Voros analysis \cite{VOROS1983,AIF_1993__43_1_163_0,aoki2005virtual}. Useful insights can also be inherited by the WKB analysis of 
the Lax connection of Hitchin systems\cite{GMN:2009hg,Gaiotto:2012rg}, with the help of a certain ``conformal limit'' \cite{Gaiotto:2014bza}.

Take $\mathrm{arg}\,\hbar$ to lie in an interval $[\vartheta - \frac{\pi}{2}, \vartheta + \frac{\pi}{2}]$. 
The ``GMN-style'' WKB analysis focusses on the complement $\overline {\cal S}_\vartheta$ of the ``spectral network'' ${\cal S}_\vartheta$ of flow lines 
\begin{equation}
e^{- i \vartheta} \sqrt{U(x)}dx \in \mathbb{R}
\end{equation}
originating from the zeroes of $U(x)$. 

We will focus on situations where $U(x)$ has at least one regular or irregular singularity, so that the flow lines generically 
end at the singularities. Each point $x_p$ away from ${\cal S}_\vartheta$ belongs to a flow line which goes from a singularity to a singularity. 
We can associate to $x_p$ two ``small solutions'' $\psi_\pm^{x_p}(x)$, defined up to rescaling, which are the parallel transport along the flow line of 
solutions which decay to zero as they approach the two singularities along the flow line. 

For convenience, denote as $\psi_a(x)$ the collection of small solutions which are selected at the singularities 
by the above procedure. These will include all the $\psi^{x_s}_{i}(x)$ at all irregular singularities and one of the two monodromy eigenvectors 
$\psi^{x_s}_{\pm}(x)$ at the regular singularities. We normalize each of the $\psi_a(x)$ once and for all in some way at each singularity. 
Then we have 
\begin{equation}
\psi_\pm^{x_p}(x) \sim \psi_{a^{x_p}_\pm}(x)
\end{equation}
with some normalization coefficients we will now estimate. 

A straightforward WKB analysis indicates that $\psi_\pm^{x_p}(x)$ are ``WKB solutions'' in the connected component of $\overline {\cal S}_\vartheta$ to which $x_p$
belongs, in the sense that 
as $\hbar \to 0$ with $\mathrm{arg}\,\hbar \in [\vartheta - \frac{\pi}{2}, \vartheta + \frac{\pi}{2}]$ one has 
\begin{equation}\label{eq:WKBsolutionatxp}
\psi_\pm^{x_p}(x) \sim \frac{1}{\sqrt{2 p(x;\hbar)}}e^{\pm \int_{x_p}^x p(x;\hbar) }
\end{equation}
where the WKB one form $p(x;\hbar) dx$ is recursive solution of 
\begin{equation}
p(x;\hbar)^2 + \frac34 \frac{(\partial_x p(x;\hbar))^2}{p(x;\hbar)^2} -\frac12 \frac{\partial^2_x p(x;\hbar)}{p(x;\hbar)} = \frac{U(x)}{\hbar^2} + \frak{t}(x)
\end{equation}
of the form 
\begin{equation}
p(x;\hbar) = \frac{\sqrt{U(x)}}{\hbar} + \hbar p_1(x) + \hbar^3 p_2(x) + \cdots\end{equation}

We can compare the relative normalization of $\psi_\pm^{x_p}(x)$ and $\psi_{a^{x_p}_\pm}(x)$
and thus compute the asymptotic behaviour of the Wronskians 
\begin{equation}
W(\psi_{a^{x_p}_+}(x), \psi_{a^{x_p}_-}(x) )_{x_p}
\end{equation} 
where the two small solutions are compared along the flow line passing by $x_p$. 

The Wronskian is controlled by the integral of $p(x;\hbar)$ along the flow line. In particular, the leading asymptotics are controlled by 
the periods of $y dx$ on the spectral curve $y^2 = U(x)$. 

We can think about $p(x;\hbar)$ as the Jacobian of a local coordinate transformation which maps the Schr\"oedinger operator to $\partial_s^2-1$.
It can be thus defined as an actual function as 
\begin{equation}
p(x;\hbar) = \frac{W(\psi_{a^{x_p}_+}(x), \psi_{a^{x_p}_-}(x) )_{x_p}}{2 \psi_{a^{x_p}_+}(x) \psi_{a^{x_p}_-}(x)}
\end{equation}

\subsection{Simple zeroes vs higher order zeroes}\label{app:HigerOrderZero}
If $U(x)$ has only simple zeroes, the GMN-style analysis is sufficient to completely characterize the 
asymptotics of the transport data, as the (cross-ratios of) Wronskians along flow lines are a complete 
collection of local coordinates on the space of flat connections. 

If $U(x)$ has higher-order zeros, then the Wronskians along flow lines are not enough and
we need to compare solutions in the neighbourhood of the zeroes by a more refined analysis, which we will develop below. Even for simple zeroes, this analysis is an important sanity check on the GMN-style analysis. 

\subsubsection{Comparison at a simple zero}
Near a simple zero $x_1$, we should be able to find a local coordinate $s$ such that the 
Schr\"odinger operator has the Airy form $\partial_s^2-s$. In such a local coordinate, the solution that fast decays along the positive real axis is given by Airy function $\sqrt{2\pi}\textrm{Ai}(s)$ with the large $s$ asymptotics
\begin{equation}\label{eq:niceSolAroundSimpleZeroAsymp}
\frac{1}{\sqrt{2}} s^{-\frac14} e^{-\frac23 s^{\frac32}}
\end{equation} 
We define the three nice solutions as
\begin{equation}\label{eq:niceSolAroundSimpleZero}
\mathrm{Ai}_a(s) \equiv \sqrt{2\pi} e^{-\frac{\pi i a}{3}} \mathrm{Ai}(e^{\frac{2 \pi i a}{3}} s)
\end{equation} 
which have Wronskian $W(\mathrm{Ai}_a(s),\mathrm{Ai}_{a+1}(s))=-i$.  

Back in the original coordinate, the corresponding solutions, defined in \eqref{eq:WKBsolutionatxp} take the form 
\begin{equation}
\psi_a^{x_1}(x) =\frac{1}{\sqrt{\partial_x s(x;\hbar)}} \textrm{Ai}_a\left(s(x;\hbar) \right)
\end{equation}

The local coordinate must solve the equation 
\begin{equation}
s(x;\hbar) \partial_x s(x;\hbar)^2 + \frac34 \frac{\partial^2_x s(x;\hbar)^2}{\partial_x s(x;\hbar)^2} -\frac12 \frac{\partial^3_x s(x;\hbar)}{\partial_x s(x;\hbar)} = \frac{U(x)}{\hbar^2} + \frak{t}(x)
\end{equation}

Using the asymptotics of the Airy functions we can match the $\psi_a^{x_1}(x)$ with the small solutions in the contiguous regions of $\overline {\cal S}_\vartheta$, i.e the three $\psi_a(x)$ associated to the three singularities reached by the flow lines originating at $x_1$. 
The Wronskians evaluated at $x_1$ obviously coincide with the Wronskians evaluated in the contiguous regions of $\overline {\cal S}_\vartheta$.

\subsubsection{Comparison at a zero of order $n$}
Near a zero $x_n$ of order $n$, we should be able to find a local coordinate $s$ such that the 
Schr\"odinger operator has the form $\partial_s^2-s^n$. In such a local coordinate, 
a solution which decays along the positive real axis takes the form 
\begin{equation}
A_n(s) = \sqrt{\frac{2 s}{\pi(n+2)}} K_{\frac{1}{n+2}}\left(\frac{2}{n+2} s^{1+\frac{n}{2}}\right)
\end{equation} 
with large $s$ asymptotics 
\begin{equation}
A_n(s) \sim \frac{1}{\sqrt{2} s^{n/4}} e^{-\frac{2}{n+2} s^{1+\frac{n}{2}}}
\end{equation} 

We can produce $n+2$ solutions by a rotation 
\begin{equation}\label{eq:niceSolAroundSecondOrderZero}
A_{n;a}(s) = e^{-\frac{\pi i}{n+2} a} A_n(e^{\frac{2 \pi i}{n+2} a} s) 
\end{equation} 
Setting $n=1$, we have $A_{1;a}(s) = \textrm{Ai}_a(s)$. Because $A_n(s)$ only involves powers $s^{k(n+2)}$ and $s^{k(n+2)+1}$, we can write 
\begin{equation}
A_{n;a}(s) = e^{-\frac{ \pi i}{n+2} a} F_n(s) + e^{\frac{ \pi i}{n+2} a} G_n(s)
\end{equation}
where $F_n$ and $G_n$ involve respectively the $s^{k(n+2)}$ and $s^{k(n+2)+1}$ powers. It follows that 
\begin{equation}
A_{n;a-1}(s) + A_{n;a+1}(s) = \left(e^{\frac{\pi i}{n+2}} + e^{- \frac{\pi i}{n+2}}\right)A_{n;a}(s)\label{eq:connectionformula}
\end{equation} 
The Wronskian of consecutive solutions is $-i$. With this, we can compute $(A_{n;a},A_{n;b})$ for any $a$ and $b$.

We have corresponding solutions 
\begin{equation}
\psi_a^{x_n}(x) =\frac{1}{\sqrt{\partial_x s(x;\hbar)}} A_{n;a}\left(s(x;\hbar) \right)
\end{equation}
which satisfy the same linear relations. We can match them to the $n+2$ solutions $\psi_a(x)$ associated to the $n+2$ singularities reached by the flow lines originating at $x_n$.

\subsection{Examples:}

\subsubsection{Quadratic potential}
\begin{equation}
\hbar^2\partial_x^2 \psi(x) = (x^2-E)\psi(x) \label{eq:app:secondorderODE}
\end{equation}
where $E>0$. There are two first order zeros at $x = \pm \sqrt{E}$ and one irregular singularity at $x = \infty$. There are four anti-Stokes lines connected to the infinity, so three are four small solutions. We will normalize (and regularize) these four solutions, according to the recipe in the general discussion above. Let's start with $\psi_0(x)$, which is defined to be, up to the overall normalization,  the unique solution that decays exponentially fast towards infinity along the positive real line. Let $I_0(x)$ be an anti-derivative of $\sqrt{x^2-E}$, in an angular sector around the infinity that contains the positive real line, then the statement of the WKB approximation is that 
\begin{equation}
\psi_0(x) \sim C_0 \frac{\sqrt{\hbar}}{\sqrt{2} (x^2-E)^{1/4}} e^{-\frac{1}{\hbar} I_0(x)},
\end{equation}
for any $x$ on the WKB flow lines connected to the positive infinity. One might want to choose be $I_0(x) = \int_{\infty}^{x} \sqrt{y^2-E}dy$. This integral clearly diverges, and we regularize by
\begin{equation}
I_0(x) = \lim\limits_{L\rightarrow \infty} \left(\int_{L}^{x} dy\sqrt{y^2-E} + \frac{L^2}{2}-\frac{1}{2} E \log L +D_0 \right)
\end{equation}
It is useful to know its asymptotics towards $x\rightarrow \infty$. 
\begin{align}
\psi_{0} &\sim C_0 \frac{\sqrt{\hbar}}{\sqrt{2} (x^2-E)^{1/4}} e^{-\frac{1}{\hbar}I_0(x;D_0)}\label{eq:secondOrderODEpsi0}\\
&\overset{x\rightarrow \infty}{\sim} C_0 \frac{\sqrt{\hbar}}{\sqrt{2x}} e^{-\frac{1}{\hbar}D_0} x^{\frac{E}{2\hbar}} e^{-\frac{1}{\hbar} \frac{x^2}{2}}
\end{align}
and similarly we have a small solution $\psi_1$ in the direction $\sqrt{\hbar}e^{-\frac{i\pi}{2}}$. 
\begin{equation}
\psi_{1} \sim C_1 \frac{-i\sqrt{\hbar}}{\sqrt{2} (x^2-E)^{1/4}} e^{\frac{1}{\hbar}I_1(x;D_1)}
\end{equation}
where $I_1(x)$ is defined in the same way as $I_0(x)$ except $L$ is taken towards infinity in the direction $\sqrt{\hbar}e^{-\frac{i\pi}{2}}$. And we get two more constants $C_1$ and $D_1$ we need to fix. We fix them once and for all by defining $C_0 = 1$, $D_0 = -\frac{1}{4}\left(E+2E\log 2-E\log E \right)$ and for $n\in \mathbb{Z}$
\begin{equation}\label{eq:secondOrderODEDefpsin}
\psi_n(x;\hbar) \equiv  \psi_0(x;\hbar e^{-i\pi n})
\end{equation}
It is easy to check that $\psi_n(x;\hbar) $ satisfy the differential equation \eqref{eq:app:secondorderODE} and decrease exponentially fast along rays of direction $\sqrt{\hbar}e^{-\frac{in\pi}{2}}$. \eqref{eq:secondOrderODEDefpsin} fixes $D_n$ such that their dependence will drop out in calculating wronskians. This amounts to fixing the ambiguity of the ground state energy in the $T$ functions. The choice of $C_i$ is determined by matching the Wronkians with the standard normalization of $T$ functions.
\begin{figure}
	\centering
	\includegraphics[width=0.7\linewidth]{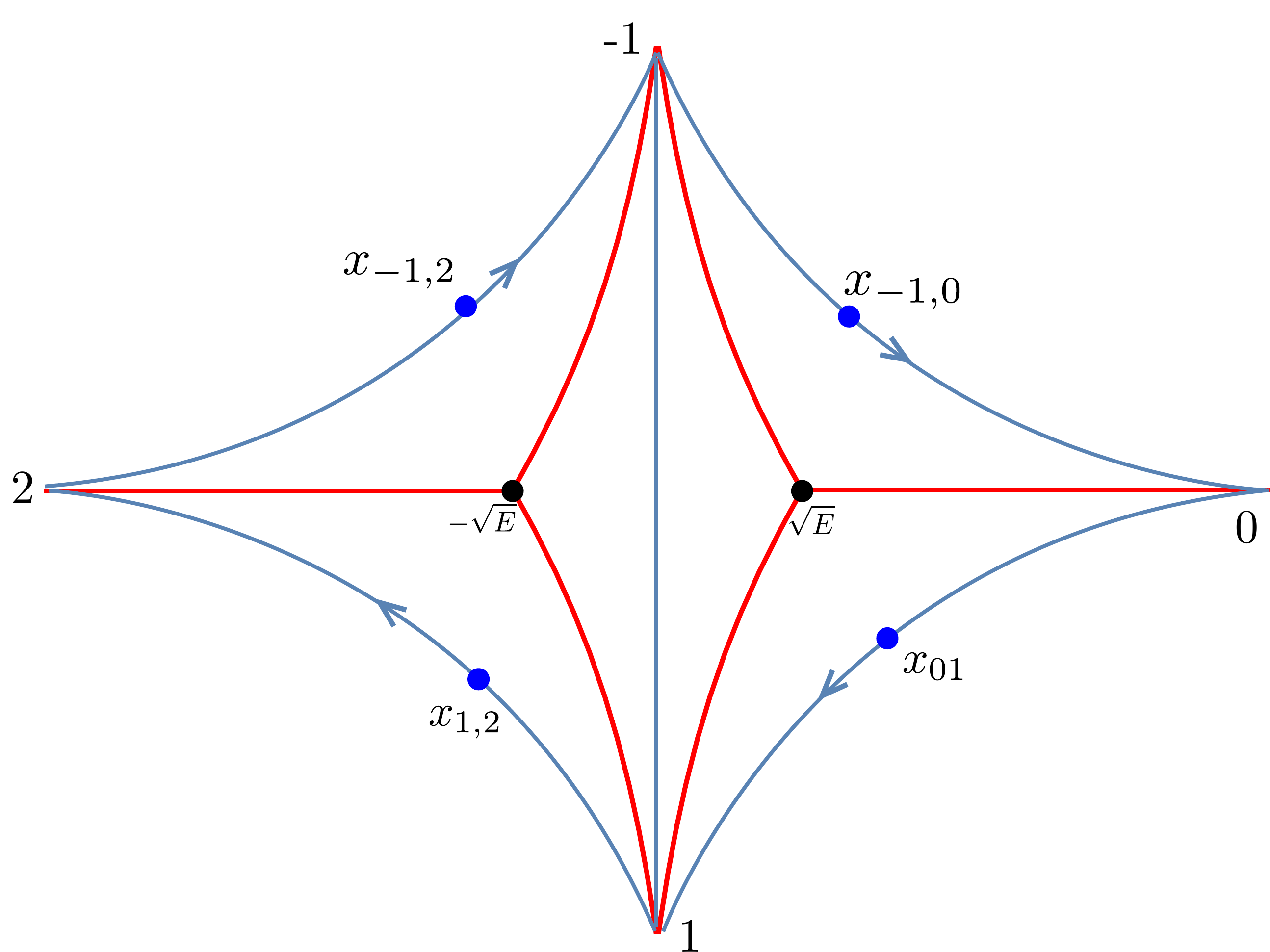}
	\caption{WKB diagram for the differential equation \eqref{eq:app:secondorderODE}. Generic flow lines and WKB lines are colored blue and red respectively.}
	\label{fig:wkbharmonic_appendix}
\end{figure}
We can then compute the asymptotics of the Wronskians. For example, 
\begin{equation}
(\psi_0,\psi_1) \sim -i e^{\frac{1}{\hbar}(I_1(x)-I_0(x))} =-i
\end{equation}
evaluated at any point in the sector $\mathcal{S}_0$, where the asymptotics of both $\psi_0$ and $\psi_1$ are valid. We therefore just obtained the central result of Wronskians: it is controlled by the (vanishing) contour integral from one asymptotic infinity to another
\begin{equation}
I_1(x)-I_0(x) = \lim\limits_{L_0\rightarrow \infty_0, L_1\rightarrow \infty_1} \left(\int_{L_1}^{L_0} dy\sqrt{y^2-E} + \frac{L_1^2}{2}-\frac{1}{2} E \log L_1 -\frac{L_0^2}{2}+\frac{1}{2} E \log L_0 \right)
\end{equation}
The Wronskian is independent of $x$ and asymptotic in $\hbar$. Similarly, we have 
\begin{equation}\label{eq:secondOrderODEpsinWronskian}
(\psi_n,\psi_{n+1}) \sim -i e^{(-1)^n\frac{1}{\hbar}(I_{n+1}(x)-I_n(x))} = -i
\end{equation}
We can now calculate the cross ratio
\begin{align}
\chi_{\gamma} \equiv& \frac{\left(\psi_{0}, \psi_{1}\right)\left(\psi_{-1}, \psi_{2}\right)}{\left(\psi_{-1}, \psi_{0}\right)\left(\psi_{1}, \psi_{2}\right)}\label{eq:secondOrderODEcrossratio}\\ =& \exp\left[{\frac{1}{\hbar}(I_{-1}(x_{-1,2})-I_{-1}(x_{-1,0})+I_{0}(x_{-1,0})-I_0(x_{01})+I_1(x_{01})-I_1(x_{12})+I_2(x_{12})-I_2(x_{-1,2})) }\right]\\
=& \exp\frac{1}{\hbar}\oint dy\ \sqrt{y^2-E} =e^{-2 \pi i \frac{E}{2\hbar} }\label{eq:secondOrderODEcrossratio1}
\end{align}
We can confirm we have the correct Wronskians from a different perspective, discussed in the previous section, namely by comparing the asymptotics with the local solutions around zeros and evaluate the Wronskians using local solutions. Let's illustrate how it works. Around the zero $x = \sqrt{E}$, the Schrodinger equation is linearized as
\begin{equation}\label{eq:linearizedODE}
	\partial_s^2 \psi(s) = s \psi(s)
\end{equation}
where $s = \alpha (x-\sqrt{E})$ and $\alpha = (2\sqrt{E}/\hbar^2)^{1/3}$. Three nice local solutions are given by $\textrm{Ai}_a(s)$ defined in \eqref{eq:niceSolAroundSimpleZero}. Compare the normalization between $\textrm{Ai}_a(s)$ and $\psi_n(x)$ in the region where both asymptotics \eqref{eq:niceSolAroundSimpleZeroAsymp} and linearized differential equation \eqref{eq:linearizedODE} are valid , we have
\begin{equation}
\psi_n(x) \sim \alpha^{-1/2}\textrm{Ai}_n(\alpha x), \quad n = -1,0,+1
\end{equation}
Since $(\textrm{Ai}_n,\textrm{Ai}_{n+1}) = -i$, we therefore arrive at the same expression \eqref{eq:secondOrderODEpsinWronskian}. The same is true around the other zero $x = -\sqrt{E}$. 

To evaluate \eqref{eq:secondOrderODEcrossratio}, we need to know the relation between $\psi_{3}$ and $\psi_{-1}$. Note that the differential equation \eqref{eq:app:secondorderODE} is regular in the whole complex plane, so its solutions are entire functions, as we have found explicitly in Appendix \ref{app:IsingODEexactsol}. \footnote{There is no problem with all the square root in the asymptotic expressions above since they are only expected to be valid in a particular angular sector. The examples in this article are \eqref{eq:niceSolAroundSimpleZeroAsymp} and \eqref{eq:app:UazAsymp}} Therefore when we do the analytic continuation, there is no true monodromy but only \emph{formal monodromy} coming from the asymptotics\footnote{One can calculate the monodromy explicitly and see it is indeed trivial.}. For example,  $\psi_n$ and $\psi_{n+4}$ are proportional to each other since both are the (unique) fast decreasing solution along the ray of $\sqrt{\hbar}e^{-\frac{in\pi}{2}}$ and the relative coefficient is given by the formal monodromy
\begin{equation}
\psi_{n+4}(x) = e^{-2\pi i \Lambda_0}\psi_n(x), \ n \in \text{odd}
\end{equation}
where $\Lambda_0 = -\frac{1}{2} - \frac{E}{2\hbar}$ is the exponent of the formal monodromy. When $n$ is even, we just replace $\hbar \rightarrow -\hbar$.

We are now ready to evaluate the spectral coordinate, 
\begin{equation}\label{eq:secondOrderODEchi}
\chi_{\gamma} \equiv \frac{\left(\psi_{0}, \psi_{1}\right)\left(\psi_{-1}, \psi_{2}\right)}{\left(\psi_{-1}, \psi_{0}\right)\left(\psi_{1}, \psi_{2}\right)}\sim -e^{2\pi i \Lambda_0} = e^{-2 \pi i \frac{E}{2\hbar}}
\end{equation}
which agree with \eqref{eq:secondOrderODEcrossratio1}.

We can also take $E$ to be zero, namely
\begin{equation}
\hbar^2\partial_x^2 \psi(x) = x^2 \psi(x) \label{eq:secondorderODEE=0}
\end{equation}
Two first order zeros collapse into a single second order zero. Our GMN-style analysis \eqref{eq:secondOrderODEpsi0}-\eqref{eq:secondOrderODEchi} still applies, namely the only spectral coordinate $\chi \sim 1$. To carry out the second perspective, we need to compare four small solutions $\psi_n$ to the local solutions $A_{2;a}$ defined in \eqref{eq:niceSolAroundSecondOrderZero} around $x=0$, a second order zero. 
\begin{equation}\label{eq:linearizedODEx2}
\partial_s^2 \psi(s) = s^2 \psi(s)
\end{equation}
where $s = \hbar^{-1/2} x$. We therefore have the identification
\begin{equation}
\psi_n(x) \sim \hbar^{1/4} \textrm{A}_{2;n}(\hbar^{-1/2} x)
\end{equation}
Since $(\textrm{A}_{2;n},\textrm{A}_{2;n+1})=-i$ and the connection formula \eqref{eq:connectionformula}, it is easy to see $\chi_\gamma = 1$.
%%%%%%%%%%%%%%%%%%%%%%%%%%%%%%%%%%%%%%%%%%%%%%%%%%%%%%%%%%%%%%%%%%%%%%%%%%%%%%%
\subsubsection{Cubic potential}
\begin{figure}
	\centering
	\includegraphics[width=0.45\linewidth]{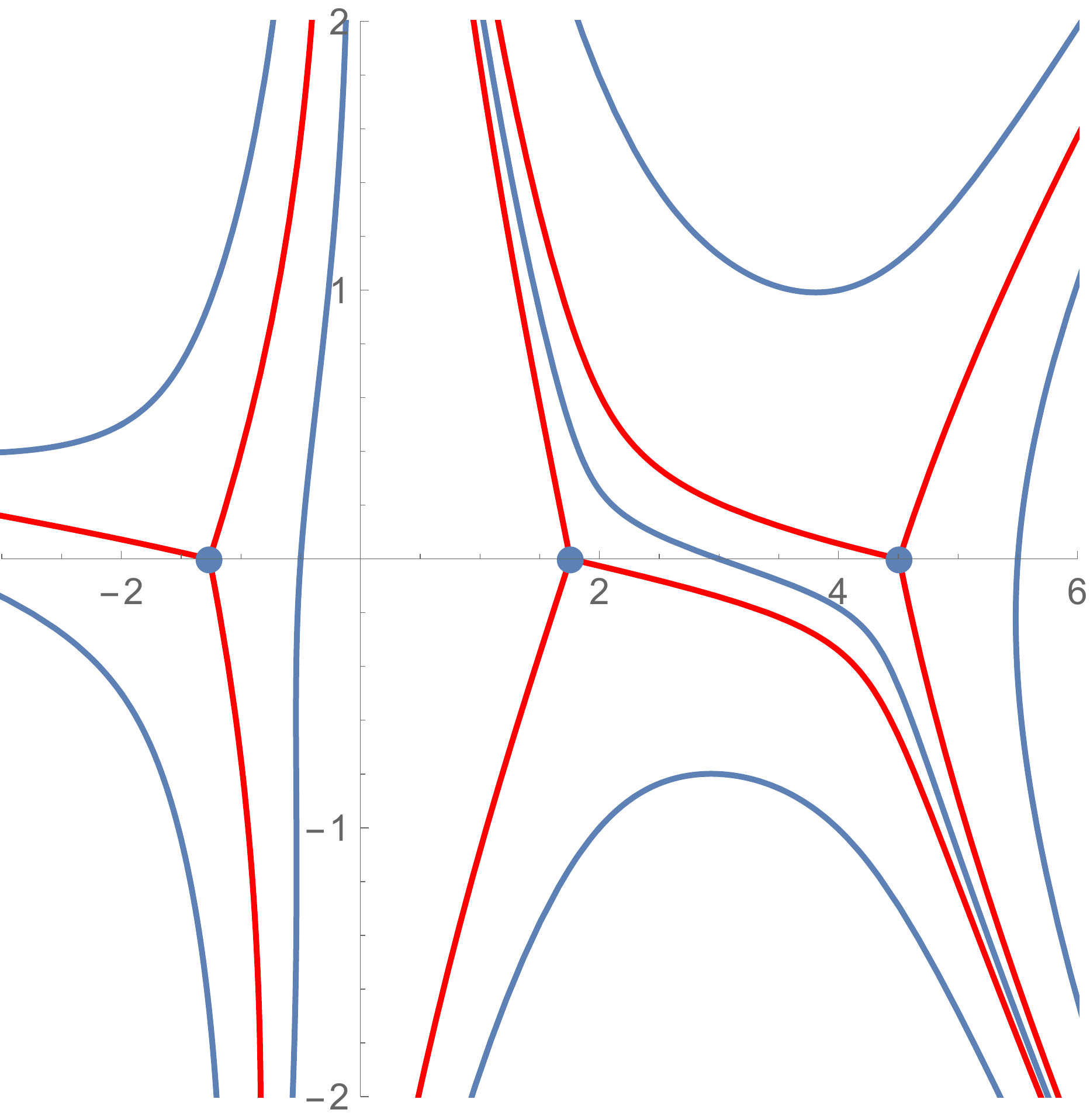}
	\includegraphics[width=0.45\linewidth]{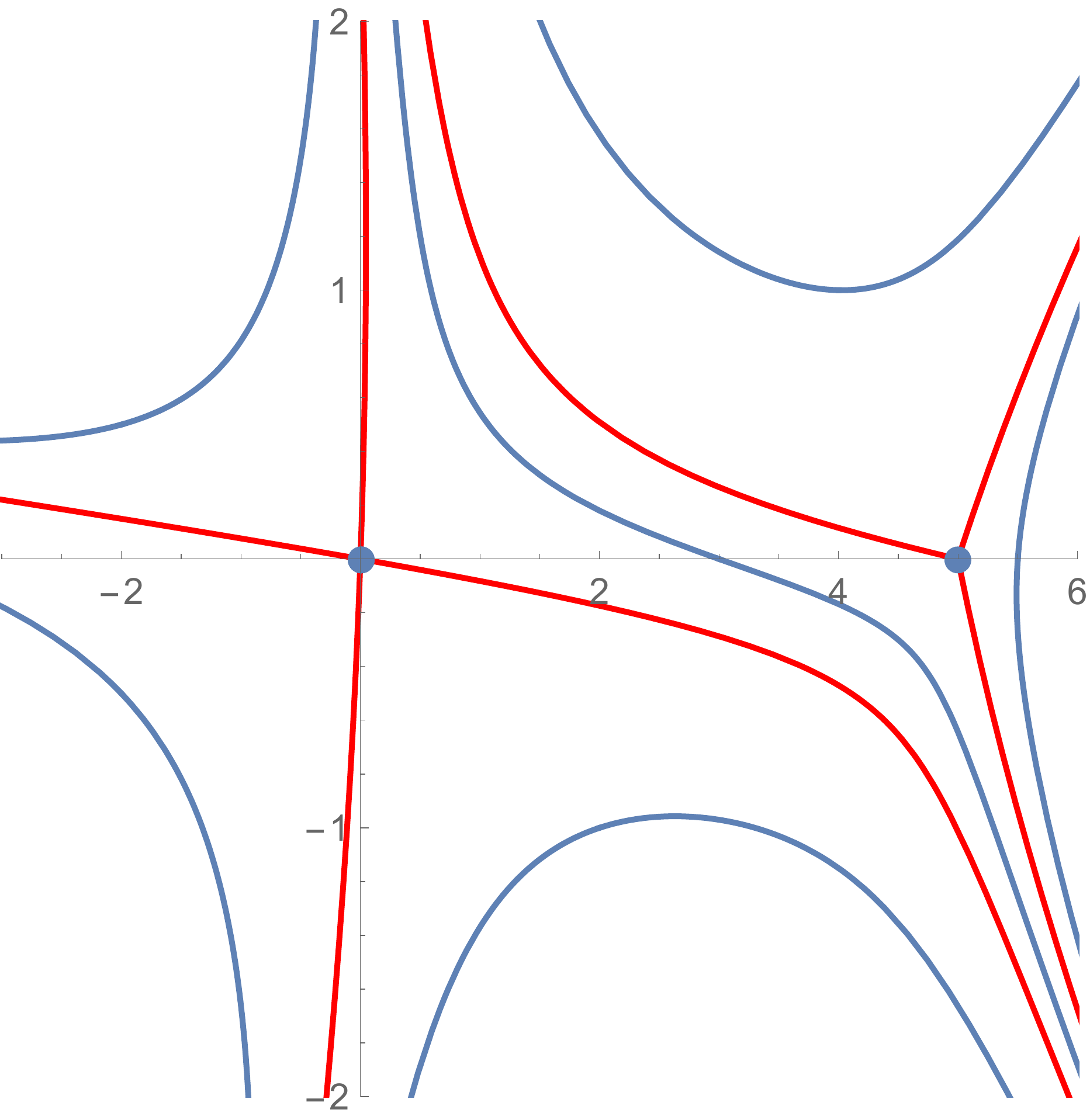}
	\caption{WKB diagram for the differential equation \eqref{eq:app:thirdOrderODE}. Generic flow lines and WKB lines are colored blue and red respectively.On the left, $E>0$ and there are three simple zeros. On the right, $E=0$ and there are one simple zero and one second order zero.}
	\label{fig:wkbcubick2}
\end{figure}
Let's study 
\begin{equation}\label{eq:app:thirdOrderODE}
\hbar^2\partial_x^2 \psi(x) = (\frac{x^2}{2}-gx^3-E) \psi(x)
\end{equation}
For simplicity, let's assume $g$ and $E$ are real positive. When $g$ is small enough, we have three zeros on the real axis denoted as $x_\pm$, $x_1$, which in small $g$ are
\begin{equation}
x_\pm \sim \pm \sqrt{2E} +\mathcal{O}(g),\quad x_1 \sim \frac{1}{2g}+\mathcal{O}(g)
\end{equation}
We have five small solutions at the infinity 
\begin{align}
\psi_{0} \sim  \frac{\sqrt{\hbar}}{\sqrt{2}Q(x)^{1 / 4}} e^{-\frac{1}{\hbar} I_{0}\left(x ; D_{0}\right)}, \qquad \psi_n(x;\hbar) \equiv  \psi_0(x;\hbar e^{-i\pi n})
\end{align}
where $Q(x) = \frac{x^2}{2}-gx^3-E$, and the regularized integral is defined as
\begin{equation}
I_0(x) = \lim\limits_{L\rightarrow \infty} \left(\int_{L}^{x} dy\sqrt{Q(y)} +\frac{2i}{5} \sqrt{g} L^{5/2}-\frac{iL^{3/2}}{6 \sqrt{g}} -\frac{iL^{1/2}}{16 g^{3/2}} +D_0  \right)
\end{equation}
Again, we have $(\psi_n,\psi_m) = -i$, whenever $\psi_n$ and $\psi_m$ are connected to the same zero and $n<m$. We have two spectral coordinates $\chi_{A}$ and $\chi_{B}$, which are controlled by the contour integral along cycles $\gamma_1$ and $\gamma_2$ for the exactly same reason in the quadratic case.

%, again from two WKB periods can be evaluated to be \cite{Alvarez_1989}
%\begin{align}
%\chi_1 = \frac{\pi}{4} \sqrt{g}\left(x_{+}-x_{-}\right)^{2}\left(x_{1}-x_{-}\right)^{1 / 2} \ _2F_1\left(-\frac{1}{2}, \frac{3}{2} ; 3 ; \frac{x_+-x_-}{x_1-x_-}\right)\\
%\chi_2 = \frac{\pi}{4} \sqrt{g}\left(x_{1}-x_{+}\right)^{2}\left(x_{1}-x_{-}\right)^{1 / 2} \ _2F_1\left(-\frac{1}{2}, \frac{3}{2} ; 3 ; \frac{x_+-x_1}{x_--x_1}\right)\\
%\end{align}

Let's see what happens if we take $E = 0$, where two zeros $x_\pm$ collide. We still have a closed contour $\gamma_1$, which can be shrunk to a point. So we have $\chi_{A} = 1$. On the other hand, we have problems evaluating the other spectral coordinate $\chi_{B}$ using GMN-style approach, since $\psi_{2}$ and $\psi_{4}$ are not connected by flow lines. This is where our second perspective is useful. $(\psi_{-1},\psi_2)$ can be evaluated from the Wronskians of local solutions around the zeros. Specifically, we choose $D_0$ in $I_0(x)$ such that it matches exactly to the nice solution around the simple zero $x_1$. We therefore have $(\psi_{-1},\psi_{0}) = (\psi_{0},\psi_{1}) = (\psi_{1},\psi_{2}) = -i$. $\psi_{-1}$ and $\psi_{2}$ will be matched with the local solutions around the double zero,
\begin{equation}
(\psi_{-1},\psi_{2}) = e^{\frac{2}{\hbar} \int_{x_1}^0dy\ \sqrt{Q(y)}} (A_{2;-1},A_{2;1}) = -i\sqrt{2}   e^{\frac{2}{\hbar} \int_{x_1}^0dy\ \sqrt{Q(y)}}
\end{equation}
Therefore we have our second spectral coordinate 
\begin{equation}
\chi_{B} \equiv \frac{\left(\psi_{0}, \psi_{1}\right)\left(\psi_{-1}, \psi_{2}\right)}{\left(\psi_{-1}, \psi_{0}\right)\left(\psi_{1}, \psi_{2}\right)}= \sqrt{2}   e^{\frac{2}{\hbar} \int_{x_1}^0dy\ \sqrt{Q(y)}}
\end{equation}

If all three zeros collide, spectral coordinate can be then evaluated using local solutions $\{A_{3,a}\}$ to be
\begin{equation}
\chi_{A} = \chi_{B} = \frac{1+\sqrt{5}}{2}
\end{equation}
The same result can also be found via symmetry consideration discussed in \cite{Hollands:2019wbr,GMN:2009hg}.

\subsection{WKB analysis for the WZW ODE/IM}\label{app:WKBBessel}
\noindent{\bf \underline{Matching around the zero:}}

Because of our choice of normalization of the small solutions, involving the integral of the WKB momentum starting from $x = - \frac{1}{g}$,
the small solution connected to the zero will match directly the small solutions of the local equation 
\begin{equation}
e^{- 2 \theta} \partial_y^2 \psi_{\mathrm{loc}}(y) = g^k e^{- \frac{2}{g}} y^k \psi_{\mathrm{loc}}(y)
\end{equation}
and the Wronskians will match asymptotically the local Wronskians. In particular, we will have WKB asymptotics 
\begin{equation}
i (\psi_n, \psi_{n'}) \sim \frac{e^{\frac{\pi i}{k+2}(n'-n)}-e^{-\frac{\pi i}{k+2}(n'-n)}}{e^{\frac{\pi i}{k+2}}-e^{-\frac{\pi i}{k+2}}} \equiv d_{n'-n}^{(k)}
\end{equation}
whenever $n_0 \leq n \leq n_0 + k + 1$ and $n_0 \leq n' \leq n_0 + k + 1$.

\vspace{5mm}

\noindent{\bf \underline{Matching around the negative infinity:}}

We will use the Lambert W function $W(z)$, which is defined to be the principal solution of $z = W(z)e^{W(z)}$.

We can expand the potential around $x = x_0$, with real part of $x_0$ assumed to be large negative:
\begin{align}
(1+gx)^k e^{2\theta + 2x} = (1+\frac{1}{\frac{1}{g}+x_0}\delta)^k e^{2\theta+2x_0+k\log(1+gx_0)}e^{2\delta}
\end{align}
where $\delta \equiv x-x_0$ is small. If $x_0$ is chosen such that $2\theta+2x_0+k\log(1+gx_0)=0$, namely
 \begin{equation}
 x_0 \sim -\theta -\frac{1}{2} k \log (-g\theta) -\frac{k^2}{4} \frac{\log(-g\theta)}{\theta}+ \mathcal{O}(\frac{1}{\theta})
 \end{equation}
the potential behaves like $e^{2\delta}$ around $x_0$. The choice of imaginary part of $x_0$ is not unique. The possible choices lie on the special WKB lines. 
Let us choose one above the real axis, and label the line on which $x_0$ lies, to be $n_{x_0}$.
 
Therefore we would like to match the solutions in the large negative region around $x_0$ to solutions of the local equation 
\begin{equation}
\partial_\delta^2\psi(x_0 + \delta) = e^{2\delta}\psi(x_0+\delta)
\end{equation}

Because we normalize the small solutions as in \eqref{eq:asy1}, the WKB parallel transport of small solutions $\{\psi_n(x;\theta)\}_{n \leq n_0}$ from positive infinity to large negative $x$ will accumulate a WKB ``phase'' 
\begin{equation}
-(-1)^n e^\theta \int^{- \infty}_{- \frac{1}{g}} e^x (1+ g x)^{\frac{k}{2}} dx 
\end{equation}
computed along a contour which passes above the real axis. That evaluates to 
\begin{equation}
(-1)^n e^\theta e^{i \frac{\pi k}{2}}e^{-\frac{1}{g}} g^{\frac{k}{2}} \Gamma(1+ \frac{k}{2}) \equiv \frac12 (-1)^n e^\theta e^{i \frac{\pi k}{2}} m_k(g)
\end{equation}
where we define the function 
\begin{equation}
m_k(g) \equiv e^{-\frac{1}{g}} g^{\frac{k}{2}} \Gamma\left(1+\frac{k}{2}\right)
\end{equation}
Thus we expect to match the small solutions to Bessel functions as 
\begin{equation}
\psi_n \sim  e^{\frac12 (-1)^n e^\theta e^{i \frac{\pi k}{2}} m_k(g)} \left( \frac{1}{\sqrt{\pi}}K_0(e^{x-x_0})- \pi i (n-n_{x_0}) \frac{1}{\sqrt{\pi}}I_0(e^{x -x_0}) \right)
\end{equation}
for a WKB line above the real axis, i.e. $n\leq n_0$, where the relative coefficients are fixed by requiring $\psi_n$ to decrease asymptotically fast along the corresponding special WKB lines. Note the shift $n_{x_0}$ in the coefficient from \eqref{eq:continuing_K0}. For a WKB line below the real axis, i.e. $n\geq n_0+k+1$, we have
\begin{equation}
\psi_n \sim  e^{\frac12 (-1)^n e^\theta e^{-i \frac{\pi k}{2}} m_k(g)} \left( \frac{1}{\sqrt{\pi}}K_0(e^{x-x_0})- \pi i (n-n_{x_0}-k) \frac{1}{\sqrt{\pi}}I_0(e^{x -x_0}) \right)
\end{equation}

Recall that $(K_0(e^x),I_0(e^x)) = 1$, we finally estimate 
\begin{equation}
i (\psi_n, \psi_{n'}) \sim e^{\frac{(-1)^n+(-1)^{n'}}{2} e^\theta e^{i \frac{\pi k}{2}} m_k(g)} (n'-n)
\end{equation}
whenever $n \leq n_0$ and $n' \leq n_0$.

Similarly, we have 
\begin{equation}
i (\psi_n, \psi_{n'}) \sim e^{\frac{(-1)^n+(-1)^{n'}}{2} e^\theta e^{-i \frac{\pi k}{2}} m_k(g)} (n'-n)
\end{equation}
whenever $n \geq n_0 + k + 1$ and $n' \geq n_0 + k + 1$. Notice the change in the phase factor in the exponential.

Finally, if $n \leq n_0$ and $n' \geq n_0 + k + 1$ we get 
\begin{equation}
i (\psi_n, \psi_{n'}) \sim   e^{\frac{(-1)^n e^{i \frac{\pi k}{2}} +(-1)^{n'} e^{-i \frac{\pi k}{2}}}{2} e^\theta  m_k(g)} (n'-n-k).
\end{equation}
Note that $n_{x_0}$ doesn't appear in any of the Wronskian, as expected.

If $n=n_0$ and $n'=n_0+k+1$ we have 
\begin{equation}
i (\psi_{n_0}, \psi_{n_0+k+1}) \sim 1 = d_{k+1}^{(k)}
\end{equation}
from both estimates. This is an useful sanity check. 

If a Wronskian does not belong to the above ranges we can use Pl\"ucker formulae to 
relate it to the ones that belong to the above ranges and obtain the WKB asymptotics. One can easily get convinced that all Wronskians can be obtained this way. For example, say that $n_0 \leq n \leq n_0 + k + 1$ and $n' \geq n_0 + k + 1$.
Then we can write 
\begin{equation}
i (\psi_n, \psi_{n'}) = - (\psi_n, \psi_{n'})(\psi_{n_0}, \psi_{n_0+k+1}) = -(\psi_{n_0}, \psi_{n})(\psi_{n_0+k+1}, \psi_{n'}) -(\psi_n, \psi_{n_0+k+1})(\psi_{n_0}, \psi_{n'}) 
\end{equation}
which can be written as  
\begin{align}
i (\psi_n, \psi_{n'}) \sim  & d_{n_0+k+2-n}^{(k)}e^{\frac{(-1)^{n_0+k+1}+(-1)^{n'}}{2}  e^{-i \frac{\pi k}{2}} e^\theta m_k(g)} (n'-n_0-k-1)+ \cr 
&+ d_{n_0+k+1-n}^{(k)}e^{\frac{(-1)^{n_0+k}  +(-1)^{n'} }{2} e^{-i \frac{\pi k}{2}}e^\theta  m_k(g)} (n'-n_0-k)
\end{align}
Notice that either of the two exponential factors is trivial, as either $(-1)^{n_0+k+1}+(-1)^{n'}=0$ or $(-1)^{n_0+k}  +(-1)^{n'}=0$.
The two summands will exchange dominance whenever $e^{-i \frac{\pi k}{2}}e^\theta$ becomes pure imaginary. 

Similarly if $n \leq n_0$ and $n_0 \leq n' \leq n_0 + k + 1$ we can write 
\begin{equation}
i (\psi_n, \psi_{n'}) = -(\psi_{n}, \psi_{n_0})(\psi_{n'}, \psi_{n_0+k+1}) -(\psi_n, \psi_{n_0+k+1})(\psi_{n_0}, \psi_{n'}) 
\end{equation}
which can be written as  
\begin{align}
i (\psi_n, \psi_{n'}) \sim  &d_{n'+1-n_0}^{(k)} e^{\frac{(-1)^n+(-1)^{n_0}}{2} e^\theta e^{i \frac{\pi k}{2}} m_k(g)} (n_0-n) + \cr 
&+ d_{n'-n_0}^{(k)} e^{\frac{(-1)^n +(-1)^{n_0+1}}{2} e^{i \frac{\pi k}{2}}  e^\theta  m_k(g)} (n_0+1-n)
\end{align}

Finally, we should specify the value of $n_0$:
\begin{itemize}
\item If $k$ is odd, the WKB analysis jumps whenever $e^\theta$ is pure imaginary. 
Real $\theta$ is not ``special'' and can be used as a starting point for the WKB analysis. Then 
$n_0 = -\frac{k+1}{2}$.  
\item If $k$ is even, then the WKB analysis jumps whenever $e^\theta$ is real. If we sit at $\theta$ with imaginary part $i \frac{\pi}{2}$ 
then $n_0 = -\frac{k}{2}-1$, while if we sit at $\theta$ with imaginary part $-i \frac{\pi}{2}$ then $n_0 = -\frac{k}{2}$.
\end{itemize}

\section{Generalities of integrable line defects} \label{app:line}

\subsection{Line defects in 2d CFTs}
A translation invariant line defect, i.e. a line defect which preserves energy conservation when 
places at some point in space, say $x^1=0$, satisfies an energy conservation relation
which controls the discontinuity in energy flux across the defect: 
\begin{equation}
\left[T - \bar T\right]_{x^1=0^+} - \left[T - \bar T\right]_{x^1=0^-} = 2 i \partial_{x^0} t^{00}
\end{equation}
where $t^{00}$ is the defect stress tensor. Notice that all four summands on the right hand side of the equation are 
well-defined defect local operators. 

The defect would be conformal invariant if and only if $t^{00}=0$. We are interested in defects which are 
{\it not} conformal invariant. 

If we act with a more general bulk conformal transformation fixing the $x^1=0$ location of the defect, 
the line defect will thus change. Infinitesimally, the deformation under a conformal transformation 
which restricts to a vector field $v^0(x^0)$ along the defect is given by the boundary action 
\begin{equation}
\int t^{00} \partial_0 v^0 dx^0
\end{equation}
In particular, $t^{00}$ can be added to the 
defect action to implement an infinitesimal scaling transformation.  

A global rescaling by a factor of $e^\theta$ will map $L$ to a new chiral line defect $L[\theta]$. 
Shifts of $\theta$ in the positive real direction correspond to RG flow of the line defect. More general conformal transformations will 
lead to a line defect with a position-dependent $\theta$ parameter. 

\subsection{Movable line defects}
It is also possible to consider defects whose correlation functions are invariant under rigid translations in a direction transverse to the line defect. 
In terms of the bulk stress tensor, this means that 
\begin{equation}
\left[T + \bar T\right]_{x^1=0^+} - \left[T + \bar T\right]_{x^1=0^-} = 2 \partial_{x^0} \tilde t^{00}
\end{equation}
for some defect operator $\tilde t^{00}$. In other words, the ``displacement operator'' is a total derivative. 

This is automatically true for translation-invariant line defects in a chiral CFT.

Now we can consider an infinitesimal conformal transformation which changes the location of the defect, followed by a 
displacement back to the original $x^1=0$ location. The result is a deformation 
\begin{equation}
\int \left[ t^{00} \partial_0 v^0 + \tilde t^{00} \partial_0 v^1 \right]dx^0
\end{equation}

In particular, a rigid rotation by an angle $\phi$ of the defect followed by a deformation back to the vertical direction 
allows us to extend the family $L[\theta]$ of integrable line defects from before to a two-parameter family $L[\theta, \phi]$. 
We can define this deformation directly for $|\phi| <\frac{\pi}{2}$ and then iterate it to reach a broader range of $\phi$. 
There is no guarantee that this is periodic in $\phi$. In general, $L[\theta, \phi+ 2 \pi] \neq L[\theta, \phi]$

If the line defect $L$ is invariant under reflections $x^0 \to - x^0$, $L[\theta, \phi]$ will break that symmetry, as $\tilde t^{00}$
is pseudo-real. If we rotate all the way by $\phi = \pi$, though, we should go back to a reflection-symmetric defect. 

Movable defects can be naturally fused. Consideration of a U-shaped configuration suggests that the fusion of 
$L[\theta, \phi+ \frac{\pi}{2}]$ and $L[\theta, \phi- \frac{\pi}{2}]$ should include the identity line defect. 

\subsection{Chiral line defects}
If the defect is chiral, so that $[\bar T]_{x^1=0^+}= [\bar T]_{x^1=0^-}$, then one has a simpler relation 
\begin{equation}
\left[T \right]_{x^1=0^+} - \left[T\right]_{x^1=0^-} = 2 i \partial_{x^0} t^{00}
\end{equation}
which implies that the line defect is also invariant under rigid translations in the $x^1$ direction, i.e. is movable. 

Furthermore, $\tilde t^{00} = i t^{00}$ and a conformal transformation deforms the line defect by 
\begin{equation}
\int \left[  \partial_0 v^0 + i \partial_0 v^1 \right]t^{00}dx^0
\end{equation}
In particular, $L[\theta, \phi] \equiv L[\theta + i \phi]$ and thus $\theta$ can be taken to be 
valued in the complex plane. 

In general, we expect $\langle|L[\theta] \rangle_R$ to be an entire function of $\theta$. 
Computing such a function is the typical objective of a calculation in this paper. 

If the line defect $L$ is invariant under reflections $x^0 \to - x^0$, the asymptotics 
\begin{equation}
\langle|L[\theta] |\rangle_R \sim e^{- 2 \pi R e^{\theta} E_0}
\end{equation}
will hold in a whole open strip of width $\frac{\pi}{2}$ around the positive real $\theta$ axis. 

As we deform all the way to $\Im \theta = \pi n$, we will reach a collection of other unitary line defects, 
with nice RG flow and asymptotics
\begin{equation}
\langle|L[\theta] |\rangle_R \sim e^{- (-1)^n 2 \pi R e^{\theta} E^{(n)}_0}
\end{equation}
which will hold in a whole open strip of width $\frac{\pi}{2}$ around the $\Im \theta = \pi n$, $\Re \theta \gg 0$ lines. 

At $\Im \theta = \pi (n+ \frac12)$ we will have wall-crossing phenomena as the IR physics of the line defect jumps. 

In the opposite limit of large negative $\theta$ we explore the UV definition of the defect. 

\subsection{Integrable line defects}
Finally, we can call a chiral line defect (or a collection of line defects) {\it integrable} if 
\begin{itemize}
\item Close line defects $L[\theta]$ for different $\theta$'s give commuting operators.
\item The identification between the compositions of line defects $L[\theta]$ and $L[\theta']$ in opposite order 
can be implemented by a R-matrix $R[\theta-\theta']$, i.e. a topological local operator interpolating 
between $L[\theta] L[\theta']$ and $L[\theta'] L[\theta]$.
\item The R-matrix satisfies Yang-Baxter relations. 
\end{itemize} 

\subsection{IR data and wallcrossing}
A typical observable of interest would be the expectation value $\langle|L[\theta] |\rangle_R$ of the line defect on a cylinder,
with some choices of states at the two ends of the cylinder, as a function of the radius $R$ of the cylinder. 
The expectation value will be a function of combination $R e^{\theta}$ and we can set $R=1$ without loss of generality.  

If the line defect $L$ is invariant under reflections $x^0 \to - x^0$, upon Wick rotation it will map to a 
defect which preserves unitarity. This property is obviously preserved by the above global rescaling 
for real $\theta$, so the whole $L[\theta]$ family is unitary. 

A unitary line defect should have a nice, monotonic RG flow landing onto some conformal-invariant line defect in the far IR. 
The expectation value on a large cylinder should thus behave as 
\begin{equation}
\langle|L[\theta] |\rangle_R \sim e^{- 2 \pi R e^{\theta} E_0}
\end{equation}
where $E_0$ is the energy of groundstate of the line defect. Subleading corrections should be suppressed by
similar exponentials with a larger real energy. 

\section{Analysis of chiral line defect in the Ising model}\label{app:Isingloopcal}
Consider the defect Lagrangian given in \cite{Casini:2016fgb}, where a Majorana fermion $\gamma$ is introduced as an auxiliary defect degree of freedom within a massless free fermion bulk. As we are interested in chiral line defects, we differ from the reference in that $\gamma$ is perturbed only chirally as
\begin{equation}
    g \int_x \psi(x,0) \gamma(x)
\end{equation}
in addition to the kinetic term of $\gamma$. We take the mode expansion on the cylinder in the NS sector
\begin{align}
    \psi(x,t) &= \sum_{n \in \mathbb{Z}+\frac{1}{2}} b_n e^{i \frac{n}{R} (x-t)} \\
    \gamma(x) &= \sum_{n \in \mathbb{Z}+\frac{1}{2}} \gamma_n e^{i \frac{n}{R} x},
\end{align}
where the modes obey $\{ b_n , b_m \} = \{ \gamma_n , \gamma_m \} =\frac{1}{2 \pi R} \delta_{n+m}$. The coupling then becomes
\begin{equation}
 g \int_x \psi(x,0) \gamma(x) = 2 \pi R g \sum_{n>0} (b_{-n} \gamma_{n} - \gamma_{-n} b_n)
\end{equation}
where $n$ are positive half-integers. Taking the vacuum expectation value of its exponential, the surviving contribution is $\exp( 2 \pi R g^2 \sum_{n>0} \gamma_{-n} \gamma_{n} )$. Combining with the kinetic term contribution $ \exp( \sum_{n>0} n \gamma_{-n} \gamma_{n} )$ (with an appropriate relative normalization constant) and integrating over the $\gamma$ modes, we get the product
\begin{equation}
 \sqrt{2} e^{2 \pi R g^2 \log e^{-1} \epsilon g^2} \prod_{m=0}^{\frac{R}{\epsilon}} \frac{ m + 1/2 + 2 \pi R g^2 }{ m + 1/2} = \frac{\sqrt{2\pi} e^{2 \pi R g^2 \log e^{-1} R g^2}}{\Gamma\left(\frac12 + 2 \pi R g^2 \right)}+ O(\epsilon).
\end{equation}
where we normalized the answer correctly in the UV and included a constant counterterm. 

Reintroducing $\theta$ and setting $2 \pi R g^2=1$ we get
\begin{equation}
    T_S(\theta) = \frac{\sqrt{2 \pi} e^{\theta e^\theta - e^\theta}}{\Gamma(\frac{1}{2} + e^\theta)}.
\end{equation}

If we evaluate vevs on states other than the vacuum, the exponential is modified as $\exp( 2 \pi R g^2 \sum_{n>0} \epsilon_n \gamma_{-n} \gamma_{n} )$
where $\epsilon_n=-1$ for occupied states. A finite collection of factors in the answer is modified to $m + 1/2 - 2 \pi R g^2$. Similarly, in the Ramond sector 
one replaces $m+\frac12$ with $m+1$ and includes a zeromode contribution. 

\section{Perturbative analysis of WZW line defects}\label{app:wzwKondoPert}

We study chiral line defects in $\mathfrak{su}(2)$ WZW models, which are defined as

\begin{equation}
\hat{T}_\mathcal{R} := \text{Tr}_\mathcal{R}{\mathcal{P} \exp \Big( i g \int_0^{2\pi R}d\sigma\ t_a J^a(\sigma,0) \Big)}
\end{equation}
where $g$ is the dimensionless coupling, $t_a$ are generators of $\mathfrak{su}(2)$ in representation $\mathcal{R}$, and $J^a$ are chiral WZW currents \footnote{The roman letters $a,b,c,d$ will be reserved for group theory indices.}. On the cylinder of radius $R$, the currents admit the mode expansion \begin{equation}
J^a(s) = \frac{1}{R}\sum_{n \in \mathbb{Z}} J_n^a e^{-n s/R}
\end{equation}
with coordinates $s=\tau + i \sigma$ on the cylinder. The modes obey the typical commutation relations of the affine Kac-Moody algebra, $[J_n^a,J_m^b] = i f_{abc} J_{n+m}^{c} + k n \delta_{ab} \delta_{n+m,0}$. We will follow the Lie algebra conventions from \cite{DiFrancesco:1997nk,Bachas:2004sy}. For completeness, we also review them in Appendix \ref{appendix:LieAlgebra}.

\subsection{Definition of the quantum operator  $\hat{T}_\mathcal{R}$}

For small $g$, the line operator $\hat{T}_\mathcal{R}$ wrapping the cylinder admits the expansion
\[
\hat{T}_\mathcal{R} = \sum_{N=0}^{\infty} (ig)^N \hat{T}_\mathcal{R}^{(N)}
\]
where
\[
\hat{T}_\mathcal{R}^{(N)} = \text{Tr}_\mathcal{R} (t^{a_1} \cdots t^{a_N}) \Big( \prod_{i=1}^{N} \int_{0}^{2 \pi R} d\sigma_i \Big) \theta_{\sigma_1>\cdots>\sigma_N} J^{a_1}(\sigma_1) \cdots J^{a_N}(\sigma_N).
\]

In the above classical expression for the line operator, the currents are not ordered. However, a quantum line operator requires an regularization scheme which prescribes an appropriate ordering for the currents which is consistent with the desired properties of $\hat{T}_\mathcal{R}$. Furthermore, the currents themselves must be regularized, which can be done by assigning a cutoff on the mode expansion of $J(\sigma)$. In doing so, we follow the regularization prescription given in \cite{Bachas:2004sy}. 

However, our treatment differs from \cite{Bachas:2004sy} in the computation of the operator, where we need to compute the full expression of the normal-ordered operator to $O(g^4)$ rather than just the leading contributions in the classical limit $\kappa \to \infty$. We also keep track of the length scale $R$ of the cylinder, which allows us, as we will describe below, to make connections, to thermodynamic Bethe ansatz, Hirota relations and computations from ODE/IM correspondence.

We now review the regularization prescription used in \cite{Bachas:2004sy}. As part of the regularization scheme, a current ordering is chosen to respects the desired symmetries of the quantum line defect. It is reasonable to assume $\hat{T}_\mathcal{R}$ to be invariant under the following transformations: (1) cyclic permutations of the inserted currents and (2) reversing the orientation of the defect combined with taking $\mathcal{R}$ to its conjugate representation $\bar{\mathcal{R}}$. A current ordering which respects cyclic invariance and orientation reversal (combined with $\mathcal{R} \to \bar{\mathcal{R}}$) is
\[
\hat{T}_\mathcal{R}^{(N)} = \text{Tr}_\mathcal{R} (t^{a_1} \cdots t^{a_N}) \Big( \prod_{i=1}^{N} \int_{0}^{2 \pi R} d\sigma_i \Big) \theta_{\sigma_1>\cdots>\sigma_N} \frac{1}{2N} \Big[ J^{a_1}(\sigma_1) \cdots J^{a_N}(\sigma_N) + \text{cyclic} + \text{reversal} \Big].
\]
We also need the regularized chiral WZW currents by imposing a short distance cutoff \cite{Bachas:2004sy}:
\[
J^a(\sigma) = \frac{1}{R} \sum_{n \in \mathbb{Z}} J_n^a e^{- i n \sigma/R -|n|\epsilon/2R}.
\]
By expanding the currents into modes, each contribution to $\hat{T}_\mathcal{R}$ becomes the product of four terms which can be independently evaluated: the group theory factor, appropriately-ordered modes, products of regulators $e^{|n_i|\epsilon/2R}$, and integrals over $\sigma_i$. The integrals over $\sigma_i$ yield delta functions on which only certain terms for which the sums of the mode numbers are equal to zero are supported. Note that this implies translation invariance of the operator along $\sigma$ direction. 

This will be our new definition of $\hat{T}_\mathcal{R}$ hereon. The expressions of the first few orders are given below
\begin{align}
&\hat{T}_\mathcal{R}^{(0)} = \dim \mathcal{R},\quad \hat{T}_\mathcal{R}^{(1)} = 0, \quad \hat{T}_\mathcal{R}^{(2)} = 2 \pi^2 \text{Tr}_\mathcal{R} (t^a t^b) J_0^a J_0^b \nonumber\\
&\hat{T}_\mathcal{R}^{(3)} = \frac{2 \pi^2}{3} \text{Tr}_\mathcal{R} (t^a t^b t^c) \Bigg[ \frac{\pi}{3} J_0^a J_0^b J_0^c + \sum_{n \neq 0} \frac{i}{n} J_{-n}^a J_{n}^b J_0^c e^{-|n|\epsilon/R} + \text{cyclic} + \text{reversal} \Bigg] \nonumber\\
&\hat{T}_\mathcal{R}^{(4)} = \frac{\pi^2}{2} \text{Tr}_\mathcal{R} (t^a t^b t^c t^d) \Bigg[  \frac{\pi^2}{6} J_0^a J_0^b J_0^c J_0^d + \sum_{n\neq 0} \frac{i \pi}{n} J_{-n}^a J_{n}^b J_0^c J_0^d e^{-|n|\epsilon/R} \nonumber \\
&+ \sum_{n \neq 0} \frac{1}{n^2} (J_{-n}^a J_{n}^b J_0^c J_0^d - J_{-n}^a J_{0}^b J_{n}^c J_0^d) e^{-|n|\epsilon/R} + \sum_{m,l,m+l \neq 0} \frac{1}{m l} J_{m}^a J_{-m-l}^b J_{l}^c J_{0}^d e^{-(|m|+|l|+|m+l|) \epsilon/ 2 R} \nonumber \\
&- \frac{1}{2} \sum_{m,n \neq 0} \frac{1}{m n} J_{-n}^a J_{n}^b J_{-m}^c J_{m}^d e^{-(|m|+|n|) \epsilon/R} + \text{cyclic} + \text{reversal} \Bigg] \label{eq:TbeforeNO}
\end{align}
This is the same expression in \cite{Bachas:2004sy}. Note that up until now, no knowledge of the representation $\mathcal{R}$, namely $ \text{Tr}_\mathcal{R} (t^a t^b t^c \dots)$ have been used, except for the cyclic properties of the trace.

The last ingredient we need is the renormalization scheme, i.e. the prescription of removing the short distance cutoff $\epsilon$ and replacing bare couplings with renormalized couplings. As in \cite{Bachas:2004sy}, there are two types of local counterterms involved, the identity operator $1$ and the marginal operator $t\cdot J$. The effects of $1$ and $t\cdot J$ are, respectively, to multiply the result by an overall factor $e^{R G(g,\epsilon)}$ and redefine the coupling $g$ to $F(g,\epsilon)$, where $G(g,\epsilon)$ and $F(g,\epsilon)$ are power series in $g$. For the reason that will be clear soon, it is very helpful to make the renormalization scheme explicit and generic, as we will do in the next section.
\subsection{Computation of normal-ordered line operator}
To facilitate our calculations later, we will first normal order the expressions \eqref{eq:TbeforeNO}. It is done by moving all positive modes to the right of the negative modes. Specifically, we need to use the commutation relation repeatedly such that the subcripts of $J_n$ are in ascending order.\footnote{The normal-ordering procedure is relatively straightforward and yet very tedious. The strategy is reorganizing and relabeling the summations so that the sums run over positive indices and subsequently applying the affine Kac-Moody commutation relations. In particular, for the terms with sums over two indices which appear at $O(g^4)$, after organizing the summations such that all indices run over positive indices $n,m>0$, the sums require additional division into the cases $\sum_{n,m>0} = \sum_{n>m>0}+\sum_{m>n>0}+\sum_{m=n>0}$ for proper normal ordering. Furthermore, at some point of the normal-ordering procedure, modes such as $J_{m-n}$ or $J_{n-m}$ as well as $J_{n}$,$J_{-n}$, $J_{m}$, or $J_{-m}$ will be present in the same term. This suggests that a further subdivision of the summation into $\sum_{n>m>0} = \sum_{n>0,n>m>n/2} + \sum_{n>0,n/2>m>0} + \sum_{m=n/2>0, n \ \text{even}}$ and similarly for $\sum_{m>n>0}$ is necessary.} In the case of equal subscripts equal $n$, we define normal ordered expression to be totally symmetric. For example, 
\begin{equation}
J_n^aJ_n^b \rightarrow J_n^{(a}J_n^{b)} +\frac{i}{2} f^{abc}J_{2n}^c
\end{equation}
Similarly for longer products $J_n^{a_1}J_n^{a_2}J_{n}^{a_3}\dots J_{n}^{a_m}$. We include $1/m!$ in the symmetrization.   

We then proceed with the renormalization by removing counterterms proportional to $R$ and performing the following redefinition in the coupling:
\[
g \rightarrow g\lambda + g^2 \lambda^2 \Big[ -2 \log{\epsilon} +C_0\Big] + g^3 \lambda^3 \Big[ + 4(\log{\epsilon})^2  -(2k+4C_0) \log{\epsilon} +D \Big]  + \cdots,
\]
where $g$ on the right hand side is the renormalized coupling. $C_0$ and $D$ are arbitrary renormalization scheme constant that depend possibly on the representation but not on $\epsilon$. We also include $\lambda$ to possibly rescale the coupling.
\allowdisplaybreaks
The renormalized and normal-ordered $SU(2)$ line operator $\hat{T}_j$ to $O(g^4)$ is
\begin{align}
\hat{T}_j^{(0)} &= 2j+1, \nonumber\\
\hat{T}_j^{(2)} &= 4 \pi^2 \lambda^2 x_j  J_0^a J_0^a, \nonumber\\
\hat{T}_j^{(3)} &= -16 i \pi^2 \lambda^3 x_j  \Bigg\{ \sum_{n>0} \frac{i}{2 n} f_{abc} J_{-n}^a J_{0}^b J_{n}^c + \sum_{n>0} \frac{2}{n} J_{-n}^a J_{n}^a - \log{R} \ J_0^a J_0^a  - \frac{k}{2} \Bigg\} \nonumber \\
\hat{T}_j^{(4)} &= 8 \pi^2 \lambda^4 x_j \Bigg\{\sum_{n,m,n+m \neq 0} \Big[ \frac{2}{3m(m+n)} :J_{-m-n}^{a} J_{0}^{b} J_{m}^{a} J_{n}^{b}: - \frac{1}{3 n m} :J_{-m-n}^{a} J_{0}^{a} J_{m}^{b} J_{n}^{b}:  \Big] \nonumber
\\ + &\sum_{n,m > 0} \frac{1}{n m} \Big[ :J_{-n}^{a} J_{-m}^{a} J_{m}^{b} J_{n}^{b}: - :J_{-n}^{a} J_{-m}^{b} J_{m}^{a} J_{n}^{b}: \Big] \nonumber
\\ + &\sum_{n>0} \frac{1}{n^2} \Big[2 J_{-n}^{a}J_{0}^{b}J_{0}^{b}J_{n}^{a} - J_{-n}^{a}J_{0}^{a}J_{0}^{b}J_{n}^{b} - J_{-n}^{a}J_{0}^{b}J_{0}^{a}J_{n}^{b} \Big] \nonumber
\\ + &\sum_{n,m>0} \frac{3 i}{n m} f_{abc} :J_{-n}^{a}J_{-m+n}^{b}J_{m}^{c}: + \sum_{n>m>0} \frac{i}{n m} f_{abc} [ J_{-n}^{a}J_{-m}^{b}J_{m+n}^{c} + J_{-m-n}^{a}J_{m}^{b}J_{n}^{c} ] \nonumber
\\ + &\sum_{n>0} \frac{6 i}{n} \Big[\log{R} - \frac{1}{3}(H_{\lfloor \frac{n}{2} \rfloor} + H_{\lfloor \frac{n-1}{2} \rfloor}) \Big] f_{abc}  J_{-n}^{a}J_{0}^{b}J_{n}^{c}+ \sum_{n>0} \frac{2i}{n^2} f_{abc} J_{-n}^{a} J_{0}^{b} J_{n}^{c} \nonumber
\\ + &\sum_{n,m>0} \frac{3}{n m} J_{-m-n}^a J_{m+n}^a - \sum_{n>m>0} \frac{6}{n m} J_{m-n}^a J_{-m+n}^a - \sum_{n>0} \frac{2}{n^2} J_{-2n}^a J_{2n}^a \nonumber
\\ + &\sum_{n>0} \frac{6}{n^2} J_{-n}^a J_{n}^a + \sum_{n>0} \frac{24}{n} \Big[\log{R} - \frac{1}{2}(H_{\lfloor \frac{n}{2} \rfloor} + H_{\lfloor \frac{n-1}{2} \rfloor}) -\frac{k}{12} \Big] J_{-n}^a J_{n}^a  \nonumber
\\ + &\frac{\pi^2}{12} (2 \alpha_j + \beta_j) (J_0^a J_0^a)^2 + \Big[2k\log{R} -6(\log{R})^2 - \frac{\pi^2}{6} \big( \beta_j - 8 j(j+1) \big) \Big] J_0^a J_0^a \nonumber
\\  + &\Big[\frac{3}{4} k^2 - 6k( 1 + \log{R}) \Big] \Bigg\},
\end{align}
where the representation of $\mathfrak{su}(2)$ are labelled by the half-integer $j$ and $x_j$ is half of the Dynkin index defined in Appendix \ref{appendix:LieAlgebra}. $: \ :$  denotes the normal ordering operation, where an equal fraction of each ambiguous combination is taken in a symmetric manner when there are ambiguities (i.e. when there exist modes with same mode numbers). One can consider the leading terms in $k$ to verify that its large $k$ limit matches with the result given in \cite{Bachas:2004sy}.
\subsection{Verification of the commutativity and Hirota relation}\label{app:HirotaSU2}
As explained above \eqref{eq:exchangeMuR} and the footnote in Section \ref{sec:PertOverviewKondo}, we identify  $2 \pi R = e^{\theta}$ and verify directly that $\hat{T}_j^{(N)}$ all commute. Therefore we have
\begin{equation}
[ \hat{T}_j[\theta] , \hat{T}_{j'}[\theta'] ] = 0,\label{eq:app:commute}
\end{equation}
We also want to verify Hirota relations \cite{KLUMPER1992304,baxter2016exactly,Kuniba:2010ir}
\begin{equation}
\hat{T}_j[\theta + \tfrac{i\pi}{2}] \hat{T}_j[\theta - \tfrac{i\pi}{2}] = 1 + \hat{T}_{j+\frac{1}{2}}[\theta] \hat{T}_{j-\frac{1}{2}}[\theta],\label{eq:app:Hirota}
\end{equation}
which can be written perturbatively in $g$ as
\begin{align}
2 \hat{T}_{j}^{(0)} \hat{T}_{j}^{(2)} &= \hat{T}_{j+\frac{1}{2}}^{(0)} \hat{T}_{j-\frac{1}{2}}^{(2)} + \hat{T}_{j-\frac{1}{2}}^{(0)} \hat{T}_{j+\frac{1}{2}}^{(2)},\\
\hat{T}_{j}^{(0)}(\hat{T}_{j}^{(3)+} + \hat{T}_{j}^{(3)-})& = \hat{T}_{j+\frac{1}{2}}^{(0)} \hat{T}_{j-\frac{1}{2}}^{(3)} + \hat{T}_{j-\frac{1}{2}}^{(0)} \hat{T}_{j+\frac{1}{2}}^{(3)},\\
\hat{T}_{j}^{(0)}(\hat{T}_{j}^{(4)+} + \hat{T}_{j}^{(4)-}) + \hat{T}_{j}^{(2)} \hat{T}_{j}^{(2)} &= \hat{T}_{j+\frac{1}{2}}^{(0)} \hat{T}_{j-\frac{1}{2}}^{(4)} + \hat{T}_{j-\frac{1}{2}}^{(0)} \hat{T}_{j+\frac{1}{2}}^{(4)} + \hat{T}_{j-\frac{1}{2}}^{(2)} \hat{T}_{j+\frac{1}{2}}^{(2)}
\end{align}
where the superscripts $\pm$ indicate shifts in the argument by $\pm\frac{i\pi}{2}$.

It turns out \eqref{eq:app:Hirota} is satisfied if 
\begin{equation}
D =  D_0 - \frac{4 \pi^2}{3}j(j+1)
\end{equation}
where $D_0$ and $C_0$ are arbitrary constants that are independent of the representation $j$ and $\epsilon$, which we choose the arbitrary constant $D_0=-\frac{5 \pi^2}{6}$ and $C_0=0$.

Note that due to commutativity \eqref{eq:app:commute}, in the common eigenspace, we can just deal with eigenvalues of $\hat{T}_j(\theta)$ and their functional relations. Nevertheless, we chose to verify the operator version of the Hirota relation, which is a stronger equation.
\subsection{Expectation values}

Let us compute the expectation value between primary states in representation $l$. We follow again the normalization in \cite{DiFrancesco:1997nk}, where $J_0^a J_0^a = 2 l(l+1)$ when acting on a primary state $|l \rangle$. The renormalized expectation value, which follows directly from the normal-ordered operator, is
\begin{align*}
\langle T_n (g, R) \rangle_{l} &= n
\\ &- g^2 \lambda^2 x_j [8 \pi^2 l(l+1)] 
\\ &+ g^3 \lambda^3 x_j [ 32 \pi^2 l(l+1) \log{R} + 8 \pi^2 k -16\pi^2 C_0 l(l+1) ]
\\ &- g^4 \lambda^4 x_j \bigg[ 96 \pi^2 l(l+1) (\log{R})^2 - 16 \pi^2 \big(k (2l(l+1)-3)+6C_0 l(l+1)\big) \log{R}
\\ &\frac{1}{15} (-2) \pi ^2 \left(4 l (l+1) \left(2 \pi ^2 \left(3 n^2 \left(l^2+l+3\right)-5 \left(l^2+l+1\right)\right)-15 C_0^2\right)\right.\\
&\qquad \left.  +180 (C_0-2) k+45 k^2\right) \bigg],
\end{align*}
We can also calculate the expectation value over excited states. This will be done in a future paper.\cite{Gaiotto:2020dhf}
\subsection{Beta function and effective coupling}
Beta function can be found to be
\begin{equation}
\beta(g) \equiv  \frac{\partial g}{\partial \log \Lambda}=  2 \lambda g^2 + 2 k \lambda^2 g^3 + \cdots.
\end{equation}
The ratio $\frac{c_1}{c_0^2}$ from $\beta(g) = c_0 g^2 + c_1 g^3 + \cdots$ is independent of the renormalization scheme and equals $\frac{k}{2}$. In accordance with the discussion in Section \ref{sec:PertOverviewKondo}, we will choose $\lambda = -\frac12$.

It is not hard to see that any constants or higher order terms in the beta function can be arbitrarily adjusted by redefining $g$. In particular, we fix it to be 
\begin{equation}
\beta(g) = -\frac{g^2}{1+\frac{k}{2}g}
\end{equation}
which give rise to a scale via dimensional transmutation.
\[
\mu = e^{-1/g} g^{k/2}
\]
through dimensional transmutation. Since $\mu$ enters into any observable computed using $\hat{T}_j$ only through the combination $R\mu$, the result is only dependent on $e^\theta e^{1/g} g^{-k/2}$. This combination can be used to define the effective coupling $g_{\text{eff}}(\theta)$ by
\[
e^{-1/g_\text{eff}(\theta)} g_\text{eff}^{k/2}(\theta) = e^\theta e^{-1/g} g^{k/2}.
\]
\subsection{generalisation to multiple $su(2)$}\label{app:multisu2}
The computations above can be easily generalized to $\prod_{i}su(2)_i$ defined by
\begin{equation}
\hat{T}_\mathcal{R} (\{g_i \}) := \Tr_\mathcal{R} \mathcal{P} \exp \left(i \int_0^{2\pi} d\sigma\ g_i t^a J^a_{i} (\sigma) \right)
\end{equation}
which admits the expansion
\begin{equation}
\hat{T}_{\mathcal{R}}\left(\left\{g_{i}\right\}\right) = \sum_{N=0}^{\infty} i^N \hat{T}_{\mathcal{R}}^{(N)}
\end{equation}
Generators in the mode expansion of the current satisfy Kac Moody algebra and commute if they belong to different $su(2)$ ,
\begin{equation}
\left[J_{i,n}^{a}, J_{j,m}^{b}\right]= \delta_{ij} (\sqrt{-1} f^{a b c} J_{i,n+m}^{c}+\kappa_i n \delta^{a b} \delta_{n+m, 0})
\end{equation}
After performing the integrals over $\sigma$, we get the operator $\boldsymbol{T}_{\mathcal{R}}^{(N)}$, which is to simply modify \eqref{eq:TbeforeNO} by summing over generators in different $su(2)$ factors, for example,
\begin{align}
&\hat{T}_\mathcal{R}^{(2)} = 2 \pi^2 \text{Tr}_\mathcal{R} (t^a t^b) \sum_{i,j} g_i g_j J_{i,0}^a J_{j,0}^b\\
&\hat{T}_\mathcal{R}^{(3)} = \frac{2 \pi^2}{3} \text{Tr}_\mathcal{R} (t^a t^b t^c) \sum_{i,j,k} g_i g_j g_k \Bigg[ \frac{\pi}{3} J_{i,0}^a J_{j,0}^b J_{k,0}^c + \sum_{n \neq 0} \frac{i}{n} J_{i,-n}^a J_{j,n}^b J_{k,0}^c e^{-|n|\epsilon/R} + \text{cyclic} + \text{reversal} \Bigg]
\end{align}
To demonstrate the computation, it is enough to take an example of $su(2)\times su(2)$. We renormalize $\hat{T}_\mathcal{R}^{(N)}$ up to $N=4$ in the same manner as in the last section\footnote{The results are too cumbersome to be presented here. Contact the author if you would like to grab a beer and drink over it.}, where renormalization is done by performing the following redefinition of the coupling
\begin{align}
&\begin{aligned}
g_1 \to & g_{1}+  g_{1}^{2}(-2 \log \epsilon + C_1) +  g_{1} g_{2} C_3 + g_{2}^{2} \mathrm{C}_5 + g_{1}^{3} \left(4 \log^{2} \epsilon -2 (k_1+2C_1) \log \epsilon+D_1\right)\\ +& 
g_{1}^{2} g_{2} (-4 C_3 \log \epsilon + D_3) + g_{1} g_{2}^{2}\big[-2(2C_5+k_2)\log\epsilon +D_5 \big]+ g_2^3 D_7+\dots
\end{aligned}\nonumber\\
&\begin{aligned}
g_2 \to & g_{2}+  g_{2}^{2}(-2 \log \epsilon + C_2) +  g_{1} g_{2} C_4 + g_{1}^{2} \mathrm{C}_6 + g_{2}^{3} \left(4 \log^{2} \epsilon -2 (k_2+2C_2) \log \epsilon+D_2\right)\\ +& 
g_{2}^{2} g_{1} (-4 C_4 \log \epsilon + D_4) + g_{2} g_{1}^{2}\big[-2(2C_6+k_1)\log\epsilon +D_6 \big]+ g_1^3 D_8+\dots
\end{aligned}
\end{align}
where $C_i$ and $D_i$ are arbitrary constants independent of the cutoff $\epsilon$. Beta function is then 
\begin{align}
	&\beta_{g_1}(g_1,g_2) = 2 \lambda g_1^2 +2 \lambda^2\big[ k_1 g_1^3 + C_3 g_1^2 g_2 -(C_3-2C_5-k_2)g_1g_2^2 -2 C_5 g_2^3\big] +\dots\\
	&\beta_{g_2}(g_1,g_2) = 2 \lambda g_2^2 +2 \lambda^2\big[ k_2 g_2^3 + C_4 g_1 g_2^2 -(C_4-2C_6-k_1)g_1^2g_2 -2 C_6 g_1^3\big] +\dots
\end{align}
As we discussed in Section \ref{sec:multichannelSU2}, DE/IM predicts that there exists a renormalization scheme such that beta functions are of the form
\begin{equation}
\beta_{g_1} = \frac{g_1^2}{1+ \frac{1}{2}\sum_{j}k_j g_j} = g_1^2 -\frac12\big[k_1 g_1^3 + k_2 g_1^2g_2  \big] + \mathcal{O}(g_1^4, g_1^3g_2,g_1^2g_2^2, g_1g_2^3, g_2^4)
\end{equation}
and a similar expression for $\beta_{g_2}$. This fixes the renormalization constants 
\begin{equation}
\lambda = -\frac12, \quad C_5 =C_6= 0, \quad C_3 = k_2,\quad C_4 = k_1
\end{equation}
The expectation value of $\hat{T}_n$ over WZW primary states can be easily computed 
\begin{align}
&\langle l_1,l_2|T_n(g_1,g_2,R)|l_1,l_2\rangle = n-8 \pi^{2} x_j \lambda^{2}\left(g_1^{2} \ell_1(1+\ell_1)+2 g_1 g_2 \ell_1 \ell_2+g_2^{2} \ell_2(1+\ell_2)\right)\nonumber\\
+ &8 \pi ^2 \lambda ^3 x_j \bigg[-2 C_1 g_1^2 l_1 (g_1 l_1+g_1+g_2 l_2) +g_1^3 k_1\bigg.\nonumber\\
&-2 g_2 (C_2 g_2 l_2 (g_1 l_1+g_2 l_2+g_2)+g_1 (g_1 l_1 (k_2+k_2 l_1+k_1 l_2)+g_2 l_2 (k_1+k_2 l_1+k_1 l_2))) \nonumber\\
&\bigg. +4 \log R \left(g_1^3 l_1 (l_1+1)+g_1^2 g_2 l_1 l_2+g_1 g_2^2 l_1 l_2+g_2^3 l_2 (l_2+1)\right)+g_2^3 k_2\bigg] + \cdots
\end{align}
where a WZW primary is labeled by two half integers $|l_1,l_2\rangle$. We do not show the full result here to fourth order in the total couplings $g_1, g_2$ but the full result in terms of slightly different coupling variables will be written in Appendix \ref{app:scropertMultiSU2}.

\section{Perturbative solutions of ODE}\label{app:schropert} 
\subsection{$SU(2)_k$ vacuum expectation value}
The Schrodinger equation for the vacuum expectation value of line defects in the $\mathfrak{su}(2)_k$ WZW model is
\[ e^{-2\theta} \partial_x^2 \psi(x) = (1 + g x)^k e^{2 x} \psi(x). \]
The T function
\[ T_n (\theta) = i \Big( \psi(x; \theta - \frac{i \pi n}{2}), \psi(x; \theta + \frac{i \pi n}{2}) \Big) \]
is defined as the Wronskian of the wavefunctions with shifted $\theta$ arguments; by general arguments this quantity is independent of $x$. The label $n$ in this section is equal to the dimension of the representation of $SU(2)$ and is related to the spin label $j$ in the direct perturbative line defect calculation by $n=2j+1$.

The differential equation can be rearranged such that it only depends on a particular combination of $g$ and $\theta$. Upon shifting $x$ by $-1/g$,
\[ g^{-k} e^{2/g} e^{-2\theta} \partial_x^2 \psi(x) = x^k e^{2 x} \psi(x), \]
so the wavefunction after the shift of x only depends on the combination $g^{k/2} e^{-1/g} e^{\theta}$. It is helpful to collect this quantity into an effecive coupling $g_{\text{eff}}(\theta)$ as
\[ g_{\text{eff}}(\theta)^{k/2} e^{-1/g_{\text{eff}}(\theta)} = g^{k/2} e^{-1/g} e^{\theta}, \]
admitting the $g$-expansion
\[ g_{\text{eff}}(\theta) = g + \theta g^2 + \theta (\theta - \frac{k}{2}) g^3 + \theta (\theta^2 - \frac{5}{4} k \theta + \frac{k^2}{4} ) g^4 + \cdots. \]

Since $T_n(\theta)$ is independent of $x$, the perturbative regime to be compared with the direct two-dimensional line defect computation can be characterized by the asymptotics of the wavefunction at large negative $x$. In this limit, the wavefunction $\psi(x;\theta)$ exhibits simple linear behavior in x and can be parametrized up to exponential corrections as
\[ \psi(x;\theta) \sim -Q(\theta) (x + \frac{1}{g}) - \tilde{Q}(\theta) \]
in terms of auxiliary functions $Q$, $\tilde Q$. This parametrization realizes the QQ relations for $T_n$:
\[ T_n(\theta) = i \bigg[ Q(\theta + \frac{i \pi n}{2}) \tilde{Q}(\theta - \frac{i \pi n}{2}) - Q(\theta - \frac{i \pi n}{2}) \tilde{Q}(\theta + \frac{i \pi n}{2}) \bigg]. \]
In the above, we included the shift of $x$ so that both $Q$ and $\tilde Q$ are functions of $g_{\text{eff}}(\theta)$ only. Then we can express $Q$ and $\tilde Q$ in the general form
\[ Q(\theta) = \frac{1}{\sqrt{\pi}} (1 + q_1 g_{\text{eff}}(\theta) + q_2 g_{\text{eff}}(\theta)^2 + \cdots ) \]
\[ \tilde{Q}(\theta) = \frac{1}{\sqrt{\pi}} (-\frac{1}{g_{\text{eff}}(\theta)} + \tilde{q}_0 + \tilde{q}_1 g_{\text{eff}}(\theta) + \cdots ) \]
such that the $\psi$ asymptotics receive $g_{\text{eff}}$-corrections to its slope as well as to its constants. In fact, normalizing the T function as $T_1 = 1$ determines the expansion coefficients of $Q$ in terms of that of $\tilde Q$. Doing so, it turns out that $T_n$ only depend on $\tilde{q}_i$ starting at $\mathcal{O}(g^4)$:
\[ T_n(\theta) = n - \frac{\pi^2}{12} k n(n^2-1) g^3 + \frac{\pi^2}{24} k n(n^2-1) (3 k - 6 \theta - 2 \tilde{q}_0 + 8 \tilde{q}_1) g^4 + \cdots.  \]
To determine the coefficients, we proceed with the systematic order-by-order solution of the Schrodinger equation.

Let us express the wavefunction $\psi(x;\theta)$ as a series $\psi = \sum_{i=0}^{\infty} g^i \psi^{{i}}$ in $g$ and perform a weak coupling expansion of the Schrodinger equation around UV fixed point $g=0$:
\[ \mathcal{O}(1): \qquad e^{-2\theta} \partial_x^2 \psi^{(0)} = e^{2 x} \psi^{(0)} \]
\[ \mathcal{O}(g): \qquad e^{-2\theta} \partial_x^2 \psi^{(1)} = e^{2 x} ( \psi^{(1)} + k x \psi^{(0)} ) \]
and so on. We only require up to $\mathcal{O}(g)$ to compare with the direct perturbative calculation. The ambiguities in the solutions $\psi^{(i)}$ coming from the integration constants are fixed by imposing that the solution decays exponentially and it does so in a very particular manner as to agree with the WKB asymptotics given in the main body of the text.

As explained in the main body of the text, the unique solution at $\mathcal{O}(1)$ satisfying these constraints is given in terms of a Bessel function
\[\psi^{(0)}(x;\theta) = \frac{1}{\sqrt{\pi}} K_0(e^{x + \theta}).\]
The large negative $x$ behavior of $\psi^{(0)}(x;\theta)$ is 
\[ \psi^{(0)}(x;\theta) \sim  -\frac{1}{\sqrt{\pi}} (x + \theta + \gamma - \log 2) \]
and so $\tilde{q}_0 = -\frac{k}{4} + \gamma - \log 2$.

The solution at $\mathcal{O}(g)$, up to an integration constant $c(\theta)$, is 
\begin{align*}
\psi^{(1)}(x;\theta) = &-\frac{k}{\sqrt{\pi}} \Bigg[ I_0(e^{x + \theta}) \int_x^\infty K_0(e^{x' + \theta})^2 x' e^{2(x' + \theta)} dx' \\
&+ K_0(e^{x + \theta}) \int_{c(\theta)}^{x}  K_0(e^{x' + \theta}) I_0(e^{x' + \theta}) x' e^{2(x' + \theta)} dx'  \Bigg]
\end{align*}
In the equation, $I_0(e^{x+\theta})$ diverges exponentially at large positive $x$, so upper limit of the first integral in the above has been chosen such that the solution decays exponentially in that limit. It is possible to fix the remaining constant $c(\theta)$ as at $\mathcal{O}(1)$ such that the total solution matches with the WKB asymptotics. However, a simpler way (which works at least at this order) is to notice that our asymptotic parametrization of $\psi$ in terms of $Q$, $\tilde Q$ picks out a coefficient multiplying $x$ which is constant and is in particular independent of $\theta$. At $\mathcal{O}(g)$, such a constant is equal to $-\frac{k}{4 \sqrt{\pi}}$. Note further that only $K_0(e^{x+\theta})$ contributes a term proportional to $x$ (more precisely, $-\frac{x}{\sqrt{\pi}}$) in the limit $x \to -\infty$. It follows that, according to our parametrization of the wavefunction, $c(\theta)$ must be chosen such that the integral multiplying $K_0(e^{x+\theta})$ is equal to $1/4$ at $x \to -\infty$. We can simply take such a condition to be the definition of $c(\theta)$, and this renders the precise form of $c(\theta)$ unnecessary.

The large negative $x$ behavior of $\psi^{(1)}$ is then
\[ \psi^{(1)}(x;\theta) \sim -\frac{k}{4 \sqrt{\pi}} (x - \theta - 2 - \gamma + \log 2) \]
and $\tilde{q}_1 = \frac{k^2}{32} - \frac{k}{2} (\frac{3}{2} + \gamma - \log 2)$.

Therefore, with our choice of parametrization at $x \to -\infty$, $T_n$ is
\[ T_n(\theta) = n - \frac{\pi^2}{12} k n(n^2-1) g^3 + \frac{\pi^2}{4} k n(n^2-1) (\frac{5}{8} k - \theta - 1 - \gamma + \log 2) g^4 + \cdots. \]
It is possible to choose a renormalization scheme in the direct two-dimensional calculation such that the vacuum expecation value of the line defect in the $SU(2)_k$ WZW model matches with the above result from the Schrodinger equation. Namely, a shift of the coupling as
\[ g \to \lambda g + \lambda^2 g^2 (-2\log \epsilon + k - 2 \gamma - 2 \log\pi) + \cdots\]
with $2 \pi R = e^\theta$ and the choice $\lambda = -1/2$, results in the above formula for $T_n$. Constants in the expectation value which are independent of $\theta$ can always be accounted for by trivial shifts of the coupling. However, it is still nontrivial that $\tilde{q}_1$ can directly be verified to be independent of $\theta$ and the precise $\theta$ dependence matches as this term is robust to local counterterms. Hirota bilinear relations are satisfied rather trivially at the level of the vev, as there is no nontrivial $n$ dependence apart from that coming from an overall Dynkin index factor. However, Hirota is nontrivial at the level of the expectations between primary states, which we now proceed to show.

\subsection{$SU(2)_k$ expectation value between primaries}
Based on evidence from existing literature \cite{Bazhanov:1998wj,dorey1999relation}, we propose that the Schrodinger equation
\[
\partial_x^2 \psi_l(x) = \Big[ e^{2\theta}e^{2x} (1+gx)^k + \frac{l(l+1)}{(x+1/g)^2} \Big] \psi_l(x)
\]
yields the solution whose Wronskian give rise to the T function $T_{n,l}(\theta) := \langle \hat{T}_n (\theta) \rangle_l = \langle l | \hat{T}_n (\theta) | l \rangle $ evaluated between primary states with level $l$. $T_{n,l}(\theta)$ are defined again in terms of the wavefunctions as
\[ T_n (\theta) = i \Big( \psi_l(x; \theta - \frac{i \pi n}{2}), \psi_l(x; \theta + \frac{i \pi n}{2}) \Big). \]

An asymptotic parametrization of $\psi_l(x;\theta)$ can be determined by analyzing the solutions to the degenerations of the Schrodinger equation at $x \to -\infty$ and at $g\to 0$, and then carefully matching the solutions in the regime of interest $1/g \gg -x \gg 0$. The resulting parametrization is
\[
\psi_l(x;\theta) \sim -\frac{x}{2l+1}\Big[ \frac{l+1}{g^l} Q_l(\theta) - g^{l+1} l \tilde{Q}_l(\theta) \Big] - \frac{1}{2l+1}\Big[ \frac{1}{g^{l+1}}Q_l(\theta) + g^l \tilde{Q}_l(\theta) \Big]
\]
where $Q_l$, $\tilde{Q}_l$ now gain an $l$-dependence in their powers of $g_{\text{eff}}(\theta)$ as
\begin{align*}
Q_l(\theta) &= \frac{g_{\text{eff}}(\theta)^l}{\sqrt{\pi}} (1 + q_{l,1} g_{\text{eff}}(\theta) + \cdots) \\
\tilde{Q}_l(\theta) &= \frac{g_{\text{eff}}(\theta)^{-l}}{\sqrt{\pi}} (-\frac{1}{g_{\text{eff}}(\theta)} + \tilde{q}_{l,0} + \tilde{q}_{l,1} g_{\text{eff}}(\theta) + \cdots ).
\end{align*}
This parametrization realizes the QQ relations for $T_{n,l}$, with an extra normalization constant:
\[ T_n(\theta) = \frac{i}{2l+1} \bigg[ Q_l(\theta + \frac{i \pi n}{2}) \tilde{Q}_l(\theta - \frac{i \pi n}{2}) - Q_l(\theta - \frac{i \pi n}{2}) \tilde{Q}_l(\theta + \frac{i \pi n}{2}) \bigg]. \]
Normalizing again as $T_{1,l}=1$, $T_{n,l}(\theta)$ now depends on $\tilde{q}_{l,i}$ starting at $\mathcal{O}(g^3)$:
\begin{align*}
T_{n,l}(\theta) = n &- \frac{\pi^2}{6}n(n^2-1)l(l+1) g^2 \\
&+ \frac{\pi^2}{12}n(n^2-1)\Big[ k(2l^2+l-1) - 4l\big( (l+1)\theta + \tilde{q}_{l,0} \big) \Big] g^3 \\
&- \frac{\pi^2}{360}n(n^2-1) \Big[ 45 k^2(l^2-1) -30k[(l+1)(8l-3)\theta+(4l-1)\tilde{q}_{l,0}] \\
&+ l(l+1)[7\pi^2 l^2+27\pi^2 l+26\pi^2-3(l^2+l+3)\pi^2n^2 + 180\theta^2] \\
&+ 120 l \tilde{q}_{l,0}(\tilde{q}_{l,0}+3\theta) + 120\tilde{q}_{l,1}(2l-1) \Big] g^4 + \cdots.
\end{align*}

Let us expand $\psi_l = \sum_{i=0}^\infty g^i \psi_l^{(i)}$ as before and obtain an order-by-order weak coupling expansion of the differential equation. As is easy to see, the $l$-dependent term drops out and we end up with the same equations up to $\mathcal{O}(g)$ as in the vev case:
\[ \mathcal{O}(1): \qquad e^{-2\theta} \partial_x^2 \psi_l^{(0)} = e^{2 x} \psi_l^{(0)} \]
\[ \mathcal{O}(g): \qquad e^{-2\theta} \partial_x^2 \psi_l^{(1)} = e^{2 x} ( \psi_l^{(1)} + k x \psi_l^{(0)} ) \]
and so on. Note that the equations do receive contributions from the $l$-dependent term starting at $\mathcal{O}(g^2)$, though we won't need them for the purposes of comparing to the line defect calculation. That the $\mathcal{O}(1)$, $\mathcal{O}(g)$ equations remain the same as the vev case indicates $\psi_{l}^{(0)}=\psi^{(0)}$ and $\psi_{l}^{(1)}=\psi^{(1)}$. The only difference then is the parametrization of the wavefunction at $x\to -\infty$ and thus the definition of the coefficients $\tilde{q}_{l,i}$. The resulting coefficients are
\begin{align*}
\tilde{q}_{l,0} &= -\frac{k}{4}+ (l+1)(\gamma - \log 2) \\
\tilde{q}_{l,1} &= \frac{1}{96} \Big[ 3k^2 - 8l(l+1)[ \pi^2 + 6(\gamma-\log 2)^2 ] -24k[3+2l+(l+2)(\gamma - \log 2)] \Big]
\end{align*}
and the expectation value is
\begin{align*}
T_{n,l}(\theta) = n &- \frac{\pi^2}{6}n(n^2-1)l(l+1) g^2 \\
&+ \frac{\pi^2}{12}n(n^2-1)\Big[ k(2l(l+1)-1) - 4l(l+1)(\theta+\gamma-\log 2) \Big] g^3 \\
&- \frac{\pi^2}{1440}n(n^2-1) \Big[ 45k^2(4l(l+1)-5) \\
&+ 4l(l+1)[ 7\pi^2 l(l+1) + 36\pi^2 - 3(l(l+1)+3)\pi^2 n^2 + 180(\theta + \gamma - \log 2)^2 ] \\
&+ 120k[ 3\theta +3 -\gamma(8l(l+1)-3) - 3\log 2 - 4l(l+1)(2\theta + 1 -2\log 2) ] \Big] g^4 + \cdots.
\end{align*}

Now we must verify from the results of the direct line defect calculation that (1) a renormalization scheme can be chosen such that the it matches the above $T_{n,l}(\theta)$ from the Schrodinger analysis and (2) the renormalized result satisfies Hirota bilinear relations. Both are nontrivial statements, respectively as local counterterms cannot depend on $\theta$ or $l$ and as $T_{n,l}(\theta)$ has a nontrivial $n$ dependence.

With some work, both (1) and (2) can be verified to hold, where the Schrodinger solution determines a unique renormalization scheme for the defect. The shift in the coupling
\begin{align*}
g \to \lambda g &+ \lambda^2 g^2 \Big[-2\log\epsilon + k - 2\gamma -2\log\pi \Big] \\
&+ \lambda^3 g^3 \Big[ 4 (\log\epsilon)^2 - 2(3k - 4\gamma - 4\log\pi)\log\epsilon + k^2 \\ 
&- 2k(2+3\gamma+3\log\pi) + \frac{\pi^2}{3}(2-n^2) + 4(\gamma+\log\pi)^2 \Big] + \cdots
\end{align*}
with $\lambda=-1/2$ in the perturbative calculation yields an expectation value matching $T_{n,l}(\theta)$, and the result satisfies Hirota.

\subsection{Multichannel expectation values} \label{app:scropertMultiSU2}

In the multichannel Kondo problem with $m$ channels, i.e. $\prod_{i=1}^{m}su(2)_{k_i}$, we are interested in studying the perturbative sector where all couplings $g_i$ with $i = 1,2,\cdots,n$ become small. Therefore a convenient thing to do is to have an overall small constant $g$ which encodes the scaling behavior of all couplings $g_i$ and parametrize the couplings as $g_i = \frac{g}{1+g z_i}$. The expansion can be done with respect to a single infinitesimal parameter $g$ and other finite parameters $z_i$ can be used to index the positions of the individual couplings.There should be, however, one constraint as there are now a total of $m+1$ parameters $(g,z_i)$. We take this to be, e.g. $\sum_i z_i = 0$. Note that we can invert the above relation to get $\frac{1}{g_i} = \frac{1}{g} + z_i$ or $\frac{1}{g} = \frac{1}{m}\sum_i \frac{1}{g_i}$ indicating that the (inverse of) $g$ is the mean of (inverses of) $g_i$.

We propose that the Schrodinger equation for the ground states of the $m$-channel Kondo problem is
\[
\partial_x^2 \psi(x) = \Big[ e^{2\theta + 2 x} \prod_{i=1}^{m} (1+ g_i x)^{k_i} + u(x)^2 - \partial_x u(x) \Big]\psi(x)
\]
where
\[
u(x) = \sum_{i=1}^{m} \frac{l_i}{x + 1/g_i}.
\]
Another choice of $u$, namely $\tilde{u}(x) = \sum_{i=1}^{m} \frac{-l_i-1}{x + 1/g_i}$, works as well but we proceed with $u$ rather than $\tilde{u}$.

Substituting for $g_i$ as described above and shifting $x \to x - 1/g$, one gets
\[
\partial_x^2 \psi(x) = \Bigg[ e^{2\theta} e^{-2/g} g^{\sum_i k_i} \prod_{i}(1+g z_i)^{-k_i} e^{2 x} \prod_{i}(x+z_i)^{k_i} + \Big( \sum_i \frac{l_i}{x + z_i} \Big)^2 + \sum_{i} \frac{l_i}{(x+z_i)^2} \Bigg]\psi(x).
\]
This indicates that the solutions only depend on $z_i$ and the following combination which can be absorbed into an effective coupling $g_{\text{eff}}(\theta)$:
\[
e^{-1/g_{\text{eff}}(\theta)} g_{\text{eff}}(\theta)^{\sum_i k_i/2} \equiv e^{\theta} e^{-1/g} g^{\sum_i k_i/2} \prod_{i}(1+g z_i)^{-k_i/2}.
\]
The effective coupling $g_{\text{eff}}(\theta)$ now depends on $m$ as well as $\theta$. Its expansion in $g$ is
\[
g_{\text{eff}}(\theta) = g + \theta g^2 + \theta(\theta-\frac{1}{2}\sum_i k_i)g^3 + \frac{1}{4}\Big[ \theta(4\theta-\sum_i k_i)(\theta-\sum_i k_i) + \sum_i z_i^2 \Big] g^4 + \cdots.
\]

The $g$-expansion of the Schrodinger equation to $\mathcal{O}(g)$ does not depend on $u$. Hence the asymptotic solutions $\psi^{(0)}$ and $\psi^{(1)}$ of $\psi = \sum_i g^i \psi^{(i)}$ are equal to that for the vevs of the single-channel Kondo problem, with $k$ substituted for $\sum_i k_i$. 

For the rest of this subsection, we focus on the case $m=2$, or $su(2)_{k_1} \times su(2)_{k_2}$, for simplicity. As was done for the single-channel primaries, the asymptotic parametrization of the multichannel solution $\psi$ in terms of $Q$,$\tilde{Q}$ can be determined by analyzing the large negative $x$ limit of the Schrodinger equation, i.e. the limit where only the $u$-dependent terms survive, and then considering the solution in the regime $1/g \gg -x \gg 0$. This yields the parametrization
\begin{align*}
\psi(x;\theta) \sim -\frac{1}{2(l_1 + l_2)+1} \Bigg\{ &\frac{Q(\theta)}{g^{l_1 + l_2 + 1}} \Big[ 1 + g x \Big( 1 + 2 (l_1 + l_2) - \frac{l_1}{1 + g z_1} - \frac{l_2}{1 + g z_2} \Big) \Big] \\
&+ \tilde{Q}(\theta) g^{l_1 + l_2} \Big[ 1 - g x \Big( \frac{l_1}{1 + g z_1} + \frac{l_2}{1 + g z_2} \Big) \Big] \Bigg\}
\end{align*}
with Q-functions
\[
Q(\theta) = \frac{g_{\text{eff}}(\theta)^{l_1+l_2}}{\sqrt{\pi}} ( 1 + q_1 g_{\text{eff}}(\theta) + \cdots )
\]
\[
\tilde{Q}(\theta) = \frac{g_{\text{eff}}(\theta)^{-l_1-l_2}}{\sqrt{\pi}} ( -\frac{1}{g_{\text{eff}}(\theta)} + \tilde{q}_0 + \tilde{q}_1 g_{\text{eff}}(\theta) + \cdots ).
\]
The multichannel function $T_n(\theta)$ is defined similarly as for the single-channel primaries, with the replacement $l \to l_1 + l_2$.

As before, the normalization $T_1 = 1$ expresses $q_i$ in terms of $\tilde{q}_i$ and perturbative solutions $\psi^{(0)}$, $\psi^{(1)}$ can be compared with the asymptotic parametrization to obtain the coefficients
\begin{align*}
\tilde{q}_0 =& - \frac{k_1 + k_2}{4} + (l_1 + l_2 + 1)(\gamma - \log 2)\\
\tilde{q}_1 =& \frac{1}{96} \Big[ 3 (k_1^2 + k_2^2) - 24(k_1 + k_2)[(\gamma - \log 2 + 2)(l_1 + l_2 + 2)-1] \\
&+ 6 k_1 k_2 - 8 (l_1 + l_2)(l_1 + l_2 + 1)[\pi^2 + 6(\gamma - \log 2)^2]\Big]
\end{align*}
These coefficients suffice to determine the multichannel $T_n(\theta)$ function
\begin{align}
    T_n(\theta) &= n - g^2 \Big[\frac{1}{6} n (n^2-1) \pi ^2 (l_1+l_2) (l_1+l_2+1)  \Big] \nonumber \\
    &+ g^3 \Big[ \frac{1}{12} \pi ^2 n (n^2-1) (k_1 (2 l_1^2+(4 l_2+2) l_1+2 l_2 (l_2+1)-1) \nonumber \\
    &+k_2 (2 l_2+2 (l_1^2+2 l_2 l_1+l_1+l_2^2)-1)-4 (l_1+l_2) (l_1+l_2+1) (t+\gamma -\log 2)) \Big] \nonumber \\
    &- g^4 \Big[ \frac{\pi ^2}{1440}n (n^2-1) [-30 k_1 (4 (l_1+l_2) (-3 k_2 (l_1+l_2+1)+8 (l_1+l_2+1) t \nonumber \\
    &+4 (l_1+l_2-(l_1+l_2+1) \log (4)))+15 k_2+16 l_1+16 l_2 \nonumber \\
    &+4 \gamma  (8 l_2+8 (l_1^2+2 l_2 l_1+l_1+l_2^2)-3)-12 t-12+\log (4096)) \nonumber \\
    &-120 k_2 (4 (l_1+l_2) (2 (l_1+l_2+1) t+l_1+l_2-2 (l_1+l_2+1) \log (2))+4 l_1+4 l_2 \nonumber \\ 
    &+\gamma  (8 l_2+8 (l_1^2+2 l_2 l_1+l_1+l_2^2)-3)-3 t-3+\log (8)) \nonumber \\
    &+45 k_1^2 (4 l_1^2+(8 l_2+4) l_1+4 l_2 (l_2+1)-5)+45 k_2^2 (4 l_1^2+(8 l_2+4) l_1+4 l_2 (l_2+1)-5) \nonumber \\
    &+4 ((l_1+l_2) (l_1+l_2+1) (\pi ^2 (-3 (l_1^2+2 l_2 l_1+l_1+l_2^2+l_2+3) n^2+7 (l_1+l_2)^2 \nonumber \\
    &+7 l_1+7 l_2+36)+180 (t+\gamma -\log (2))^2 ] \Big].
\end{align}
Comparing with the perturbative defect calculations, the following renormalization scheme with the shifts
\begin{align*}
g_1 &\to \lambda g_1 + \lambda^2 g_1^2 [-2\log\epsilon - 2 \gamma + k_1 - 2\log\pi - 2 z_1] + \lambda^2 g_1 g_2 [k_2] \\ 
&+ \lambda^3 g_1^3 [4(\log\epsilon)^2 - 2(-4\gamma + 3k_1 - 4\log\pi - 4 z_1)\log\epsilon + D_1] \\
&+ \lambda^3 g_1^2 g_2 [-4k_2 \log\epsilon + D_3] + \lambda^3 g_1 g_2^2 [-2k_2 \log\epsilon + D_5] + \lambda^3 g_2^3 [D_7] \\
g_2 &\to \lambda g_2 + \lambda^2 g_2^2 [-2\log\epsilon - 2 \gamma + k_2 - 2\log\pi - 2 z_2] + \lambda^2 g_2 g_1 [k_1] \\ 
&+ \lambda^3 g_2^3 [4(\log\epsilon)^2 - 2(-4\gamma + 3k_2 - 4\log\pi - 4 z_2)\log\epsilon + D_2] \\
&+ \lambda^3 g_2^2 g_1 [-4k_1 \log\epsilon + D_4] + \lambda^3 g_2 g_1^2 [-2k_1 \log\epsilon + D_6] + \lambda^3 g_1^3 [D_8]
\end{align*}
with $\lambda = -1/2$ and the conditions
\begin{align*}
D_1 + D_3 + D_5 + D_7 =& -4 k_1 z_1-2 k_2 z_1-2 k_2 z_2+\gamma  (-6 k_1-6 k_2+8 (z_1+\log \pi )) \\
&+k_1^2+2 k_2 k_1-4 k_1+k_2^2-4 k_2-6 k_1 \log \pi -6 k_2 \log \pi \\
&-\frac{1}{3} \pi ^2 \left(n^2-2\right)+4 z_1^2+2 z_1+2 z_2+8 z_1 \log \pi +4 \gamma ^2+4 \log ^2\pi \\
D_2 + D_4 + D_6 + D_8 =& -4 k_2 z_2-2 k_1 z_2-2 k_1 z_1+\gamma  (-6 k_2-6 k_1+8 (z_2+\log \pi )) \\
&+k_2^2+2 k_1 k_2-4 k_2+k_1^2-4 k_1-6 k_2 \log \pi -6 k_1 \log \pi \\
&-\frac{1}{3} \pi ^2 \left(n^2-2\right)+4 z_2^2+2 z_2+2 z_1+8 z_2 \log \pi +4 \gamma ^2+4 \log ^2\pi 
\end{align*}
matches multichannel Schrodinger analysis and the result satisfies Hirota bilinear relations. In the above, the two auxiliary variables $z_1$ and $z_2$ can be identified by the condition $z_1 + z_2 =0$.

\section{Lie algebra conventions}\label{appendix:LieAlgebra}
We will follow the convention from \cite{DiFrancesco:1997nk,Bachas:2004sy}. We choose orthonormal basis $\{t^a\}$, namely Killing form $K(t^a,t^b) = \delta^{a,b}$, so adjoint indices can be raised and lowered freely. Note that we define the Killing form  with a normalization constant,
\begin{equation}
K(X,Y) \equiv \frac{1}{h^\vee \psi^2} \Tr(\mathrm{ad}X \mathrm{ad}Y)
\end{equation}
so that, 
\begin{equation}
\sum_{a, b} f^{a b c} f^{a b d}=h^{\vee} \psi^{2} \delta^{a b}
\end{equation}
where $\psi^2$ is the length squared of the longest root, which account for the arbitrary normalization of the generators. We will choose $\psi^2=2$, unless otherwise stated, and structure constant $f^{abc}$ is defined in 
\begin{equation}
[t^a,t^b] = if^{abc}t^c
\end{equation}

Using the definitions above, we are ready to list some useful identities for $\mathfrak{su}(2)$. Representations of $\mathfrak{su}(2)$ are labelled by nonnegative half integer $j$, denoted as $R_j$, we have the following
\begin{align}
    \Tr_{R_j}(t^a t^b) &= I_{R_j} \delta^{ab}, \\
    \Tr_{R_j}\left(t^{a} t^{b} t^{c}\right) &=\frac{i}{2} f^{a b c} I_{R_j}\\
    \Tr_{R_j}\left(t_{a} t_{b} t_{c} t_{d}\right) &= \frac{1}{2}\alpha_{j} I_{R_j}\left(\delta_{a b} \delta_{c d}+\delta_{a d} \delta_{b c}\right)+ \frac12 \beta_{j} I_{R_j}\left(\delta_{a c} \delta_{b d}\right),
\end{align}
with
\begin{align}
    & C_2(R_j) = j(j+1) \psi^2, \quad \mathrm{dim}(R_j) = 2j+1\\
    & I_{R_j} = \frac13 j(j+1)(2j+1)\psi^2, \quad f_{abc} = \sqrt{2} \epsilon_{abc}\\
    &\alpha_{j}=\frac{4}{5}\left(j(j+1)+\frac{1}{2}\right), \quad \beta_{j}=\frac{4}{5}(j(j+1)-2).
\end{align}

\bibliographystyle{JHEP}

\bibliography{mono}

\end{document}